\documentclass[%
 reprint,longbibliography,
 amsmath,amssymb,
 aps,superscriptaddress
]{revtex4-1}

\usepackage{graphicx}
\usepackage{dcolumn}
\usepackage{bm}
\usepackage{hyperref}
\usepackage{xcolor}
\usepackage{amsmath, amssymb, amsfonts}
\usepackage{mathtools}
\usepackage{subfigure}
\usepackage{mathrsfs}
\usepackage{multirow}

\definecolor{purple1}{rgb}{128,0,128}

\newcommand{\bea}{\begin{eqnarray}}
\newcommand{\ea}{\end{eqnarray}}

\begin{document}

\preprint{APS/123-QED}

\title{Quantum metrology with ultracold chemical reactions}
\author{Seong-Ho Shinn}

\affiliation{%
  Seoul National University, Department of Physics and Astronomy, Center for Theoretical Physics, 08826 Seoul, Korea}%

\author{Uwe R. Fischer}
\email[]{Corresponding author. E-mail uwerfi@gmail.com}
\affiliation{%
Seoul National University, Department of Physics and Astronomy, Center for Theoretical Physics, 08826 Seoul, Korea}%

\author{Daniel Braun}

\affiliation{%
Eberhard-Karls-Universit\"at T\"ubingen, Institut f\"ur Theoretische Physik, 72076 T\"ubingen, Germany}

\date{\today}

\begin{abstract}
\end{abstract}

\maketitle

\noindent{\bf 
  Chemical chain reactions are known to enable extremely sensitive detection schemes in chemical, biological, and medical analysis, and have even been used in the search for dark matter.   Here we show that coherent, ultracold chemical reactions harbor great potential for quantum metrology:  In an atom-molecule Bose-Einstein condensate (BEC), a weak external perturbation can modify the reaction dynamics and lead to the coherent creation of molecules in an atom-dominant regime which can be selectively detected with modern spectroscopic techniques.  This
  promises to substantially improve the viability of 
  previously proposed BEC-based 
  sensors for acceleration, gravitational waves, and other physical quantities, including the detection of dark matter, 
  that so far relied on the detection of the tiny density modulations caused by the creation of single phonons. 
}

Chemical chain reactions, such as the polymerase chain reaction that has become popular in COVID-19 tests, have developed to one of the most important, highly sensitive analysis and detection schemes not only in medicine, but more generally in biological and chemical analysis \cite{augspurger_chemical_2018}.  Within physics, a ``DNA detector'' for dark matter particles was proposed \cite{drukier_new_2014,ohare_particle_2021}.  However, the potential of {\em ultracold coherent chemical reactions} for quantum metrology has not been studied so far. 

The conventional picture of chemical reactions being due to (classical) collisions gets modified drastically at ultralow temperatures.
Specifically, when bosonic atoms and molecules lose their individual character and condense in coherent matter waves
\cite{Balakrishnan2016,NatureChemReview,Drummond1998,Superchemistry}, chemical reactions can lead to coherent macroscopic, non-linear oscillatory dynamics even for simple reactions such as $A+A \to A_2$ for which no classical chemical oscillations are possible.  
Such a collectively enhanced ``superchemistry"  was demonstrated experimentally by coherent Rabi oscilations between atomic and molecular fields driven by magnetic Feshbach resonances or Raman transitions \cite{Wynar}.  
The collective chemical reactions of bosonic matter fields lead to a novel type of quasiparticle, which we 
coin 
the {\it reacton}. It is analogous to phonons in single-component BECs, but represents  an elementary excitation of a quantized time- and space-dependent chemical reaction field 
with vector 
character 
that fully describes the reaction dynamics, rather than a single scalar density perturbation.
Its composition in terms of atomic and molecular amplitudes and the corresponding holes can be largely tuned through system parameters.

In what follows, we show that in this regime counting individual molecules facilitates the detection of small 
  perturbations applied to the ultracold gas. 
  We demonstrate that a lower bound of the 
  classical Fisher information from counting molecules approaches the quantum Fisher information which sets the upper bound on the sensitivity with which the perturbation parameter can be estimated \cite{HELSTROM1967,Braunstein94,Holevo1982}.
Corresponding sensors have been proposed e.g.~for measuring acceleration \cite{PhysRevD.89.065028}, the detection of gravitational waves (GW) \cite{sabin_phonon_2014,Dennis,Robbins_2019,singh_detecting_2017}, for gravimeters and measuring the gravitational field gradient on a millimetre scale \cite{bravo_phononic_2020}, and the detection of dark matter \cite{howl_quantum_2021}. 
However, the classical Fisher information for specific and realizable measurements, and hence the sensitivity in practical realizations for these proposals, is unknown so far.  Only the theoretically achievable maximum sensitivity based on the quantum Fisher information (QFI) was established.   
Ultimately, these proposed sensors require measuring density modulations from a small number of phonons in a BEC, but 
single phonon detection in condensates has been achieved experimentally so far only in the 
superfluid helium II \cite{Harris}
and there is no report of achieving single phonon detection in BECs yet. 
Since, on the other hand, molecules in a BEC can be detected on a single-particle level with spectroscopic   
techniques \cite{Julienne,Zwierlein} 
(the coherent formation of single molecules was demonstrated in \cite{He}), 
our scheme presents an 
important step forward
towards the realization of quantum sensors based on density perturbations in a condensate.

Molecular condensates were demonstrated experimentally via magnetic Feshbach resonances and photoassociation~\cite{RevModPhys.78.1311,RevModPhys.78.483,Yan2013}, and coherent oscillations between atomic and molecular condensates were observed~\cite{Donley2002,Chin,zhang_many-body_2022}. 
Effective quantum field theories were proposed~\cite{Drummond1998,Javanainen1999,Javanainen2000,Naidon2008,Richter_2015} for photoassociation and direct reactions, 
leading to predictions of non-linear effects known from non-linear optics 
such as solitons~\cite{Drummond1998}, and classical wave chaos~\cite{Dey2020}. 

To set the stage of our analysis, 
we first develop the full quantum field theory of ultracold chemical reactions. Most 
calculations  are delegated to the Supplemental Material.\\

\vspace{0.5cm}
\noindent{\bf Quantized chemical reactions}\\
In the Heisenberg picture, 
the reaction $A+A \to A_2$ of two ultracold bosonic atoms $A$ 
in a condensate
to an  ultracold bosonic molecule $A_2$ is described in quantum field theory by the Hamiltonian~\cite{Javanainen1999,Javanainen2000,Richter_2015} 
\begin{eqnarray}
\hat{H} && \; = 
\int d^3 r \; 
\hat{\psi}^{\dagger}_{a} \left( \boldsymbol{r}, t \right) 
\left\{ 
- \frac{\hbar^2}{2 m_{a}} 
\nabla^2 
+ V_{a} \left( \boldsymbol{r}, t \right) 
\right\} 
\hat{\psi}_{a} \left( \boldsymbol{r}, t \right) 
\nonumber\\
&& 
+ 
\frac{g_{a}}{2} 
\int d^3 r \; 
\hat{\psi}^{\dagger}_{a} \left( \boldsymbol{r}, t \right) 
\hat{\psi}^{\dagger}_{a} \left( \boldsymbol{r}, t \right) 
\hat{\psi}_{a} \left( \boldsymbol{r}, t \right) 
\hat{\psi}_{a} \left( \boldsymbol{r}, t \right) 
\nonumber\\
&& + 
\int d^3 r \; 
\hat{\psi}^{\dagger}_{m} \left( \boldsymbol{r}, t \right) 
\left\{ 
- \frac{\hbar^2}{2 m_{m}} 
\nabla^2 
+ V_{m} \left( \boldsymbol{r}, t \right) 
+ \epsilon 
\right\} 
\hat{\psi}_{m} \left( \boldsymbol{r}, t \right) 
\nonumber\\
&& 
+ 
\frac{g_{m}}{2} 
\int d^3 r \; 
\hat{\psi}^{\dagger}_{m} \left( \boldsymbol{r}, t \right) 
\hat{\psi}^{\dagger}_{m} \left( \boldsymbol{r}, t \right) 
\hat{\psi}_{m} \left( \boldsymbol{r}, t \right) 
\hat{\psi}_{m} \left( \boldsymbol{r}, t \right) 
\nonumber\\
&& + 
g_{am} 
\int d^3 r \; 
\hat{\psi}^{\dagger}_{a} \left( \boldsymbol{r}, t \right) 
\hat{\psi}_{a} \left( \boldsymbol{r}, t \right) 
\hat{\psi}^{\dagger}_{m} \left( \boldsymbol{r}, t \right) 
\hat{\psi}_{m} \left( \boldsymbol{r}, t \right) 
\nonumber\\
&& 
+ 
\frac{\alpha}{\sqrt{2}} 
\int d^3 r \; 
\left\{ 
\hat{\psi}^{\dagger}_{m} \left( \boldsymbol{r}, t \right) 
\hat{\psi}_{a} \left( \boldsymbol{r}, t \right) 
\hat{\psi}_{a} \left( \boldsymbol{r}, t \right) 
+ 
h.c. 
\right\} ,
\label{Ham}
\end{eqnarray}
where $\hat{\psi}_{a} \left( \boldsymbol{r}, t \right)$ and $\hat{\psi}_{m} \left( \boldsymbol{r}, t \right)$ are 
position- and time-dependent 
annihilation operators of 
a bosonic atom and molecule, respectively. 
Furthermore, 
$m_{a}$ and $m_{m} \left( \simeq 2 m_{a} \right)$ are atom and molecule mass, 
respectively; $V_{a} \left( \boldsymbol{r}, t \right)$ and $V_{m} \left( \boldsymbol{r}, t \right)$ are the trap potentials for atoms and molecules, and
$g_{a}$, $g_{m}$, and $g_{am}$ are
 the contact-interaction 
couplings between two atoms, between two molecules, and between atoms and molecules, respectively. 
Finally, the 
 parameter $\epsilon$ denotes the energy difference between two 
atoms and 
a molecule, $\alpha$ is the 
coupling coefficient that determines the coherent conversion rate of two atoms to a 
molecule, and $h.c.$ means hermitian conjugate. 
Eq.~\eqref{Ham} is a real-space version of the experimentally realizable 
momentum-space hamiltonian of photo-association for vanishing photon-momentum \cite{Javanainen1999,Javanainen2000}.

The operator for the total number of atoms (including the ones bound in molecules),
\begin{eqnarray}
\!\!\!\!\!\!\!\!\!
\hat{N} \coloneqq 
\!\int \! d^3 r \; 
\left\{ 
\hat{\psi}^{\dagger}_{a} \left( \boldsymbol{r}, t \right) 
\hat{\psi}_{a} \left( \boldsymbol{r}, t \right) 
+ 
2 
\hat{\psi}^{\dagger}_{m} \left( \boldsymbol{r}, t \right) 
\hat{\psi}_{m} \left( \boldsymbol{r}, t \right) 
\right\}\,,
\label{hat_N_def}
\end{eqnarray}
commutes with the Hamiltonian in Eq.~\eqref{Ham},  
is not explicitly time-dependent, and hence conserved; we denote by  
$N$ 
the average value of $\hat N$.

\vspace{0.5cm}
\noindent{\bf Mean-field description}\\ 
We expand $\hat{\psi}_{j} \left( \boldsymbol{r}, t \right) = \psi_{j} \left( \boldsymbol{r}, t \right) + \delta \hat{\psi}_{j} \left( \boldsymbol{r}, t \right)$ where $\psi_{j} \left( \boldsymbol{r}, t \right) \coloneqq \left\langle \hat{\psi}_{j} \left( \boldsymbol{r}, t \right) \right\rangle$ is the mean-field value of 
$\hat{\psi}_{j} \left( \boldsymbol{r}, t \right)$ for $j = a, m$, 
for small corrections $\delta \hat{\psi}_{j} \left( \boldsymbol{r}, t \right) \ll \psi_{j} \left( \boldsymbol{r}, t \right)$ where most of atoms and 
molecules are condensed, respectively. 
For a box trap \cite{Navon}, 
where $V_{a} \left( \boldsymbol{r}, t \right) = 0$ and $V_{m} \left( \boldsymbol{r}, t \right) = 0$, let $V$ be the system volume and $n \coloneqq N / V$ the 
total number density. 
$\psi_{a} \left( \boldsymbol{r}, t \right) = \left\vert \psi_{a} \left( t \right) \right\vert 
\textrm{exp} \left\lbrack i \left\{ \varphi_{a} \left( t \right) - \mu t / \hbar \right\} \right\rbrack
$ 
and $\psi_{m} \left( \boldsymbol{r}, t \right) = \left\vert \psi_{m} \left( t \right) \right\vert 
\textrm{exp} \left\lbrack i \left\{ \varphi_{m} \left( t \right) - 2 \mu t / \hbar \right\} \right\rbrack$ 
satisfy the mean field Eqs.~\eqref{psi_eqofmot} \footnote{Equation numbers in the form 
(S$\cdots$) refer to equations in the Supplement},  
and the mean-field Hamiltonian $H_{0}$ reads
\begin{eqnarray}
H_{0} && \; \left( t \right) / \left( N g_{a} n \right) = 
\frac{1}{2} x^4 \left( t \right) 
+ \frac{1}{2} \tilde{g}_{m} y^4 \left( t \right) 
+ \tilde{\epsilon} y^2 \left( t \right) 
\nonumber\\
&& 
+ \tilde{g}_{am} x^2 \left( t \right) y^2 \left( t \right) 
+ \tilde{\alpha} x^2 \left( t \right) y \left( t \right) 
\cos \varphi_{am} \left( t \right) , 
\quad 
\label{H0_const}
\end{eqnarray}
where we have normalized the couplings as follows, 
$\tilde{g}_{m} \coloneqq g_{m} / g_{a}$, 
$\tilde{g}_{am} \coloneqq g_{am} / g_{a}$, 
$\tilde{\alpha} \coloneqq \alpha \sqrt{2 n} / \left( g_{a} n \right)$, 
$\tilde{\epsilon} \coloneqq \epsilon / \left( g_{a} n \right)$, 
$\varphi_{am} \left( t \right) \coloneqq \varphi_{m} \left( t \right) - 2 \varphi_{a} \left( t \right) $, 
$x \left( t \right) \coloneqq \left\vert \psi_a \left( t \right) \right\vert / \sqrt{n}$, 
and $y \left( t \right) \coloneqq \left\vert \psi_m \left( t \right) \right\vert / \sqrt{n}$. 
Note that Eq.~\eqref{H0_const} is minimized for
$\alpha \cos \varphi_{am} \left( t \right) = - \left\vert \alpha \right\vert$ since $x^2 \left( t \right) \ge 0$.

We use below in \eqref{Bog_H_2nd} a Bogoliubov expansion to go beyond mean field theory. 
In order to not mask the effects of the time-dependent perturbation (whose amplitude we want to probe),  by a time-dependent mean-field background, we impose vanishing mean-field Josephson oscillations between atoms and molecules \cite{Lin}  
(cf.~Eq.~\eqref{psi_eqofmot} in the Supplement). This can be achieved as follows: 
From Eq.~\eqref{varphi_eqofmot}, $\varphi_{am} \left( t \right)$ can be set to be a constant,
 and hence, when $\alpha \cos \varphi_{am} \left( t \right) = - \left\vert \alpha \right\vert$, from Eq.~\eqref{xy_eqofmot}, $x \left( t \right)$ and $y \left( t \right)$ are also constant in time. 
Rendering $\varphi_{am} \left( t \right)$ constant in time with $\alpha \cos \varphi_{am} = - \left\vert \alpha \right\vert$ is possible 
by tuning $\tilde\epsilon$  to 
\begin{eqnarray}
\tilde{\epsilon} = && \; 
2 x^2
+ \tilde{g}_{am} \left( 2 y^2 - x^2 \right) 
- \tilde{g}_{m} y^2 
- \frac{\left\vert \tilde{\alpha} \right\vert}{2} \left( 4 y - \frac{x^2}{y} \right) . 
\quad \;
\label{epsilon_constraint}
\end{eqnarray}
Then $\varphi_{a} \left( t \right)$ and $\varphi_{m} \left( t \right)$ become constant in time, and 
by minimizing $H_0 - \mu N$, $x$ and $y$ follow from Eq.~\eqref{eq_y} and $x = \sqrt{1 - 2 y^2}$ from Eq.~\eqref{hat_N_def} (see supplemental material~\ref{hom_xy_gen_sol} for derivations). 
For simplicity, we will set $\varphi_{a} = 0$ and $\tilde{g}_{m} = 0$ in what follows.

\begin{figure} [b]
\centering
\includegraphics[width=0.49\textwidth]{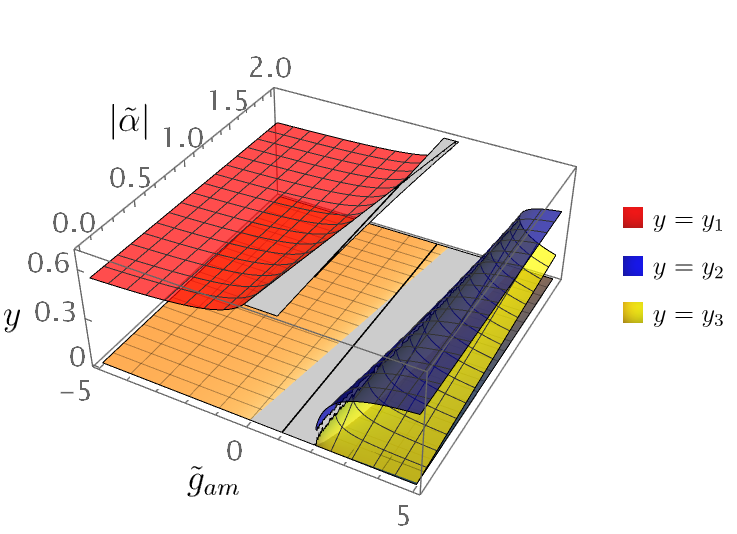}
\caption{Physical values of $y$ ($y\in\mathbb R$ and $0 \le y \le 1 / \sqrt{2}$) as a function of $\tilde{g}_{am}$ and $\left\vert \tilde{\alpha} \right\vert$. 
$y_j$ is the $j$-th solution of Eq.~\eqref{eq_y}. 
For $\tilde{g}_{am} < 0$, $y_1$ is the physical value of $y$, whereas for $\tilde{g}_{am} > 1$, 
$y_2$ or $y_3$ are physical values of $y$. 
In the gray zones, no physical solution of $y$ exists 
(upper zone: $y > 1 / \sqrt{2}$ and $x$ is imaginary; lower zone: no real solutions $x$ and $y$ exist). 
The orange plane is a projection of the physical 
values of $y_1$. 
} 
\label{y-range-fig}
\end{figure}

Fig~\ref{y-range-fig} shows the physical parameter regimes of $\tilde{g}_{am}$ and $\left\vert \tilde{\alpha} \right\vert$, 
defined by $x,y\ge 0$, hence $0 \le y \le 1 / \sqrt{2}$. 
We focus on the atom-dominated  case, so for $\tilde{g}_{am} > 1$, we choose $y_3$, 
the third solution of Eq.~\eqref{eq_y}. For $\tilde{g}_{am} < 0$, Fig.~\ref{y-range-fig} shows that only the first solution $y_1$ of Eq.~\eqref{eq_y} is physical and hence we choose that one.

\vspace{0.5cm}
\noindent{\bf Quasiparticle excitations: Reactons}\\
A Bogoliubov expansion~\cite{Uedareview} of
 the grand canonical Hamiltonian corresponding to \eqref{Ham} 
 in Fourier space gives (up to $O \left( \delta \hat{\Psi}^3_{j} \right)$)
\begin{eqnarray}
&& \hat{H} - \mu \hat{N} = 
H_0 - \mu N 
- \frac{g_{a} n}{2} 
\sum_{\boldsymbol{k} \neq 0} 
\left\{ 
M_{11} \left( k \right) + M_{22} \left( k \right) 
\right\} 
\nonumber\\
&& \quad 
+ \frac{g_{a} n}{2} 
\sum_{p = 1, 2} 
\sum_{\boldsymbol{k} \neq 0} 
\left\lbrack 
\begin{array}{c}
\tilde{\omega}_p \left( k \right) 
\hat{b}^{\dagger}_p \left( \boldsymbol{k}, t \right) 
\hat{b}_p \left( \boldsymbol{k}, t \right) 
\qquad \quad 
\\
+ 
\tilde{\omega}^{*}_p \left( k \right) 
\hat{b}_p \left( - \boldsymbol{k}, t \right) 
\hat{b}^{\dagger}_p \left( - \boldsymbol{k}, t \right) 
\end{array}
\right\rbrack 
,  
\label{Bog_H_2nd}
\end{eqnarray}
where $\tilde{\omega}_p \left( k \right) \coloneqq \hbar \omega_p \left( k \right) / \left( g_{a} n \right)$, 
$\tilde{\omega}^{*}_p \left( k \right)$ is the complex conjugate of $\tilde{\omega}_p \left( k \right)$, 
and we label $p=1,2$ such that 
$\textrm{Re} 
\left\{ 
\omega_2 \left( 0 \right) 
\right\} 
\ge 
\textrm{Re} 
\left\{ 
\omega_1 \left( 0 \right) 
\right\}$. 
Note that Eq.~\eqref{Bog_H_2nd} gives 
$\hat{b}_{p} \left( \boldsymbol{k}, t \right) 
= 
\exp \left\lbrack 
- i 
\textrm{Re} 
\left\{ 
\omega_p \left( k \right) 
\right\} t 
\right\rbrack 
\hat{b}_{p} \left( \boldsymbol{k}, 0 \right)$, 
which shows that our Bogoliubov Hamiltonian in Eq.~\eqref{Bog_H_2nd} is constant in time $t$.

The $\hat{b}_p \left( \boldsymbol{k}, t \right)$ are bosonic 
annihilation operators of the $p$th Bogoliubov mode 
($p = 1, 2$), 
i.e.~linear combinations of the Fourier transforms of $\delta \hat{\psi}_a \left( \boldsymbol{r}, t \right)$ and $\delta \hat{\psi}_m \left( \boldsymbol{r}, t \right)$ 
and their hermitian conjugates 
via Eq.~\eqref{Bog_trans}. 
These linear combinations are
``reaction fields'' whose time-evolution due to
the coherent, collective chemical reactions determines the dynamics of the spatiotemporal dependence of the atomic and molecular concentrations. 
The $\hat b_p\left( \boldsymbol{k}, t \right)$ describe quantized excitations of the 
reaction fields, and $M_{11}(k), M_{22}(k)$ are defined in Eqs.~\eqref{M11_hom_gen} and \eqref{M22_hom_gen}.

The dimensionless eigenvalues $\tilde{\omega}_p \left( k \right)$ of the $p$th Bogoliubov mode define the two dispersion branches of reactons. They come in pairs of opposite signs, where, however, only the positive-frequency branches are physically meaningful \cite{Uedareview}. 
We additionally define the dimensionless gap $\Delta \tilde{\omega} \left( 0 \right) \coloneqq \tilde{\omega}_2 \left( 0 \right) - \tilde{\omega}_1 \left( 0 \right)$ for convenience.

\begin{figure} [b]
\begin{minipage}{0.5\textwidth}
\centering
(a)
\includegraphics[width=0.415\textwidth]{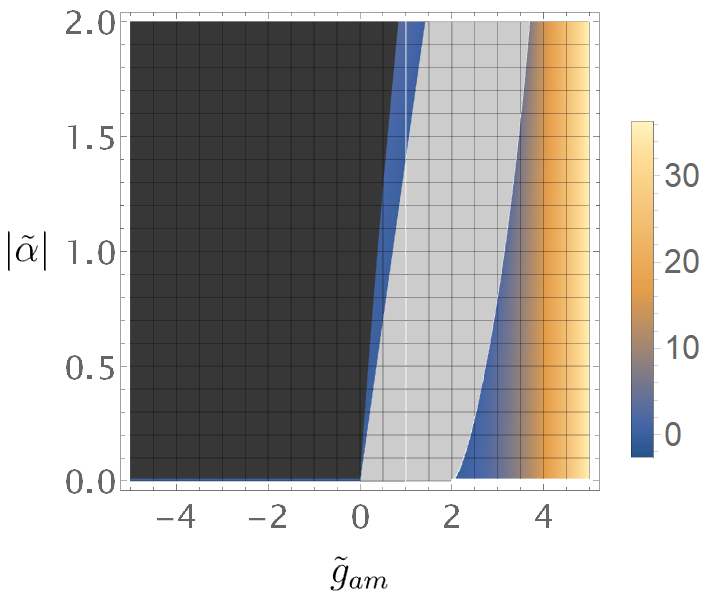}
\quad 
(b)
\includegraphics[width=0.415\textwidth]{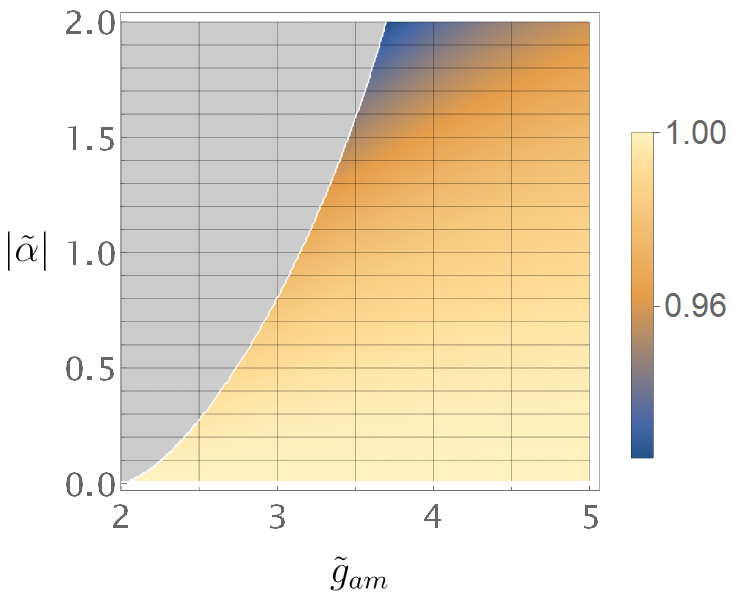}
\end{minipage}
\begin{minipage}{0.25\textwidth}
\centering
(c)
\includegraphics[width=0.8\textwidth]{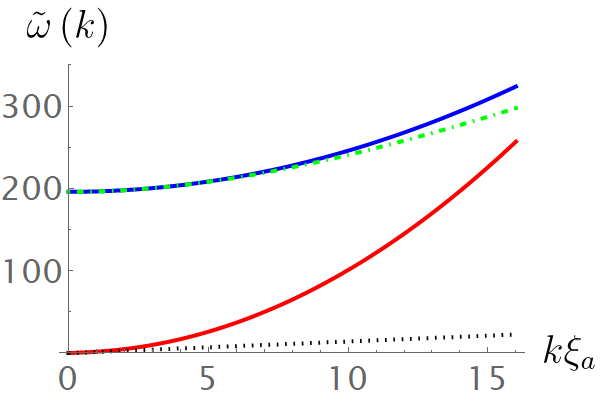}
\end{minipage}
\quad 
\begin{minipage}{0.2\textwidth}
\centering
(d) 
\begin{ruledtabular}
\begin{tabular}{cc}
Parameters
&
Value
\\
\hline
$\tilde{g}_{am} \coloneqq g_{am} / g_a$ & 100 \\
$\tilde{g}_{m} \coloneqq g_m / g_a$ & 0 \\
$\tilde{\alpha} \coloneqq \alpha \sqrt{2n} / \left( g_a n \right)$ & 1 \\
$y \coloneqq \left\vert \psi_m \right\vert / \sqrt{n}$ & 0.0026 \\
$L / \xi_a$ & 161.3 \\
$c_s$ (in mm/s) & 4.23
\end{tabular}
\end{ruledtabular}
\end{minipage}
\caption{
\label{w_2D-fig}
(a)
: Excitation gap $\Delta \tilde{\omega} \left( 0 \right)$ as a function of $\tilde{g}_{am}$ and $\left\vert \tilde{\alpha} \right\vert$. The frequency $\tilde{\omega}_{1} \left( k \right)$ becomes imaginary in the black regions, whereas $y$ does not have physical values in the grey regions. 
The gap increases with $\tilde{g}_{am}$ and $\left\vert \tilde{\alpha} \right\vert$. 
(b): $c_r / c_s$ as a function of $\tilde{g}_{am}$ and $\left\vert \tilde{\alpha} \right\vert$. 
(c)
: Dispersion relations $\tilde{\omega}_p \left( k \right)$ as a function of $k \xi_{a}$ of the two reacton species: 
Massless reacton (red continuous line) $\tilde{\omega}_1$, massive reacton (blue 
line) $\tilde{\omega}_2$, with 
expansion of $\tilde{\omega}_1 \left( k \right)$ up to $O \left( k \right)$ (black dotted line), 
and expansion of $\tilde{\omega}_2 \left( k \right)$ up to ${O \left( k^2 \right)}$
(green dot-dashed line);
$\tilde{g}_{am} = 100$ and $\tilde{\alpha} = 1$, 
which renders $y = 0.0026$. 
Here, $\xi_{a} = \hbar / \sqrt{2 m_{a} g_{a} n}$ is the atomic healing length. 
(d) Parameters used in our numerical calculations. Our system is a 1D BEC with length $L$. 
}
\end{figure}

Using the approach of~\cite{Lin} facilitates analytic expressions for $\tilde{\omega}_{p} \left( k \right)$ in Eqs.~\eqref{eigenvalue_omega_p}.
We  find that $\tilde{\omega}_{p} \left( k \right)$ depends on $\left\vert \tilde{\alpha} \right\vert$, and 
expand $\tilde{\omega}_{p} \left( k \right)$ up to $O \left( k^2 \right)$ in Eqs.~\eqref{omega_12_expression}. 
Two important features reveal themselves ($A_1$, $B_1$, $B_2$, and $B_3$ are defined in 
Eqs.~\eqref{A1_def} to~\eqref{B3_def}): 
(1) $\tilde{\omega}_{1} \left( 0 \right)$ becomes imaginary if $A_1 < 0$. 
(2) When $A_1 \ge 0$, $\tilde{\omega}_{1} \left( k \right) = k \xi_{a} \sqrt{B_2 - B_3 / \left( 4 A_1 \right)}$. 
Thus, if $B_2 - B_3 / \left( 4 A_1 \right) < 0$, $\tilde{\omega}_{1} \left( k \right)$ becomes purely imaginary. 
Fig~\ref{w_2D-fig} (a) shows the gap $\Delta \tilde{\omega} \left( 0 \right)$, where grey regions represent that the gap has nonzero imaginary value and black regions that $\tilde{\omega}_{1} \left( k \right)$ becomes imaginary. In both regions, the system is unstable and we will thus exclude them. 

Sources of damping for collective excitations in single-component BECs are 
Landau damping, which is dominant at higher temperatures and is $\propto k_B T$ for 
$k_B T \gg m {c_s^2}$,  
and Beliaev damping at low temperatures ($k_B T \ll m {c_s^2}$) and 
if $\hbar k \ll m {c_s}$ ~\cite{Chung_2009}.
For $\hbar \omega_1 \left( k
\right)$ linear in $k$  (corresponding to 
Nambu-Goldstone modes), 
in a large range, Landau and Beliaev dissipation and quasiparticle damping are negligible ~\cite{Nagao2016}. We will  focus 
on an atom-rich regime, and hence expect that these damping mechanisms can also be neglected in our system. 

For $\omega_{p} \left( k \right) \in \mathbb R$, since 
$i \partial \hat{b}_{p} \left( \boldsymbol{k}, t \right) / \partial t 
= \omega_{p} \left( k \right) \hat{b}_{p} \left( \boldsymbol{k}, t \right)$, 
from Eq.~\eqref{Bog_H_2nd}, if $\omega_{1} \left( k \right) = {c_r} k + O \left( k^2 \right)$, 
we have 
$k^2 \hat{b}_{1} \left( \boldsymbol{k}, t \right) 
+ \left( 1 / c^2_r \right) \partial^2 \hat{b}_{1} \left( \boldsymbol{k}, t \right) / \partial t^2 
= O \left( k^3 \right)$, 
which is a wave equation for $\hat{b}_1 \left( \boldsymbol{k}, t \right)$ 
in the small $k$ limit, with the reacton propagation speed $c_r$. 
We have $c_r = \left( \xi_{a} g_{a} n / \hbar \right) \sqrt{B_2 - B_3 / \left( 4 A_1 \right)} 
= \sqrt{B_2 - B_3 / \left( 4 A_1 \right)} \sqrt{g_{a} n / \left( 2 m_{a} \right)}$, which leads to 
$c_r / c_s = \sqrt{ \left\{ B_2 - B_3 / \left( 4 A_1 \right) \right\} / 2}$ shown in Fig.~\ref{w_2D-fig} (b). 

Since our Hamiltonian in Eq.~\eqref{Bog_H_2nd} is constant in time, 
from the Bogoliubov expansion in Eq.~\eqref{Bog_trans},  
using the notation of Eq.~\eqref{Bog_matrix_eig}, the annihilation operator 
for reacton
branch $p$ can be decomposed as  
\begin{eqnarray}
\hat{b}_p \left( \boldsymbol{k}, t \right) = && \; 
u^{*}_{p1} \left( k \right) 
\delta \hat{\Psi}_{a} \left( \boldsymbol{k}, t \right) 
- 
v^{*}_{p1} \left( k \right) 
\delta \hat{\Psi}^{\dagger}_{a} \left( - \boldsymbol{k}, t \right) 
\nonumber\\
&& 
+ 
u^{*}_{p2} \left( k \right) 
\delta \hat{\Psi}_{m} \left( \boldsymbol{k}, t \right) 
- 
v^{*}_{p2} \left( k \right) 
\delta \hat{\Psi}^{\dagger}_{m} \left( - \boldsymbol{k}, t \right) 
. \quad 
\label{BogoExpansion}
\end{eqnarray}
We define a cutoff $k_c$ and choose parameters $\tilde{g}_{am}$ and $\tilde{\alpha}$ such that 
$\omega_{1} \left( k \right) \ll \omega_{2} \left( 0 \right)$ for $k < k_c$ implying that excitations of the second reacton branch can be neglected, and such that for $k > k_c$,  $v_{11} \left( k \right)$ and $v_{12} \left( k \right) \rightarrow 0$.  
Then in summations involving $v_{11} \left( k \right)$ or $v_{12} \left( k \right)$ over all $k$, the effect of the massive reacton branch is negligible. 
From now on, we set for concreteness $\tilde{g}_{am} = 100$ and $\tilde{\alpha} = 1$ where $\tilde{\omega}_{1} \left( k_c \right) = 0.1 \tilde{\omega}_{2} \left( 0 \right) = 19.6$ with $k_c \xi_{a} = 4.32$. 
Then the mean-field state is an ``atom-rich'' state as 
$y^2 = 6.5 \times 10^{-6}$. 
For these 
and $\varphi_{a} = 0$ and $\alpha \cos \varphi_{am} = - \left\vert \alpha \right\vert$, 
the eigenvectors $\left\lbrack \begin{array}{cccc} u_{p1} \left( k \right) & u_{p2} \left( k \right) & v_{p1} \left( k \right) & v_{p2} \left( k \right) \end{array} \right\rbrack^{T}$ 
are all real.

\begin{figure} [t]
\centering
\includegraphics[width=0.49\textwidth]{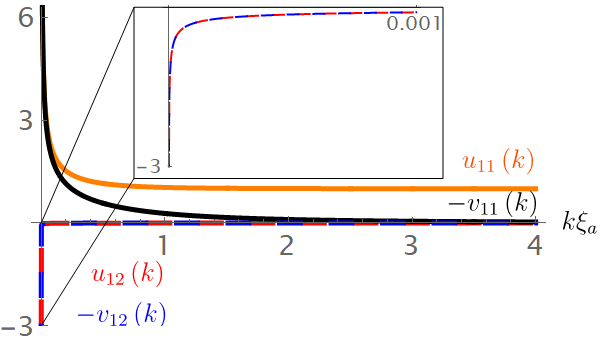}
\caption{
\label{u1_v1_gam_100-fig}
Bogoliubov mode amplitudes in Eq.~\eqref{BogoExpansion}, 
$u_{11}$ (Orange line), $- v_{11}$ (Black line), $u_{12}$ (Red dashed), and $- v_{12}$ (Blue dashed) as a function of $k \xi_{a}$ at $\tilde{g}_{am} = 100$ and $\tilde{\alpha} = 1$. 
Inset: $u_{12}$ (Red dashed) and $- v_{12}$ (Blue dashed) with $0 < k \xi_{a} < 0.001$.
As $k \rightarrow 0$, $u_{1p} / \left( - v_{1p} \right) \rightarrow 1$, which implies that atomic (or molecular) creation- and annihilation operators contribute equally to $\hat{b}_1$ near $k = 0$. 
}
\end{figure}

Fig.~\ref{u1_v1_gam_100-fig} shows that for $k < k_c$, $\hat{b}_{1}$ is more ``atomic'' in the sense that the coefficient of $\delta \hat{\Psi}_{a}$ is bigger than that of $\delta \hat{\Psi}_{m}$, i.e.~also in the reacton excitations atoms dominate over molecules.  
Also, $u_{1q} \left( k \right) \simeq - v_{1q} \left( k \right)$ for $q = 1, 2$, which means that $\hat{b}_{1}$ has approximate particle-hole
symmetry near $k = 0$ as for a 
single-component BEC.

\begin{figure}[b]
\centering
(a) 
\subfigure{\includegraphics[width=0.19\textwidth]{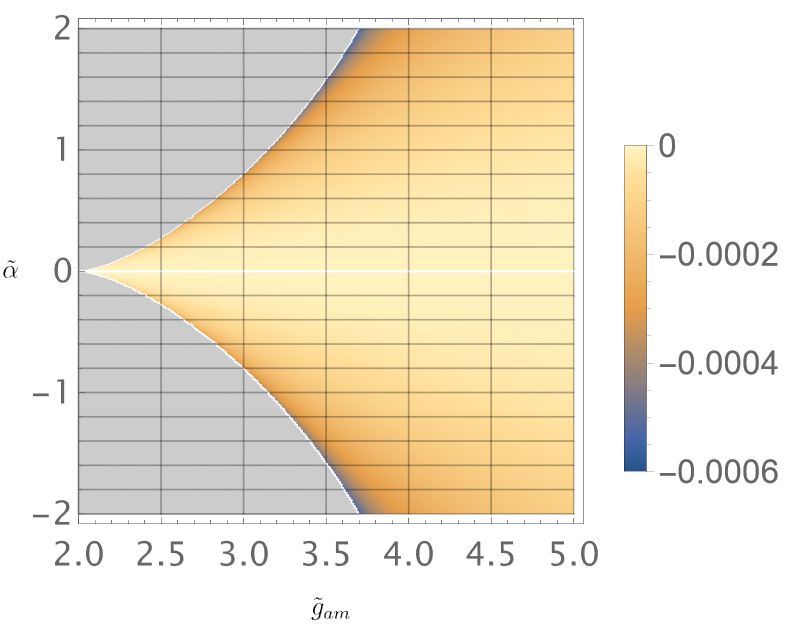}}
(b) 
\subfigure{\includegraphics[width=0.19\textwidth]{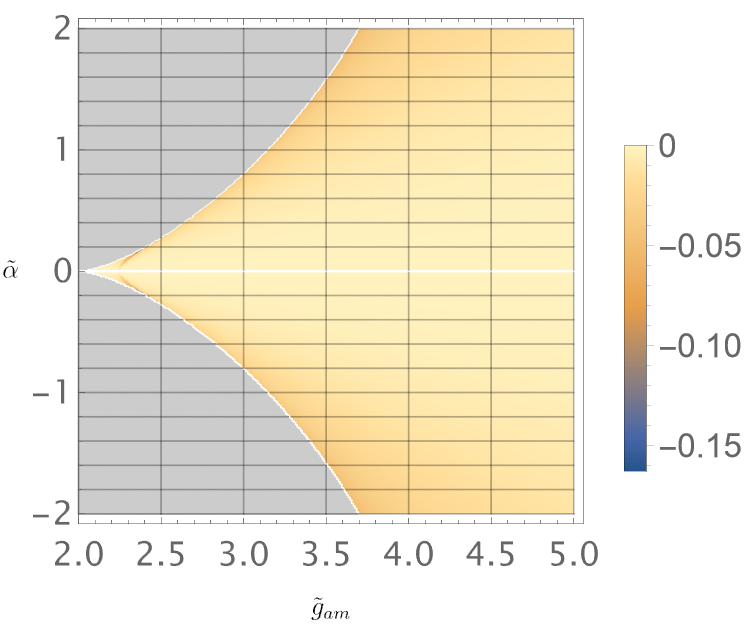}}
\caption{
\label{reaction-fig}
$\alpha \Xi_{13} \left( k \right)$ which determines reaction rates (see Eqs.~\eqref{op_R_up_to_2nd},~\eqref{reaction_squeezed_coherent},~\eqref{Reaction_pairstate}) as a function of $\tilde{g}_{am}$ and $\tilde{\alpha}$. 
Left figure is for $k = 2 \pi / L$, and right figure is for $k = 20 \pi / L$ with parameters in the table (d) of~\ref{w_2D-fig}. 
$y$ does not have physical values in the grey regions. 
}
\end{figure}

In the Supplement~\ref{Reaction_Op}, we show that 
creation and annihilation of pairs of reactons, 
$\hat{b}^{\dagger}_1 \left( \boldsymbol{k}, t \right) 
\hat{b}^{\dagger}_1 \left( - \boldsymbol{k}, t \right)$ and 
$\hat{b}_1 \left( \boldsymbol{k}, t \right) 
\hat{b}_1 \left( - \boldsymbol{k}, t \right)$ over 
any nonzero $\boldsymbol{k}$ converts atoms to molecules and vice versa. This can be formalized in a reaction rate operator $\hat{R}_2 \left( t \right)$ 
defined in Eq.~\eqref{op_R_up_to_2nd}, which shows that $\alpha \Xi_{13} \left( k \right)$ 
determines the reaction rate, up to the expectation values of $\hat{b}^{\dagger}_1 \left( \boldsymbol{k}, t \right) 
\hat{b}^{\dagger}_1 \left( - \boldsymbol{k}, t \right) 
- 
h. c. $ that depend on the specific quantum state of the reactons. $\Xi_{13}$ is defined in \eqref{Xi_def_12}.
Fig.~\ref{reaction-fig} shows that $\alpha \Xi_{13} \left( k \right)$ is 
symmetric 
in $\tilde{\alpha}$ and its absolute value increases as $k$ increases. 
It also increases as one approaches  the boundary of the grey regions where $y$ does not have physical values.

\vspace{0.5cm}
\noindent{\bf 
Metrology with quantized chemical reactions
}\\

We now 
demonstrate the metrological usefulness of ultracold chemical reactions
for which small molecule concentrations can be detected \cite{Zwierlein}. 
To that end, we compare a lower bound of the classical Fisher information (CFI) based on molecule counting with the quantum Fisher information (QFI) for various small, 
time-dependent perturbations 
$\hat{V} \left( t \right)$ 
turned on at time $t = 0$. 
As until $t = 0$ our Hamiltonian is time-independent (see Eq.~\eqref{Bog_H_2nd}), we use the interaction picture with state $\left\vert \Psi_{I} \left( t \right) \right\rangle$ and initial state $\left\vert \Psi_{I} \left(0 \right) \right\rangle = \left\vert \textrm{vac} \right\rangle$   to calculate the response to $\hat{V} \left( t \right)$. 
Here, $\left\vert \textrm{vac} \right\rangle$ is the Bogoliubov vacuum, 
$\hat{b}_{1, I} \left( \boldsymbol{k}, t \le 0 \right) \left\vert
  \textrm{vac} \right\rangle = 0$, and $\hat{b}_{1, I} \left(
  \boldsymbol{k}, t \right)$ are the Bogoliubov operators in the interaction picture.  Hence, 
before time-dependent perturbations are applied, the system has lowest energy up to $O \left( \delta \hat{\Psi}^3_j \right)$ (see Eq.~\eqref{Bog_H_2nd}).

We calculate the QFI and a lower bound to the CFI derived in \cite{Stein2014}; for 
details see the Supplement~\ref{CFI_QFI_Formalism}. 
The inverse
of the QFI defines the smallest uncertainty with which a parameter can
be estimated, optimized over all possible (``positive operator-valued
measure'', POVM) measurements and data analysis schemes with unbiased
estimators. This so-called quantum Cram\'er-Rao bound \cite{Rao1945,Helstrom1968} 
can be 
saturated in the limit of a large number of measurements and hence 
constitutes an important fundamental benchmark for the sensitivity 
with which a parameter can be estimated 
that is in principle 
achievable once all technical 
noise 
has been eliminated and only noise inherent in the quantum state remains. 
While our scheme is general, 
we illustrate it at the hand of one example: the amplitude of a metric perturbation that arises e.g.~from a gravitational wave on top of a flat Minkowski space-time background.

A metric perturbation $h_{\xi\xi}(\xi,\eta,\zeta,t)$ on top of a flat Minkowski spacetime, created by a gravitational wave propagating in $\zeta$-direction, leads to a Newtonian 
tidal 
force in $\xi$ direction on non-relativistic particles of mass $m$ given by (see eq.(8.2) in \cite{maggiore2008gravitational}) 
$F_\xi=(m/2)\xi \ddot h_{\xi\xi}$, where $h$ is 
in transverse traceless gauge, $\bm r=(\xi,\eta,\zeta)$ and time $t$  are coordinates in the proper detector frame 
of a free-falling observer, and a dot signifies time-derivative.  In the simplest mono-chromatic case, $h_{\xi\xi}=\tilde{h}_{\xi\xi}\sin(k_\zeta \zeta-\omega_\text{GW} t)$. 
The force derives from a potential $U=-(m/4)\xi^2\ddot h_{\xi\xi}$, which translates to an interaction hamiltonian between the GW and the BEC located at $\zeta=0$ given in second quantization by 
$\hat{V}_S \left( t \right)$ with 
\begin{eqnarray}
-\frac{
{L^2} 
\omega^2_\text{GW}}{48}\tilde{h}_{\xi\xi}\sin(\omega_\text{GW} t) 
\sum_{l=a, m} m_l 
\sum_{\bm k\ne 0}
\delta\hat{\psi}_l^\dagger(\bm k) \delta\hat{\psi}_l(\bm k) 
, \qquad 
\label{eq:hintgw}
\end{eqnarray}
where we assumed a 1D BEC aligned in $\xi$ direction, 
and approximated $(1/L)\int_{-L/2}^{L/2}d\xi\,e^{i(k_\xi'-k_\xi)\xi}\xi^2\simeq L^2\delta_{k_\xi,k_{\xi}'}/12$, where 
$L$ is the length of the quasi-1D BEC.
The latter approximation means that the GW couples purely to mass density and scattering of reactons due to the spatial structure of the GW is neglected on the length-scale of the BEC. The parts linear in $\delta \hat\psi_l(\bm r,t)$ vanish 
as we set our background mean-field to be homogenenous in space (i.e.~it has only a $\bm k=0$ component). 

The perturbation in Eq.~\eqref{eq:hintgw} applied to the system with box trap potentials satisfies $V_m \left( t \right) = 2 V_a \left( t \right)$ and hence, from Eqs.~\eqref{xy_eqofmot} and~\eqref{varphi_eqofmot}, the 
background mean field remains homogeneous in space. 
One can write $\hat V_S(t)$ as a ``density perturbation'' of the general form 
$\hat{V}_{S} \left( t \right)=
V_{ex} f \left( t \right) \int d^3 r \; 
\left\{ 
\delta\hat{\psi}^{\dagger}_{a} \left( \boldsymbol{r} \right) 
\delta\hat{\psi}_{a} \left( \boldsymbol{r} \right) 
+ 
2 
\delta\hat{\psi}^{\dagger}_{m} \left( \boldsymbol{r} \right) 
\delta\hat{\psi}_{m} \left( \boldsymbol{r} \right) 
\right\} 
$ with 
$V_{ex} = - L^2 \omega^2_\text{GW} \tilde{h}_{\xi\xi} m_a / 48$
and $f(t)=\sin(\omega_\text{GW} t)$. 
Throughout, we assume that the perturbations are small such that $V_{ex} \ll \hbar \omega_{1} \left( k \right)$ for $k \neq 0$.

A large class of sensing applications, including the one 
mentioned above, is covered by 
the following generic perturbation, bilinear in the lower reacton branch basis,
which acts on the system from $t = 0$: 
\begin{eqnarray}
\!\!\!\!\!\!\!\!\!\!\!\!\!
\hat{V}_{S} \left( t \right) & = & 
V_{ex} f \left( t \right) 
\sum_{\boldsymbol{k} \neq 0} 
\mathbb{V}_1 \left( k \right) 
\hat{b}^{\dagger}_{1} \left( \boldsymbol{k} \right) 
\hat{b}_{1} \left( \boldsymbol{k} \right) 
\nonumber\\
&& + 
V_{ex} f \left( t \right) 
\sum_{\boldsymbol{k} \neq 0} 
\left\{ 
\mathbb{V}_2 \left( k \right) 
\hat{b}^{\dagger}_{1} \left( \boldsymbol{k} \right) 
\hat{b}^{\dagger}_{1} \left( - \boldsymbol{k} \right) 
+ 
h.c. 
\right\} 
, \label{Vst}
\end{eqnarray}
where $V_{ex}$, $f \left( t \right)$, 
and 
$\mathbb{V}_1 \left( k \right)$ 
are real.  $V_{ex} > 0$ has units of energy and we define the dimensionless $\tilde{V}_{ex} \coloneqq V_{ex} / \left( g_{a} n \right)$.  Also, for periodic perturbations with angular frequency $\omega_a$, we set $\tilde{\omega}_{a} \coloneqq \hbar \omega_{a} / \left( g_{a} n \right)$.  We assume $\left\vert f \left( t \right) \right\vert \le 1\; \forall\,t$ 
and focus on a 1D system with parameters given in table (d) in 
Fig.~\ref{w_2D-fig}
using values of a ${}^{87} \textrm{Rb}$ BEC experiment~\cite{Burt1997} 
and define $k_{n_1} \coloneqq 2 \pi n_1 / L$ where 
the summation over 
$k$ becomes a summation over 
integers $n_1$. 
We define the cutoff $n_c$ 
such that $\tilde{\omega}_{1} \left( k_{n_c} \right) = 0.1 \tilde{\omega}_{2} \left( 0 \right)$ in order not to excite any massive reactons, as this would complicate the dynamics substantially.  
Since $k_c \xi_{a} = 4.32$, the closest integer $n_c$ is 110 where $k_{n_c} = 4.28$, which satisfies $v_{11} \left( k_{n_c} \right)$ and $v_{12} \left( k_{n_c} \right) \rightarrow 0$, cf.~
left panel of Fig.~\ref{u1_v1_gam_100-fig}. 

\begin{figure} [t]
\centering
\subfigure{
\includegraphics[width=0.45\textwidth]{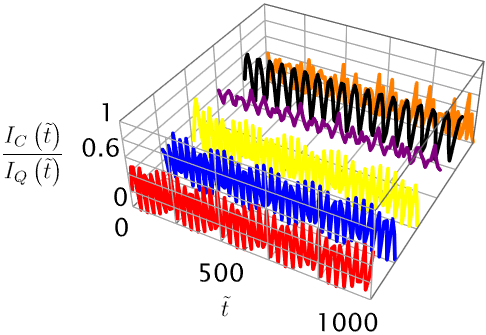}}
\caption{
\label{pert_CFI_QFI-fig}
The lower bound of the CFI ($I_{C} \left( \tilde{t} \right)$) over QFI ($I_{Q} \left( \tilde{t} \right)$) for density perturbations as a function of $\tilde{t}$ with perturbation profiles $f \left( t \right) = \sin \left( \omega_a t \right)$ for $\tilde{t} \ge 0$ (Red line), 
$f \left( t \right) = \cos \left( \omega_a t \right)$ for $\tilde{t} \ge 0$ (Blue line), 
$f \left( t \right) = \left\{ 1 - \cos \left( \omega_a t \right) \right\} / 2$ for $\tilde{t} \ge 0$ (Yellow line), 
$f \left( t \right) = \left\{ 1 + \cos \left( \omega_a t \right) \right\} / 2$ for $\tilde{t} \ge 0$ (Purple line), 
$f \left( t \right) = \theta \left( t \right)$ for $\tilde{t} \ge 0$ (Black 
line), 
and
$f \left( t \right) = \delta \left( t \right)$ (Orange 
line). 
}
\end{figure}

We give the result for the QFI and the CFI based on molecule counting for a general perturbation \eqref{Vst} in 
Eqs.~\eqref{QFI_symplectic} and \eqref{CFI_symplectic}. 
In Fig.~\ref{pert_CFI_QFI-fig}
we show the ratio of the lower bound $I_C(\tilde t)$ of the CFI and the QFI $I_Q(\tilde t)$ for 6 different temporal perturbation profiles for density perturbations 
when estimating $\left\vert \tilde{V}_{ex} \right\vert$. 
In the Supplement~\ref{comp_Dennis_sec}, we show that our QFI calculation for $t \gg 1$ and the perturbation \eqref{eq:hintgw} is consistent with~\cite{Dennis} when there are no BEC molecules in the system. 
While the functional form of the ratio $I_C(\tilde t)/I_Q(\tilde t)$ depends on the form of the signal, the most important message is that CFI and QFI are on the same order of magnitude. 
Counting molecules therefore constitutes a close-to-optimal measurement for 
homogeneous density perturbations. 
This is 
a central result of this paper that goes 
beyond previous proposals of sensors based on density perturbations in BECs~\cite{PhysRevD.89.065028,sabin_phonon_2014,Dennis,Robbins_2019,singh_detecting_2017,bravo_phononic_2020,howl_quantum_2021} for which so far no practical measurement schemes with corresponding sensitivities were known. 

\vspace{0.5cm}
\noindent{\bf Conclusion and outlook}\\ 
In summary, we explored quantized chemical reactions in ultracold gases of atoms and molecules 
for their use in quantum metrology. Both the atom/molecule ratio in the mean-field ground state and the composition of the ``reacton'' quasiparticles, the quantized excitation of the fields that describe the chemical reactions in space and time, can be widely tuned with the parameters of the system. We identified an atom-rich regime, where the mean-field dynamics is shut off, 
and the creation of molecules dominated by the response to an external perturbation. 
We showed that a measurement of the number of molecules in this regime is very close to reaching maximum possible sensitivity for 
perturbations that couple to the total density of atoms (including the molecules). 
Since single-molecule counting can be achieved with established spectroscopic techniques, our scheme offers an attractive alternative to proposed quantum metrological schemes that require the detection of single phonons in BECs \cite{sabin_phonon_2014,quantumlab}, which is an extremely challenging task experimentally \cite{Ralf,ThreeBody}. 
With spatially resolved molecule detection one might even be able to identify the modes in which the molecules are created and hence acquire additional information about an external perturbation of the system.

Numerous other avenues of quantum metrology based on ultracold chemistry 
can be explored, such as 
the exploration of unstable regions, or more complex reactions. Atomic analogues of superconducting transition-edge detectors \cite{edge} might be envisaged, and chemical chain reactions in the regime of ultracold quantum gases explored (see also \cite{fiderer2018} for the benefits of chaotic dynamics for metrology). 
Quantized reaction rates could also be studied on a 
self-consistent level \cite{Alon}, creating an avenue to many-body 
quantum chemical oscillations and their metrology beyond Bogoliubov theory.

\vspace*{1em}

\noindent{\bf Acknowledgments}\\
The work of SHS 
was supported by the National Research Foundation of Korea (NRF), Grant No. NRF-2015-033908 (Global PhD Fellowship Program). SHS also acknowledges 
the hospitality of the University of T\"ubingen during his stay in the spring and summer of 2021. 
URF has been supported by the NRF under Grant No.~2017R1A2A2A05001422 
and Grant No.~2020R1A2C2008103. 
\\

\vspace*{1em}
\noindent{\bf Author contributions}\\
SHS did main calculations (details in the supplemental material), part of numerical calculations, and made figures.

URF defined major research directions, provided the cold atoms knowledge and Bogoliubov theory background, including (a) the clarification of analogies and differences to two-component BECs (b) established the vanishing mean-field oscillations condition, (c) identified the necessary form of the Bogoliubov spectrum and (d) identified 
key publications for the Bogoliubov theory of atom-molecule mixtures.

DB conceived the general idea, defined major research directions and identified key publications, 
co-supervised the research,  performed parts of the calculations, and 
wrote the first version of the manuscript. 

All authors discussed all results and the manuscript at all times.

\vspace{0.5cm}
\noindent{\bf Competing interests}\\
The authors declare no competing interests.

\bibliography{qcr21}

\newpage
\begin{widetext}
\begin{center}
\textbf{\large Supplemental Material}
\end{center}
\renewcommand\thesubsection{\Alph{subsection}}
\renewcommand{\theequation}{S\arabic{equation}}
\setcounter{subsection}{0}
\setcounter{equation}{0}
\setcounter{figure}{0}
\setcounter{table}{0}
\setcounter{page}{1}
\renewcommand{\theequation}{S\arabic{equation}}
\renewcommand{\thefigure}{S\arabic{figure}}

\section{\label{hom_xy_gen_sol}Mean Field Solutions}

From the Hamiltonian in Eq.~\eqref{Ham}, the Heisenberg equations of motion are 
\begin{eqnarray}
i \hbar \frac{\partial \hat{\psi}_{a} \left( \boldsymbol{r}, t \right)}{\partial t} & = & 
\left\{ 
- \frac{\hbar^2}{2 m_{a}} 
\nabla^2 
+ V_{a} \left( \boldsymbol{r}, t \right) 
+ g_{a} 
\hat{\psi}^{\dagger}_{a} \left( \boldsymbol{r}, t \right) 
\hat{\psi}_{a} \left( \boldsymbol{r}, t \right) 
+ g_{am} 
\hat{\psi}^{\dagger}_{m} \left( \boldsymbol{r}, t \right) 
\hat{\psi}_{m} \left( \boldsymbol{r}, t \right) 
\right\} 
\hat{\psi}_{a} \left( \boldsymbol{r}, t \right) 
\nonumber\\
&&
+ \alpha \sqrt{2} 
\hat{\psi}^{\dagger}_{a} \left( \boldsymbol{r}, t \right) 
\hat{\psi}_{m} \left( \boldsymbol{r}, t \right) , 
\nonumber\\
i \hbar \frac{\partial \hat{\psi}_{m} \left( \boldsymbol{r}, t \right)}{\partial t} & = & 
\left\{ 
- \frac{\hbar^2}{2 m_{m}} 
\nabla^2 
+ V_{m} \left( \boldsymbol{r}, t \right) 
+ \epsilon 
+ g_{m} 
\hat{\psi}^{\dagger}_{m} \left( \boldsymbol{r}, t \right) 
\hat{\psi}_{m} \left( \boldsymbol{r}, t \right) 
+ g_{am} 
\hat{\psi}^{\dagger}_{a} \left( \boldsymbol{r}, t \right) 
\hat{\psi}_{a} \left( \boldsymbol{r}, t \right) 
\right\} 
\hat{\psi}_{m} \left( \boldsymbol{r}, t \right) 
\nonumber\\
&&
+ 
\frac{\alpha}{\sqrt{2}} 
\left\{ 
\hat{\psi}_{a} \left( \boldsymbol{r}, t \right) 
\right\}^2 .
\label{hat_psi_eqofmot}
\end{eqnarray}
By writing 
$\hat{\psi}_{a} \left( \boldsymbol{r}, t \right) 
= 
\hat{\psi}_{a, 0} \left( \boldsymbol{r}, t \right) 
e^{- i \int^{t} d t_1 V_a \left( \boldsymbol{r}, t_1 \right) / \hbar} 
e^{- i \mu t / \hbar}
$ and 
$\hat{\psi}_{m} \left( \boldsymbol{r}, t \right) 
= 
\hat{\psi}_{m, 0} \left( \boldsymbol{r}, t \right) 
e^{- i \int^{t} d t_1 V_m \left( \boldsymbol{r}, t_1 \right) / \hbar} 
e^{- 2 i \mu t / \hbar}
$, Eq.~\eqref{hat_psi_eqofmot} can be expressed as

\begin{eqnarray}
i \hbar \frac{\partial \hat{\psi}_{a, 0} \left( \boldsymbol{r}, t \right)}{\partial t} 
& = & 
\left\{ 
- \frac{\hbar^2}{2 m_{a}} 
\nabla^2 
- \mu 
+ g_{a} 
\hat{\psi}^{\dagger}_{a, 0} \left( \boldsymbol{r}, t \right) 
\hat{\psi}_{a, 0} \left( \boldsymbol{r}, t \right) 
+ g_{am} 
\hat{\psi}^{\dagger}_{m, 0} \left( \boldsymbol{r}, t \right) 
\hat{\psi}_{m, 0} \left( \boldsymbol{r}, t \right) 
\right\} 
\hat{\psi}_{a, 0} \left( \boldsymbol{r}, t \right) 
\nonumber\\
&&
+ \alpha \sqrt{2} 
\hat{\psi}^{\dagger}_{a, 0} \left( \boldsymbol{r}, t \right) 
\hat{\psi}_{m, 0} \left( \boldsymbol{r}, t \right) 
e^{- i \int^{t} d t_1 \left\{ 
V_m \left( \boldsymbol{r}, t_1 \right) 
- 
2 V_a \left( \boldsymbol{r}, t_1 \right) 
\right\} / \hbar} 
\nonumber\\
&& 
+ \frac{\hbar^2}{2 m_{a}} 
\left\lgroup 
\left\{ 
\frac{1}{\hbar} 
\int^{t} d t_1 \; 
\nabla V_{a} \left( \boldsymbol{r}, t_1 \right) 
\right\}^2 
+ 
\frac{i}{\hbar} \int^{t} d t_1 \; 
\left\lbrack 
\nabla^2 V_{a} \left( \boldsymbol{r}, t_1 \right) 
+ 
2 
\left\{ 
\nabla V_{a} \left( \boldsymbol{r}, t_1 \right) 
\right\} 
\cdot 
\nabla 
\right\rbrack 
\right\rgroup 
\hat{\psi}_{a, 0} \left( \boldsymbol{r}, t \right) 
, 
\nonumber\\
i \hbar \frac{\partial \hat{\psi}_{m, 0} \left( \boldsymbol{r}, t \right)}{\partial t} 
& = & 
\left\{ 
- \frac{\hbar^2}{2 m_{m}} 
\nabla^2 
+ \epsilon 
- 2 \mu 
+ g_{m} 
\hat{\psi}^{\dagger}_{m, 0} \left( \boldsymbol{r}, t \right) 
\hat{\psi}_{m, 0} \left( \boldsymbol{r}, t \right) 
+ g_{am} 
\hat{\psi}^{\dagger}_{a, 0} \left( \boldsymbol{r}, t \right) 
\hat{\psi}_{a, 0} \left( \boldsymbol{r}, t \right) 
\right\} 
\hat{\psi}_{m, 0} \left( \boldsymbol{r}, t \right) 
\nonumber\\
&&
+ 
\frac{\alpha}{\sqrt{2}} 
\left\{ 
\hat{\psi}_{a, 0} \left( \boldsymbol{r}, t \right) 
\right\}^2 
e^{i \int^{t} d t_1 \left\{ 
V_m \left( \boldsymbol{r}, t_1 \right) 
- 
2 V_a \left( \boldsymbol{r}, t_1 \right) 
\right\} / \hbar} 
\nonumber\\
&& 
+ \frac{\hbar^2}{2 m_{m}} 
\left\lgroup 
\left\{ 
\frac{1}{\hbar} 
\int^{t} d t_1 \; 
\nabla V_{m} \left( \boldsymbol{r}, t_1 \right) 
\right\}^2 
+ 
\frac{i}{\hbar} \int^{t} d t_1 \; 
\left\lbrack 
\nabla^2 V_{m} \left( \boldsymbol{r}, t_1 \right) 
+ 
2 
\left\{ 
\nabla V_{m} \left( \boldsymbol{r}, t_1 \right) 
\right\} 
\cdot 
\nabla 
\right\rbrack 
\right\rgroup 
\hat{\psi}_{m, 0} \left( \boldsymbol{r}, t \right) 
. \qquad \qquad 
\label{hat_psi_eqofmotSt1}
\end{eqnarray}
Note that, according to Eqs.~\eqref{hat_psi_eqofmotSt1}, changing $V_j \left( \boldsymbol{r}, t \right)$ from 0 to $V_j \left( t \right)$ ($j = a, m$) does not affect $\hat{\psi}_{j, 0} \left( \boldsymbol{r}, t \right)$ if $V_m \left( t \right) = 2 V_a \left( t \right)$.

By following~\cite{CastinDum1998}, in the Heisenberg picture, we split $\hat{\psi}_j \left( \boldsymbol{r}, t \right)$ as 
$\hat{\psi}_j \left( \boldsymbol{r}, t \right) 
= 
\psi_j \left( \boldsymbol{r}, t \right) 
\hat{\Phi}_{j, c} \left( t \right) 
+ 
\delta \hat{\psi}_j \left( \boldsymbol{r}, t \right)$ where 
$\int d^3 r \; 
\psi^{*}_j \left( \boldsymbol{r}, t \right) 
\delta \hat{\psi}_j \left( \boldsymbol{r}, t \right) 
\coloneqq 0$. 
Then $\hat{\Phi}_{j, c} \left( t \right) 
= 
\int d^3 r \; 
\psi^{*}_j \left( \boldsymbol{r}, t \right) 
\hat{\psi}_j \left( \boldsymbol{r}, t \right) 
/ 
\int d^3 r \; 
\left\vert 
\psi_j \left( \boldsymbol{r}, t \right) 
\right\vert^2$ and from this result, one can get 
$\left\lbrack 
\hat{\Phi}_{j, c} \left( t \right) 
, 
\hat{\Phi}^{\dagger}_{j, c} \left( t \right) 
\right\rbrack 
= 
1 / 
\int d^3 r \; 
\left\vert 
\psi_j \left( \boldsymbol{r}, t \right) 
\right\vert^2$, 
$\left\lbrack 
\hat{\Phi}_{j, c} \left( t \right) 
, 
\delta \hat{\psi}_j \left( \boldsymbol{r}, t \right) 
\right\rbrack 
= 
\left\lbrack 
\hat{\Phi}_{j, c} \left( t \right) 
, 
\delta \hat{\psi}^{\dagger}_j \left( \boldsymbol{r}, t \right) 
\right\rbrack 
= 
0$, 
$\left\lbrack 
\delta \hat{\psi}_j \left( \boldsymbol{r}, t \right) 
, 
\delta \hat{\psi}^{\dagger}_j \left( \boldsymbol{r}', t \right) 
\right\rbrack 
= 
\delta \left( \boldsymbol{r} - \boldsymbol{r}' \right) 
- 
\psi_j \left( \boldsymbol{r}, t \right) 
\psi^{*}_j \left( \boldsymbol{r}', t \right) 
/ 
\int d^3 r \; 
\left\vert 
\psi_j \left( \boldsymbol{r}, t \right) 
\right\vert^2$, 
and 
$\left\langle 
\hat{\Phi}^{\dagger}_{j, c} \left( t \right) 
\hat{\Phi}_{j, c} \left( t \right) 
\right\rangle 
= 
N_{j, c} \left( t \right) 
/ 
\int d^3 r \; 
\left\vert 
\psi_j \left( \boldsymbol{r}, t \right) 
\right\vert^2$ 
where $N_{a, c} \left( t \right)$ is the number of BEC atoms at time $t$ and $N_{m, c} \left( t \right)$ is the number of BEC molecules at time $t$. 
From now on, we will set 
$\int d^3 r \; 
\left\vert 
\psi_j \left( \boldsymbol{r}, t \right) 
\right\vert^2 
= 
N_{j, c} \left( t \right)$ so that $\psi_j \left( \boldsymbol{r}, t \right)$ is the mean-field value of the $\hat{\psi}_j \left( \boldsymbol{r}, t \right)$. 
Then we have 
$\left\langle 
\hat{\Phi}^{\dagger}_{j, c} \left( t \right) 
\hat{\Phi}_{j, c} \left( t \right) 
\right\rangle 
= 1$ and for $N_{j, c} \left( t \right) \gg 1$, 
$\left\lbrack 
\hat{\Phi}_{j, c} \left( t \right) 
, 
\hat{\Phi}^{\dagger}_{j, c} \left( t \right) 
\right\rbrack 
= 
1 / N_{j, c} \left( t \right) 
\rightarrow 0$. 
We will additionally substitute $\hat{\Phi}_{j, c} \left( t \right) \rightarrow 1$ for calculational convenience, which gives the usual non-number conserving approach $\hat{\psi}_{j} \left( \boldsymbol{r}, t \right) = \psi_{j} \left( \boldsymbol{r}, t \right) + \delta \hat{\psi}_{j} \left( \boldsymbol{r}, t \right)$.

Let $N$ be the total number of BEC atoms (including the ones bound in molecules), 
$V$ 
the volume of the system, and $n \coloneqq N / V$ 
the mean total number density of BEC atoms. 
For $x \left( \boldsymbol{r}, t \right) 
\coloneqq 
\left\vert \psi_{a} \left( \boldsymbol{r}, t \right) \right\vert / \sqrt{n}$ 
and 
$y \left( \boldsymbol{r}, t \right) 
\coloneqq 
\left\vert \psi_{m} \left( \boldsymbol{r}, t \right) \right\vert / \sqrt{n}$, 
we set 
\begin{eqnarray}
\psi_{a} \left( \boldsymbol{r}, t \right) = 
x \left( \boldsymbol{r}, t \right) \sqrt{n} 
e^{- i \int^{t} d t_1 V_a \left( \boldsymbol{r}, t_1 \right) / \hbar} 
e^{i \left\{ \varphi_{a} \left( \boldsymbol{r}, t \right) - \mu t / \hbar \right\}}
, 
\quad 
\psi_{m} \left( \boldsymbol{r}, t \right) = 
y \left( \boldsymbol{r}, t \right) \sqrt{n} 
e^{- i \int^{t} d t_1 V_m \left( \boldsymbol{r}, t_1 \right) / \hbar} 
e^{i \left\{ \varphi_{m} \left( \boldsymbol{r}, t \right) - 2 \mu t / \hbar \right\}} 
, \qquad 
\end{eqnarray}
which satisfy (note that $m_m \simeq 2 m_a$)

\begin{eqnarray}
i \frac{\partial x \left( \boldsymbol{r}, t \right)}{\partial \tilde{t}} 
& = & 
\left\{ 
\frac{\partial \varphi_a \left( \boldsymbol{r}, t \right)}{\partial \tilde{t}} 
+ 
x^2 \left( \boldsymbol{r}, t \right) 
+ 
\tilde{g}_{am} 
y^2 \left( \boldsymbol{r}, t \right) 
- 
\tilde{\mu} 
\right\} 
x \left( \boldsymbol{r}, t \right) 
+ 
\tilde{\alpha} 
x \left( \boldsymbol{r}, t \right) 
y \left( \boldsymbol{r}, t \right) 
e^{i \varphi_{am} \left( \boldsymbol{r}, t \right)} 
e^{- i \int^{t} d t_1 \left\{ 
V_m \left( \boldsymbol{r}, t_1 \right) 
- 
2 V_a \left( \boldsymbol{r}, t_1 \right) 
\right\} / \hbar} 
\nonumber\\
&& 
- \xi_a^2 \left\lgroup 
\left\{ \nabla^2 x \left( \boldsymbol{r}, t \right) \right\} 
- 
x \left( \boldsymbol{r}, t \right) 
\left\{ 
\nabla \varphi_a \left( \boldsymbol{r}, t \right) 
\right\}^2 
+ 
\frac{i}{x \left( \boldsymbol{r}, t \right)} 
\nabla \cdot 
\left\lbrack 
x^2 \left( \boldsymbol{r}, t \right) 
\left\{ 
\nabla \varphi_a \left( \boldsymbol{r}, t \right) 
\right\} 
\right\rbrack 
\right\rgroup 
\nonumber\\
&& 
+ \xi^2_a  
\left\lbrack 
\left\{ 
\frac{1}{\hbar} 
\int^{t} d t_1 \; 
\nabla V_{a} \left( \boldsymbol{r}, t_1 \right) 
\right\}^2 
x \left( \boldsymbol{r}, t \right) 
+ 
\frac{1}{x \left( \boldsymbol{r}, t \right)} 
\frac{i}{\hbar} \int^{t} d t_1 \; 
\nabla \cdot 
\left\{ 
x^2 \left( \boldsymbol{r}, t \right) 
\nabla V_{a} \left( \boldsymbol{r}, t_1 \right) 
\right\} 
\right\rbrack 
\nonumber\\
&& 
- 2 \xi^2_a 
\left\{ 
\nabla \varphi_a \left( \boldsymbol{r}, t \right) 
\right\} 
\cdot 
\left\{ 
\frac{1}{\hbar} 
\int^{t} d t_1 \; 
\nabla V_{a} \left( \boldsymbol{r}, t_1 \right) 
\right\} 
x \left( \boldsymbol{r}, t \right) 
, 
\nonumber\\ 
i \frac{\partial y \left( \boldsymbol{r}, t \right)}{\partial \tilde{t}} 
& = & 
\left\{ 
\frac{\partial \varphi_m \left( \boldsymbol{r}, t \right)}{\partial \tilde{t}} 
+ 
\tilde{g}_{m} 
y^2 \left( \boldsymbol{r}, t \right) 
+ 
\tilde{g}_{am} 
x^2 \left( \boldsymbol{r}, t \right) 
+ 
\tilde{\epsilon} 
- 
2 \tilde{\mu} 
\right\} 
y \left( \boldsymbol{r}, t \right) 
+ 
\frac{\tilde{\alpha}}{2} 
x^2 \left( \boldsymbol{r}, t \right) 
e^{- i \varphi_{am} \left( \boldsymbol{r}, t \right)} 
e^{i \int^{t} d t_1 \left\{ 
V_m \left( \boldsymbol{r}, t_1 \right) 
- 
2 V_a \left( \boldsymbol{r}, t_1 \right) 
\right\} / \hbar} 
\nonumber\\
&& 
- \frac{\xi_a^2}{2} 
\left\lgroup 
\left\{ \nabla^2 y \left( \boldsymbol{r}, t \right) \right\} 
- 
y \left( \boldsymbol{r}, t \right) 
\left\{ 
\nabla \varphi_m \left( \boldsymbol{r}, t \right) 
\right\}^2 
+ 
\frac{i}{y \left( \boldsymbol{r}, t \right)} 
\nabla \cdot 
\left\lbrack 
y^2 \left( \boldsymbol{r}, t \right) 
\left\{ 
\nabla \varphi_m \left( \boldsymbol{r}, t \right) 
\right\} 
\right\rbrack 
\right\rgroup 
\nonumber\\
&& 
+ \frac{\xi_a^2}{2} 
\left\lbrack 
\left\{ 
\frac{1}{\hbar} 
\int^{t} d t_1 \; 
\nabla V_{m} \left( \boldsymbol{r}, t_1 \right) 
\right\}^2 
y \left( \boldsymbol{r}, t \right) 
+ 
\frac{1}{y \left( \boldsymbol{r}, t \right)} 
\frac{i}{\hbar} \int^{t} d t_1 \; 
\nabla \cdot 
\left\{ 
y^2 \left( \boldsymbol{r}, t \right) 
\nabla V_{m} \left( \boldsymbol{r}, t_1 \right) 
\right\} 
\right\rbrack 
\nonumber\\
&& 
- \xi^2_a 
\left\{ 
\nabla \varphi_m \left( \boldsymbol{r}, t \right) 
\right\} 
\cdot 
\left\{ 
\frac{1}{\hbar} 
\int^{t} d t_1 \; 
\nabla V_{m} \left( \boldsymbol{r}, t_1 \right) 
\right\} 
y \left( \boldsymbol{r}, t \right) 
, 
\label{psi_eqofmot}
\end{eqnarray}
which give

\begin{eqnarray}
\int d^3 r \; 
x \left( \boldsymbol{r}, t \right) 
\frac{\partial x \left( \boldsymbol{r}, t \right)}{\partial \tilde{t}} 
& = & 
\tilde{\alpha} 
\int d^3 r \; 
x^2 \left( \boldsymbol{r}, t \right) 
y \left( \boldsymbol{r}, t \right) 
\sin \left\lbrack 
\varphi_{am} \left( \boldsymbol{r}, t \right) 
- 
\frac{1}{\hbar} 
\int^{t} d t_1 \left\{ 
V_m \left( \boldsymbol{r}, t_1 \right) 
- 
2 V_a \left( \boldsymbol{r}, t_1 \right) 
\right\} 
\right\rbrack 
, \nonumber\\
\int d^3 r \; 
y \left( \boldsymbol{r}, t \right) 
\frac{\partial y \left( \boldsymbol{r}, t \right)}{\partial \tilde{t}} 
& = & 
- \frac{\tilde{\alpha}}{2} 
\int d^3 r \; 
x^2 \left( \boldsymbol{r}, t \right) 
y \left( \boldsymbol{r}, t \right) 
\sin \left\lbrack 
\varphi_{am} \left( \boldsymbol{r}, t \right) 
- 
\frac{1}{\hbar} 
\int^{t} d t_1 \left\{ 
V_m \left( \boldsymbol{r}, t_1 \right) 
- 
2 V_a \left( \boldsymbol{r}, t_1 \right) 
\right\} 
\right\rbrack 
, 
\label{xy_eqofmot}
\end{eqnarray}
and

\begin{eqnarray}
\frac{\partial \varphi_{am} \left( \boldsymbol{r}, t \right)}{\partial \tilde{t}} 
& = & 
\frac{\xi_a^2}{2} 
\left\lbrack 
\frac{\nabla^2 y \left( \boldsymbol{r}, t \right)}{y \left( \boldsymbol{r}, t \right)} 
- 
\left\{ 
\nabla \varphi_m \left( \boldsymbol{r}, t \right) 
\right\}^2 
- 
4 \frac{\nabla^2 x \left( \boldsymbol{r}, t \right)}{x \left( \boldsymbol{r}, t \right)} 
+ 
4 \left\{ 
\nabla \varphi_a \left( \boldsymbol{r}, t \right) 
\right\}^2 
\right\rbrack 
\nonumber\\
&& 
- \frac{\xi_a^2}{2} 
\left\lbrack 
\frac{1}{\hbar} 
\int^{t} d t_1 \; 
\nabla 
\left\{ 
V_{m} \left( \boldsymbol{r}, t_1 \right) 
+ 
2 V_{a} \left( \boldsymbol{r}, t_1 \right) 
\right\} 
\right\rbrack 
\cdot 
\left\lbrack 
\frac{1}{\hbar} 
\int^{t} d t_1 \; 
\nabla 
\left\{ 
V_{m} \left( \boldsymbol{r}, t_1 \right) 
- 
2 V_{a} \left( \boldsymbol{r}, t_1 \right) 
\right\} 
\right\rbrack 
\nonumber\\
&& 
+ 
\xi^2_a 
\left\lgroup 
\left\{ 
\nabla \varphi_m \left( \boldsymbol{r}, t \right) 
\right\} 
\cdot 
\left\lbrack 
\frac{1}{\hbar} 
\int^{t} d t_1 \; 
\nabla 
\left\{ 
V_{m} \left( \boldsymbol{r}, t_1 \right) 
- 
2 V_{a} \left( \boldsymbol{r}, t_1 \right) 
\right\} 
\right\rbrack 
+ 
2 \left\{ 
\nabla \varphi_{am} \left( \boldsymbol{r}, t \right) 
\right\} 
\cdot 
\left\{ 
\frac{1}{\hbar} 
\int^{t} d t_1 \; 
\nabla V_{a} \left( \boldsymbol{r}, t_1 \right) 
\right\} 
\right\rgroup 
\nonumber\\
&& + 
2 
x^2 \left( \boldsymbol{r}, t \right) 
+ 
\tilde{g}_{am} 
\left\{ 
2 
y^2 \left( \boldsymbol{r}, t \right) 
- 
x^2 \left( \boldsymbol{r}, t \right) 
\right\} 
- 
\tilde{g}_{m} 
y^2 \left( \boldsymbol{r}, t \right) 
- 
\tilde{\epsilon} 
\nonumber\\
&& + 
\frac{\tilde{\alpha}}{2} 
\left\{ 
4 y \left( \boldsymbol{r}, t \right) 
- 
\frac{x^2 \left( \boldsymbol{r}, t \right)}{y \left( \boldsymbol{r}, t \right)} 
\right\} 
\cos \left\lbrack 
\varphi_{am} \left( \boldsymbol{r}, t \right) 
- 
\frac{1}{\hbar} 
\int^{t} d t_1 \left\{ 
V_m \left( \boldsymbol{r}, t_1 \right) 
- 
2 V_a \left( \boldsymbol{r}, t_1 \right) 
\right\} 
\right\rbrack 
\,. \qquad \quad 
\label{varphi_eqofmot}
\end{eqnarray}
We normalize to atomic condensate units as follows: 
$\tilde{t} \coloneqq g_{a} n t / \hbar$, 
$\tilde{g}_{am} \coloneqq g_{am} / g_{a}$, 
$\tilde{g}_{m} \coloneqq g_{m} / g_{a}$, 
$\tilde{\alpha} \coloneqq \alpha \sqrt{2 n} / \left( g_{a} n \right)$, 
$\tilde{\mu} \coloneqq \mu / \left( g_{a} n \right)$, 
$\tilde{\epsilon} \coloneqq \epsilon / \left( g_{a} n \right)$, 
$\xi_a \coloneqq \hbar / \sqrt{2 m_a g_a n}$ is the atomic healing length, 
and 
$\varphi_{am} \left( \boldsymbol{r}, t \right) 
\coloneqq 
\varphi_{m} \left( \boldsymbol{r}, t \right) 
- 
2 \varphi_{a} \left( \boldsymbol{r}, t \right)$.

When the system is in box traps ($V_a \left( \boldsymbol{r}, t \right) = V_m \left( \boldsymbol{r}, t \right) = 0$) or $V_m \left( \boldsymbol{r}, t \right) = 2 V_a \left( \boldsymbol{r}, t \right) = F_{\rm tr} \left( t \right)$ where $F_{\rm tr} \left( t \right)$ is some real function which is homogeneous in space, 
Eqs.~\eqref{xy_eqofmot} and~\eqref{varphi_eqofmot} can have solutions where $x \left( \boldsymbol{r}, t \right)$, 
$y \left( \boldsymbol{r}, t \right)$, and 
$\varphi_{am} \left( \boldsymbol{r}, t \right)$ do not depend on space.

  This leads to the following conclusion: Suppose we add a spatially homogeneous density perturbation at $t \ge 0$ to the system with atoms and molecules trapped in box potentials.
Let $\psi_{j, 0} \left( t \right)$ be the homogeneous mean-field of this system for $t < 0$ and $\psi_{j, 1} \left( \boldsymbol{r}, t \right)$ be the mean-field of the system for $t \ge 0$ ($j = a, m$). 
Then, from Eqs.~\eqref{xy_eqofmot} and~\eqref{varphi_eqofmot}, 
$\psi_{j, 1} \left( \boldsymbol{r}, t \right) 
= 
\psi_{j, 0} \left( t \right) 
e^{- i \int_{0}^{t} d t_1 V_j \left( t_1 \right) / \hbar}$. 
From now on, we will consider these homogeneous solutions whose mean-field Hamiltonian $H_{0} \left( t \right)$ is 
\begin{eqnarray}
H_{0} \left( t \right) / (N g_{a} n) 
& = & 
\frac{1}{2} x^4 \left( t \right) 
+ \frac{\tilde{g}_{m}}{2} y^4 \left( t \right) 
+ \tilde{\epsilon} y^2 \left( t \right) 
+ \tilde{g}_{am} x^2 \left( t \right) y^2 \left( t \right) 
+ \tilde{\alpha} x^2 \left( t \right) y \left( t \right) 
\cos \varphi_{am} \left( t \right) 
\nonumber\\
&& + 
\frac{V_a \left( t \right) x^2 \left( t \right) + V_m \left( t \right) y^2 \left( t \right)}{g_a n} 
, 
\label{H0_const_app}
\end{eqnarray}

Thus, by assuming $g_{a} > 0$, $H_{0} \left( t \right)$ is minimized when 
$\alpha \cos \varphi_{am} \left( t \right) = - \left\vert \alpha \right\vert$. 
Since $\alpha \cos \varphi_{am} \left( t \right) = - \left\vert \alpha \right\vert$ makes $\sin \varphi_{am} = 0$, which forces  $x \left( t \right)$ and $y \left( t \right)$ to be constant according to Eqs.~\eqref{xy_eqofmot}, 
from now on, we will write $x \left( t \right)$ as $x$ and $y \left( t \right)$ as $y$.

From Eq.~\eqref{varphi_eqofmot}, 
$\alpha \cos \varphi_{am} \left( t \right) = - \left\vert \alpha \right\vert$ can be achieved when

\begin{eqnarray}
\tilde{\epsilon} = && \; 
2 x^2
+ \tilde{g}_{am} \left( 2 y^2 - x^2 \right) 
- \tilde{g}_{m} y^2 
- \frac{\left\vert \tilde{\alpha} \right\vert}{2} \left( 4 y - \frac{x^2}{y} \right) . 
\qquad
\label{epsilon_for_constant_sol}
\end{eqnarray}

Using Eqs.~\eqref{H0_const_app} and~\eqref{epsilon_for_constant_sol}, 
one can get 
\begin{eqnarray}
\frac{H_0}{N g_{a} n} = 
\frac{V_a \left( t \right) x^2 + V_m \left( t \right) y^2}{g_a n} 
+ \frac{1}{2} x^4 
- \frac{1}{2} \left( \tilde{g}_{m} - 4 \tilde{g}_{am} \right) y^4 
+ 2 x^2 y^2 
- 2 \left\vert \tilde{\alpha} \right\vert y^3 
- \frac{\left\vert \tilde{\alpha} \right\vert}{2} x^2 y , 
\label{H0_lagrangian}
\end{eqnarray}
where $x^2 + 2 y^2 = 1$ from number conservation in Eq.~\eqref{hat_N_def}.

Now, let $\varphi_{a} \left( t \right)$ and $\varphi_{m} \left( t \right)$ to be also constant in time $t$. 
From Eq.~\eqref{psi_eqofmot}, these conditions can be achieved when 
\begin{eqnarray}
\tilde{\mu} &=& 
x^2 + \tilde{g}_{am} y^2 - \left\vert \tilde{\alpha} \right\vert y . 
\label{mu_cond}
\end{eqnarray}

With Eqs.~\eqref{hat_N_def},~\eqref{H0_lagrangian}, and~\eqref{mu_cond}, 
\begin{eqnarray}
\frac{H_0 - \mu N}{N g_{a} n} 
= 
- \frac{1}{2} 
- \left\{ 2 + \frac{1}{2} \left( \tilde{g}_{m} - 4 \tilde{g}_{am} \right) \right\} y^4 
+ \left( 2 - \tilde{g}_{am} \right) y^2 
- \left\vert \tilde{\alpha} \right\vert y^3 
+ \frac{\left\vert \tilde{\alpha} \right\vert}{2} y 
+ 
\frac{V_a \left( t \right) + \left\{ V_m \left( t \right) - 2 V_a \left( t \right) \right\} y^2}{g_a n} 
. \qquad \quad 
\label{H0_mu}
\end{eqnarray}

For the case of the box traps or $V_m \left( t \right) = 2 V_a \left( t \right)$, 
the energy in Eq.~\eqref{H0_mu} is minimized when 
\begin{eqnarray}
&& \left( 8 + 2 \tilde{g}_{m} - 8 \tilde{g}_{am} \right) y^3 
+ 3 \left\vert \tilde{\alpha} \right\vert y^2 
- 2 \left( 2 - \tilde{g}_{am} \right) y 
- \frac{\left\vert \tilde{\alpha} \right\vert}{2} 
= 0 , 
\label{eq_y}
\end{eqnarray}
with $y \ge 0$ and $x = \sqrt{1 - 2 y^2}$ from number conservation.

\section{\label{BovExp_Sup}Bogoliubov Expansion}

In the Heisenberg picture, 
by writing $\hat{\psi}_{j} \left( \boldsymbol{r}, t \right) = \psi_{j} \left( \boldsymbol{r}, t \right) + \delta \hat{\psi}_{j} \left( \boldsymbol{r}, t \right)$ with 
\begin{eqnarray}
\delta \hat{\psi}_{a} \left( \boldsymbol{r}, t \right) 
& = & 
e^{- i \int^{t} d t_1 V_a \left( \boldsymbol{r}, t_1 \right) / \hbar} 
e^{- i \mu t / \hbar} 
\frac{1}{\sqrt{V}} 
\sum_{\boldsymbol{k} \neq 0} 
e^{i \boldsymbol{k} \cdot \boldsymbol{r}}
\delta \hat{\Psi}_{a} \left( \boldsymbol{k}, t \right) 
, \nonumber\\
\delta \hat{\psi}_{m} \left( \boldsymbol{r}, t \right) 
& = & 
e^{- i \int^{t} d t_1 V_m \left( \boldsymbol{r}, t_1 \right) / \hbar} 
e^{- 2 i \mu t / \hbar} 
\frac{1}{\sqrt{V}} 
\sum_{\boldsymbol{k} \neq 0} 
e^{i \boldsymbol{k} \cdot \boldsymbol{r}}
\delta \hat{\Psi}_{m} \left( \boldsymbol{k}, t \right) 
, 
\label{homogeneous_box_delta_hat_psi_am}
\end{eqnarray}
from the commutation relations 
$\left\lbrack 
\hat{\psi}_{j} \left( \boldsymbol{r}_1, t \right) , 
\hat{\psi}^{\dagger}_{l} \left( \boldsymbol{r}_2, t \right) 
\right\rbrack 
= 
\delta_{j, l} 
\delta \left( \boldsymbol{r}_1 - \boldsymbol{r}_2 \right) 
$
and 
$\left\lbrack 
\hat{\psi}_{j} \left( \boldsymbol{r}_1, t \right) , 
\hat{\psi}_{l} \left( \boldsymbol{r}_2, t \right) 
\right\rbrack 
= 0
$
for $j = a, m$ and $l = a, m$, one can get

\begin{eqnarray}
\left\lbrack 
\delta \hat{\Psi}_{j} \left( \boldsymbol{k}_1, t \right) , 
\delta \hat{\Psi}^{\dagger}_{l} \left( \boldsymbol{k}_2, t \right) 
\right\rbrack 
= 
\delta_{j, l} 
\delta_{\boldsymbol{k}_1, \boldsymbol{k}_2} , \quad 
\left\lbrack 
\delta \hat{\Psi}_{j} \left( \boldsymbol{k}_1, t \right) , 
\delta \hat{\Psi}_{l} \left( \boldsymbol{k}_2, t \right) 
\right\rbrack 
= 0 .
\label{delta_psi_psi_k}
\end{eqnarray}

For box trap case ($V_{a} \left( \boldsymbol{r}, t \right) = 0$ and $V_{m} \left( \boldsymbol{r}, t \right) = 0$) 
or $V_m \left( t \right) = 2 V_a \left( t \right)$, 
with solutions 
$\psi_{a} \left( \boldsymbol{r}, t \right) 
= 
x \sqrt{n} 
e^{- i \int^{t} d t_1 V_a \left( \boldsymbol{r}, t \right) / \hbar} 
e^{i \left\{ \varphi_{a} - \mu t / \hbar \right\}}$ and 
$\psi_{m} \left( \boldsymbol{r}, t \right) 
= 
y \sqrt{n} 
e^{- i \int^{t} d t_1 V_m \left( \boldsymbol{r}, t \right) / \hbar} 
e^{i \left\{ \varphi_{m} - 2 \mu t / \hbar \right\}}$ 
where 
$x$, $y$, 
$\varphi_{a}$, and $\varphi_{m}$ are constant in time $t$ by satisfying Eq.~\eqref{mu_cond} and 
$\alpha \cos \varphi_{am} \left( t \right) = - \left\vert \alpha \right\vert$ by satisfying Eq.~\eqref{epsilon_for_constant_sol}, up to second order in $\delta \hat{\Psi}_{j}$ where $j = a, m$, $\hat{H} \left( t \right) - \mu \hat{N}$ can be written as

\begin{eqnarray}
&& \hat{H} \left( t \right) - \mu \hat{N} = 
H_0 - \mu N 
- \frac{g_{a} n}{2} 
\sum_{\boldsymbol{k} \neq 0} 
\left\{ 
M_{11} \left( k \right) + M_{22} \left( k \right) 
\right\} 
+ O \left( \delta \hat{\Psi}^3_{j} \right) 
\nonumber\\
&& + 
\frac{g_{a} n}{2} 
\sum_{\boldsymbol{k} \neq 0} 
\left\lbrack 
\begin{array}{cccc}
\delta \hat{\Psi}^{\dagger}_{a} \left( \boldsymbol{k}, t \right) 
& 
\delta \hat{\Psi}^{\dagger}_{m} \left( \boldsymbol{k}, t \right) 
&
\delta \hat{\Psi}_{a} \left( - \boldsymbol{k}, t \right) 
&
\delta \hat{\Psi}_{m} \left( - \boldsymbol{k}, t \right) 
\end{array}
\right\rbrack 
\left\lbrack 
\begin{array}{cccc}
M_{11} \left( k \right) 
& 
M_{12} 
& 
M_{13} 
& 
M_{14} 
\\
M^{*}_{12} 
& 
M_{22} \left( k \right) 
& 
M_{14} 
& 
M_{24} 
\\
M^{*}_{13} 
& 
M^{*}_{14} 
& 
M_{11} \left( k \right) 
& 
M^{*}_{12} 
\\
M^{*}_{14} 
& 
M^{*}_{24} 
& 
M_{12} 
& 
M_{22} \left( k \right) 
\end{array}
\right\rbrack 
\left\lbrack 
\begin{array}{c} 
\delta \hat{\Psi}_{a} \left( \boldsymbol{k}, t \right) 
\\
\delta \hat{\Psi}_{m} \left( \boldsymbol{k}, t \right) 
\\
\delta \hat{\Psi}^{\dagger}_{a} \left( - \boldsymbol{k}, t \right) 
\\
\delta \hat{\Psi}^{\dagger}_{m} \left( - \boldsymbol{k}, t \right) 
\end{array}
\right\rbrack 
\nonumber\\
&& + 
V_a \left( t \right) 
\sum_{\boldsymbol{k} \neq 0} 
\left\{ 
\delta \hat{\Psi}^{\dagger}_{a} \left( \boldsymbol{k}, t \right) 
\delta \hat{\Psi}_{a} \left( \boldsymbol{k}, t \right) 
+ 
2 
\delta \hat{\Psi}^{\dagger}_{m} \left( \boldsymbol{k}, t \right) 
\delta \hat{\Psi}_{m} \left( \boldsymbol{k}, t \right) 
\right\} 
, \qquad \qquad 
\label{H_muN_2nd_hom_gen}
\end{eqnarray}
where 
$\mu$ is in Eq.~\eqref{mu_cond}, $H_0 - \mu N$ in Eq.~\eqref{H0_mu},

\begin{eqnarray}
M_{11} \left( k \right) \coloneqq && \; 
\frac{\hbar^2 k^2}{2 m_{a} g_{a} n} 
+ 2 x^2 
+ \tilde{g}_{am} y^2 
- \tilde{\mu} 
= k^2 \xi^2_{a} + 1 - 2 y^2 + \left\vert \tilde{\alpha} \right\vert y ,
\label{M11_hom_gen}
\end{eqnarray}

\begin{eqnarray}
M_{22} \left( k \right) \coloneqq && \; 
\frac{\hbar^2 k^2}{2 m_{m} g_{a} n} 
+ \tilde{\epsilon} 
+ 2 \tilde{g}_{m} y^2 
+ \tilde{g}_{am} x^2 
- 2 \tilde{\mu} 
= \frac{1}{2} k^2 \xi^2_{a} + \tilde{g}_{m} y^2 
- \left\vert \tilde{\alpha} \right\vert 
\left( y - \frac{1}{2 y} \right) , 
\label{M22_hom_gen}
\end{eqnarray}

\begin{eqnarray}
M_{12} \coloneqq 
\left\{ 
\tilde{\alpha} 
- \textrm{sign} \left( \tilde{\alpha} \right) \tilde{g}_{am} y 
\right\} 
\sqrt{1 - 2 y^2} ,
\quad 
M_{13} \coloneqq 
1 - 2 y^2  
- \left\vert \tilde{\alpha} \right\vert y , 
\quad 
M_{14} \coloneqq && \; 
- \textrm{sign} \left( \tilde{\alpha} \right) \tilde{g}_{am} y \sqrt{1 - 2 y^2} , 
\quad 
M_{24} \coloneqq
\tilde{g}_{m} y^2 . 
\qquad \quad 
\label{M12_to_24_hom_gen}
\end{eqnarray}
Here, for convenience, we set $\varphi_{a} = 0$ and define $\textrm{sign} \left( \alpha \right) = 1$ if $\alpha \ge 0$ and $\textrm{sign} \left( \alpha \right) = -1$ if $\alpha < 0$.

We define the $4 \times 4$ matrix $M \left( k \right)$ as 
\begin{eqnarray}
M \left( k \right) \coloneqq && \; 
\left\lbrack 
\begin{array}{cccc}
M_{11} \left( k \right) 
& 
M_{12} 
& 
M_{13} 
& 
M_{14} 
\\
M^{*}_{12} 
& 
M_{22} \left( k \right) 
& 
M_{14} 
& 
M_{24} 
\\
M^{*}_{13} 
& 
M^{*}_{14} 
& 
M_{11} \left( k \right) 
& 
M^{*}_{12} 
\\
M^{*}_{14} 
& 
M^{*}_{24} 
& 
M_{12} 
& 
M_{22} \left( k \right) 
\end{array}
\right\rbrack 
= 
\left\lbrack 
\begin{array}{cc} 
M_{B, 1} \left( k \right) 
& 
M_{B, 2} 
\\
M^{*}_{B, 2} 
& 
M^{*}_{B, 1} \left( k \right) 
\end{array}
\right\rbrack ,
\qquad \quad 
\label{matrix_M}
\end{eqnarray}
with $2 \times 2$ matrices $M_{B, \zeta}$ ($\zeta = 1, 2$).

According to Eq.~\eqref{H_muN_2nd_hom_gen}, 
a system with $V_m \left( t \right) = 2 V_a \left( t \right) \neq 0$ can be interpreted as  a system in a box trap with an additional perturbation $\hat{V}_{H} \left( t \right) 
= 
V_a \left( t \right) 
\int d^3 r \; 
\left\{ 
\hat{\psi}^{\dagger}_a \left( \boldsymbol{r}, t \right) 
\hat{\psi}_a \left( \boldsymbol{r}, t \right) 
+ 
2 
\hat{\psi}^{\dagger}_m \left( \boldsymbol{r}, t \right) 
\hat{\psi}_m \left( \boldsymbol{r}, t \right) 
\right\}$.  In the following we first focus on the case $\hat{V}_{H} =0$, and consider perturbations later in section~\ref{CFI_QFI_Formalism}. 
Then, 
from Bogoliubov expansion~\cite{pethick,Uedareview}, we may write 
\begin{eqnarray}
\left\lbrack 
\begin{array}{c} 
\delta \hat{\Psi}_{a} \left( \boldsymbol{k}, t \right) 
\\
\delta \hat{\Psi}_{m} \left( \boldsymbol{k}, t \right) 
\\
\delta \hat{\Psi}^{\dagger}_{a} \left( - \boldsymbol{k}, t \right) 
\\
\delta \hat{\Psi}^{\dagger}_{m} \left( - \boldsymbol{k}, t \right) 
\end{array}
\right\rbrack 
= 
\left\lbrack 
\begin{array}{cc} 
U \left( k \right) 
& 
V^{*} \left( k \right) 
\\
V \left( k \right) 
& 
U^{*} \left( k \right) 
\end{array}
\right\rbrack 
\left\lbrack 
\begin{array}{c} 
\hat{b}_1 \left( \boldsymbol{k}, t \right) 
\\ 
\hat{b}_2 \left( \boldsymbol{k}, t \right) 
\\
\hat{b}^{\dagger}_1 \left( - \boldsymbol{k}, t \right) 
\\ 
\hat{b}^{\dagger}_2 \left( - \boldsymbol{k}, t \right) 
\end{array}
\right\rbrack , 
\nonumber\\
\label{Bog_trans}
\end{eqnarray}
where $U \left( k \right)$ and $V \left( k \right)$ are $2 \times 2$ matrices 
and $\hat{b}_p \left( \boldsymbol{k}, t \right)$ are bosonic annihilation operators $(p = 1, 2)$. 

If $
\left\lbrack 
\hat{b}_{p} \left( \boldsymbol{k}_1, t \right) , 
\hat{b}^{\dagger}_{q} \left( \boldsymbol{k}_2, t \right)
\right\rbrack = 
\delta_{p, q} \delta_{\boldsymbol{k}_1, \boldsymbol{k}_2}$ and 
$
\left\lbrack 
\hat{b}_p \left( \boldsymbol{k}_1, t \right) , 
\hat{b}_q \left( \boldsymbol{k}_2, t \right)
\right\rbrack = 0$ for $p, q = 1, 2$,

\begin{eqnarray}
\delta_{\boldsymbol{k}_1, \boldsymbol{k}_2} 
\left\{ 
U \left( k_1 \right) 
U^{\dagger} \left( k_1 \right) 
- 
V^{*} \left( k_1 \right) 
V^{T} \left( k_1 \right) 
\right\} 
= \delta_{\boldsymbol{k}_1, \boldsymbol{k}_2} I , 
\quad 
\delta_{\boldsymbol{k}_1, - \boldsymbol{k}_2} 
\left\{ 
U \left( k_1 \right) 
V^{\dagger} \left( k_1 \right) 
- 
V^{*} \left( k_1 \right) 
U^{T} \left( k_1 \right) 
\right\} 
= 0 . 
\end{eqnarray}
Thus, we will set 
\begin{eqnarray}
&& 
U \left( k \right) 
U^{\dagger} \left( k \right) 
- 
V^{*} \left( k \right) 
V^{T} \left( k \right) 
\coloneqq I , 
\quad 
U \left( k \right) 
V^{\dagger} \left( k \right) 
- 
V^{*} \left( k \right) 
U^{T} \left( k \right) 
\coloneqq 0 . \quad
\label{UV_cond1}
\end{eqnarray}

Then one can get 
\begin{eqnarray}
U^{\dagger} \left( k \right) 
U \left( k \right) 
- 
V^{\dagger} \left( k \right) 
V \left( k \right) 
= I , 
\quad 
U^{\dagger} \left( k \right) 
V^{*} \left( k \right) 
- 
V^{\dagger} \left( k \right) 
U^{*} \left( k \right) 
= 0 . \quad
\label{UV_cond2}
\end{eqnarray}

By following the approach in~\cite{Uedareview}, suppose that $M_{B, p}$, $U \left( k \right)$, and $V \left( k \right)$ satisfy 
\begin{eqnarray}
&& g_{a} n \left\lbrack 
\begin{array}{cc} 
M_{B, 1} \left( k \right) 
& 
M_{B, 2} 
\\
- M^{*}_{B, 2} 
& 
- M^{*}_{B, 1} \left( k \right) 
\end{array}
\right\rbrack 
\left\lbrack 
\begin{array}{c} 
U \left( k \right) 
\\
V \left( k \right) 
\end{array}
\right\rbrack 
= 
\left\lbrack 
\begin{array}{c} 
U \left( k \right) 
\\
V \left( k \right) 
\end{array}
\right\rbrack 
\mathcal{E} \left( k \right) , 
\quad \textrm{where } 
\mathcal{E} \left( k \right) \coloneqq 
\hbar 
\left\lbrack 
\begin{array}{cc} 
\omega_1 \left( k \right) & 0 
\\ 
0 & \omega_2 \left( k \right) 
\end{array}
\right\rbrack .
\label{eigenmatrix_UV_def}
\end{eqnarray}

Let 
$
\sigma_{z} \coloneqq 
\left\lbrack 
\begin{array}{cc}
1 & 0
\\
0 & -1
\end{array}
\right\rbrack 
$. 
Then, from Eq.~\eqref{eigenmatrix_UV_def}, 
$
\mathcal{V} \left( k \right) \coloneqq 
\left\lbrack 
\begin{array}{c} 
U \left( k \right) 
\\
V \left( k \right) 
\end{array}
\right\rbrack 
$
is the $4 \times 2$ eigenmatrix of $\sigma_z M \left( k \right)$ and 
Eq.~\eqref{UV_cond2} can be written as
\begin{eqnarray}
\mathcal{V}^{\dagger} \left( k \right) 
\left\lbrack 
\begin{array}{cc} 
I & 0 
\\ 
0 & - I 
\end{array}
\right\rbrack 
\mathcal{V} \left( k \right) 
= I .
\label{eigenmatrix_cond1}
\end{eqnarray}

Note that if Eq.~\eqref{eigenmatrix_UV_def} is satisfied, 
\begin{eqnarray}
&& g_{a} n \left\lbrack 
\begin{array}{cc} 
M_{B, 1} \left( k \right) 
& 
M_{B, 2} 
\\
- M^{*}_{B, 2} 
& 
- M^{*}_{B, 1} \left( k \right) 
\end{array}
\right\rbrack 
\left\lbrack 
\begin{array}{c} 
V^{*} \left( k \right) 
\\
U^{*} \left( k \right) 
\end{array}
\right\rbrack 
= 
- 
\left\lbrack 
\begin{array}{c} 
V^{*} \left( k \right) 
\\
U^{*} \left( k \right) 
\end{array}
\right\rbrack 
\mathcal{E}^{*} \left( k \right) ,
\nonumber\\
\end{eqnarray}
so $\left\lbrack 
\begin{array}{cc} 
V^{\dagger} \left( k \right) 
& 
U^{\dagger} \left( k \right) 
\end{array}
\right\rbrack^{T}$ is also eigenmatrix of $\sigma_z M \left( k \right)$.

However, 
\begin{eqnarray}
&& \left\lbrack 
\begin{array}{c} 
V^{*} \left( k \right) 
\\
U^{*} \left( k \right) 
\end{array}
\right\rbrack^{\dagger} 
\left\lbrack 
\begin{array}{cc} 
I & 0 
\\ 
0 & - I 
\end{array}
\right\rbrack 
\left\lbrack 
\begin{array}{c} 
V^{*} \left( k \right) 
\\
U^{*} \left( k \right) 
\end{array}
\right\rbrack 
= 
- 
\left\{ 
U^{\dagger} \left( k \right) 
U \left( k \right) 
- 
V^{\dagger} \left( k \right) 
V \left( k \right) 
\right\}^{*} 
 = - I ,
\end{eqnarray}
and thus it does not satisfy Eq.~\eqref{eigenmatrix_cond1}.
Therefore, half of the eigenvalues are unphysical values (in the sense that their eigenmatrices $\mathcal{V} \left( k \right)$ do not satisfy Eq.~\eqref{eigenmatrix_cond1} and hence bosonic commutation relations are not satisfied).

After discarding those unphysical eigenvalues, Eq.~\eqref{H_muN_2nd_hom_gen} can be written as 
\begin{eqnarray}
\hat{H} \left( t \right) - \mu \hat{N} & = & 
H_0 - \mu N 
+ \frac{1}{2} 
\sum_{\boldsymbol{k} \neq 0} 
\left\lbrack 
\hbar 
\left\{ 
\omega_1 \left( k \right) 
+ \omega_2 \left( k \right) 
\right\}^{*} 
- g_{a} n 
\left\{ 
M_{11} \left( k \right) + M_{22} \left( k \right) 
\right\} 
\right\rbrack 
\nonumber\\
&& 
+ \hbar 
\sum_{\boldsymbol{k} \neq 0} 
\left\lbrack 
\textrm{Re} 
\left\{ 
\omega_1 \left( k \right) 
\right\} 
\hat{b}^{\dagger}_1 \left( \boldsymbol{k}, t \right) 
\hat{b}_1 \left( \boldsymbol{k}, t \right) 
+ 
\textrm{Re} 
\left\{ 
\omega_2 \left( k \right) 
\right\} 
\hat{b}^{\dagger}_2 \left( \boldsymbol{k}, t \right) 
\hat{b}_2 \left( \boldsymbol{k}, t \right) 
\right\rbrack 
+ O \left( \delta \hat{\Psi}^3_{j} \right) . 
\end{eqnarray}

From~\cite{Uedareview}, by writing 
\begin{eqnarray}
U \left( k \right) 
\coloneqq 
\left\lbrack 
\begin{array}{cc} 
u_1 \left( k \right) 
& 
u_2 \left( k \right) 
\end{array}
\right\rbrack , \quad 
V \left( k \right) 
\coloneqq 
\left\lbrack 
\begin{array}{cc} 
v_1 \left( k \right) 
& 
v_2 \left( k \right) 
\end{array}
\right\rbrack , 
\label{def_ujvj}
\end{eqnarray}
where $u_p \left( k \right) = \left\lbrack \begin{array}{cc} u_{p1} \left( k \right) & u_{p2} \left( k \right) \end{array} \right\rbrack^{T}$ and $v_p \left( k \right) = \left\lbrack \begin{array}{cc} v_{p1} \left( k \right) & v_{p2} \left( k \right) \end{array} \right\rbrack^{T}$ are $2 \times 1$ matrices $(p = 1, 2)$, 
Eq.~\eqref{eigenmatrix_UV_def} can be written as 
\begin{eqnarray}
\left\lbrack 
\begin{array}{cc} 
M_{B, 1} \left( k \right) 
& 
M_{B, 2} 
\\
- M^{*}_{B, 2} 
& 
- M^{*}_{B, 1} \left( k \right) 
\end{array}
\right\rbrack 
\left\lbrack 
\begin{array}{c} 
u_p \left( k \right) 
\\ 
v_p \left( k \right) 
\end{array}
\right\rbrack 
= 
\tilde{\omega}_{p} \left( k \right) 
\left\lbrack 
\begin{array}{c} 
u_p \left( k \right) 
\\ 
v_p \left( k \right) 
\end{array}
\right\rbrack , 
\nonumber\\
\label{Bog_Exc_j}
\end{eqnarray}
where $\tilde{\omega}_p \left( k \right) \coloneqq \hbar \omega_p \left( k \right) / \left( g_{a} n \right)$. 
Note that, if $M_{B, 1} \left( k \right)$, $M_{B, 2}$, and $\tilde{\omega}_p \left( k \right)$ are all real, then $u_p \left( k \right)$ and $v_p \left( k \right)$ should be also real from Eq.~\eqref{Bog_Exc_j}. 
In terms of the above eigenvectors, the $4\times 4$ eigenproblem reads 
\begin{eqnarray}
&& \left\lbrack 
\begin{array}{cccc}
M_{11} \left( k \right) & M_{12} & M_{13} & M_{14} \\
M^{*}_{12} & M_{22} \left( k \right) & M_{14} & M_{24} \\
- M^{*}_{13} & - M^{*}_{14} & - M_{11} \left( k \right) & - M^{*}_{12} \\
- M^{*}_{14} & - M^{*}_{24} & - M_{12} & - M_{22} \left( k \right) 
\end{array}
\right\rbrack 
\left\lbrack 
\begin{array}{c} 
u_{p1} \left( k \right) \\
u_{p2} \left( k \right) \\
v_{p1} \left( k \right) \\
v_{p2} \left( k \right) 
\end{array}
\right\rbrack 
= 
\tilde{\omega}_p \left( k \right) 
\left\lbrack 
\begin{array}{c} 
u_{p1} \left( k \right) \\
u_{p2} \left( k \right) \\
v_{p1} \left( k \right) \\
v_{p2} \left( k \right) 
\end{array}
\right\rbrack 
\label{Bog_matrix_eig}
\end{eqnarray}
where $M_{\zeta \eta}$ is defined from Eqs.~\eqref{M11_hom_gen} to~\eqref{M12_to_24_hom_gen} ($\zeta, \eta = 1, 2$).

Now, to get $\tilde{\omega}_p \left( k \right)$ and its eigenvector $\left\lbrack \begin{array}{cccc} u_{p1} \left( k \right) & u_{p2} \left( k \right) & v_{p1} \left( k \right) & v_{p2} \left( k \right) \end{array} \right\rbrack^{T}$, 
we will follow~\cite{Lin} and introduce 
\begin{eqnarray}
\left\lbrack 
\begin{array}{c} 
\delta \hat{\Psi}_{a} \left( \boldsymbol{k}, t \right) 
\\
\delta \hat{\Psi}_{m} \left( \boldsymbol{k}, t \right) 
\\
\delta \hat{\Psi}^{\dagger}_{a} \left( - \boldsymbol{k}, t \right) 
\\
\delta \hat{\Psi}^{\dagger}_{m} \left( - \boldsymbol{k}, t \right) 
\end{array}
\right\rbrack 
= 
\left\lbrack 
\begin{array}{cccc} 
\cos \phi \left( k \right) & - \sin \phi \left( k \right) & 0 & 0 
\\
\sin \phi \left( k \right) & \cos \phi \left( k \right) & 0 & 0 
\\
0 & 0 & \cos \phi \left( k \right) & - \sin \phi \left( k \right) 
\\
0 & 0 & \sin \phi \left( k \right) & \cos \phi \left( k \right) 
\end{array} 
\right\rbrack 
\left\lbrack 
\begin{array}{c} 
\delta \hat{\Psi}_{R, 1} \left( \boldsymbol{k}, t \right) 
\\
\delta \hat{\Psi}_{R, 2} \left( \boldsymbol{k}, t \right) 
\\
\delta \hat{\Psi}^{\dagger}_{R, 1} \left( - \boldsymbol{k}, t \right) 
\\
\delta \hat{\Psi}^{\dagger}_{R, 2} \left( - \boldsymbol{k}, t \right) 
\end{array}
\right\rbrack , 
\label{Rot_UV}
\end{eqnarray}
and define 
$\Theta \left( k \right) \coloneqq 
\left\lbrack 
\begin{array}{cc} 
\cos \phi \left( k \right) & - \sin \phi \left( k \right) 
\\
\sin \phi \left( k \right) & \cos \phi \left( k \right) 
\end{array}
\right\rbrack
$. 
Note that $\Theta \left( k \right)$ is real $2 \times 2$ matrix and 
$\Theta^{-1} \left( k \right) = \Theta^{T} \left( k \right)$. 
It is trivial to check that $\delta \hat{\Psi}_{R, 1} \left( \boldsymbol{k}, t \right)$ and $\delta \hat{\Psi}_{R, 2} \left( \boldsymbol{k}, t \right)$ are independent bosonic operators.

From the second line in Eq.~\eqref{H_muN_2nd_hom_gen}, $M \left( k \right)$ is transformed to

\begin{eqnarray}
\left\lbrack 
\begin{array}{cc} 
\Theta \left( k \right) & 0 
\\
0 & \Theta \left( k \right) 
\end{array} 
\right\rbrack^{-1} 
M \left( k \right) 
\left\lbrack 
\begin{array}{cccc} 
\Theta \left( k \right) & 0 
\\
0 & \Theta \left( k \right) 
\end{array} 
\right\rbrack 
= 
\left\lbrack 
\begin{array}{cccc} 
A_{11} \left( k \right) & A_{12} \left( k \right) & A_{13} \left( k \right) & A_{14} \left( k \right) 
\\
A_{12} \left( k \right) & A_{22} \left( k \right) & A_{14} \left( k \right) & A_{24} \left( k \right) 
\\
A_{13} \left( k \right) & A_{14} \left( k \right) & A_{11} \left( k \right) & A_{12} \left( k \right) 
\\
A_{14} \left( k \right) & A_{24} \left( k \right) & A_{12} \left( k \right) & A_{22} \left( k \right) 
\end{array}
\right\rbrack , 
\end{eqnarray}
where

\begin{eqnarray}
A_{13} \left( k \right) \coloneqq && \; 
M_{13} \cos^2 \phi \left( k \right) 
+ M_{24} \sin^2 \phi \left( k \right) 
+ M_{14} \sin \left\{ 2 \phi \left( k \right) \right\} , 
\nonumber\\
= && \; 
\frac{1 - 2 y^2 - \left\vert \tilde{\alpha} \right\vert y + \tilde{g}_{m} y^2}{2} 
+ \frac{1 - 2 y^2 - \left\vert \tilde{\alpha} \right\vert y - \tilde{g}_{m} y^2}{2} \cos \left\{ 2 \phi \left( k \right) \right\} 
- \textrm{sign} \left( \alpha \right) \tilde{g}_{am} y \sqrt{1 - 2 y^2} \sin \left\{ 2 \phi \left( k \right) \right\} , 
\qquad \quad 
\end{eqnarray}

\begin{eqnarray}
A_{14} \left( k \right) \coloneqq && \; 
M_{14} \cos \left\{ 2 \phi \left( k \right) \right\} 
- \frac{1}{2} \left( M_{13} - M_{24} \right) \sin \left\{ 2 \phi \left( k \right) \right\} , 
\nonumber\\
= && \; 
- \textrm{sign} \left( \alpha \right) \tilde{g}_{am} y \sqrt{1 - 2 y^2} \cos \left\{ 2 \phi \left( k \right) \right\} 
- \frac{1}{2} 
\left( 
1 - 2 y^2 
- \left\vert \tilde{\alpha} \right\vert y 
- \tilde{g}_{m} y^2 
\right) 
\sin \left\{ 2 \phi \left( k \right) \right\} , 
\end{eqnarray}

\begin{eqnarray}
A_{24} \left( k \right) \coloneqq && \; 
M_{24} \cos^2 \phi \left( k \right) 
+ M_{13} \sin^2 \phi \left( k \right) 
- M_{14} \sin \left\{ 2 \phi \left( k \right) \right\} , 
\nonumber\\
= && \; 
\frac{1 - 2 y^2 - \left\vert \tilde{\alpha} \right\vert y + \tilde{g}_{m} y^2}{2} 
- \frac{1 - 2 y^2 - \left\vert \tilde{\alpha} \right\vert y - \tilde{g}_{m} y^2}{2} \cos \left\{ 2 \phi \left( k \right) \right\} 
+ \textrm{sign} \left( \alpha \right) \tilde{g}_{am} y \sqrt{1 - 2 y^2} \sin \left\{ 2 \phi \left( k \right) \right\} , 
\qquad \quad 
\end{eqnarray}

\begin{eqnarray}
A_{11} \left( k \right) \coloneqq && \; 
M_{11} \left( k \right) 
\cos^2 \phi \left( k \right) 
+ M_{22} \left( k \right) 
\sin^2 \phi \left( k \right) 
+ \textrm{Re} \left( M_{12} \right) 
\sin \left\{ 2 \phi \left( k \right) \right\} 
\nonumber\\
= && \; 
\tilde{\omega}_{c} \left( k \right) \cos^2 \phi \left( k \right) 
+ \tilde{\omega}_{d} \left( k \right) \sin^2 \phi \left( k \right) 
+ \tilde{\alpha} \sqrt{1 - 2 y^2} \sin \left\{ 2 \phi \left( k \right) \right\} 
+ A_{13} \left( k \right) , 
\end{eqnarray}

\begin{eqnarray}
A_{22} \left( k \right) \coloneqq && \; 
M_{22} \left( k \right) 
\cos^2 \phi \left( k \right) 
+ M_{11} \left( k \right) 
\sin^2 \phi \left( k \right) 
- \textrm{Re} \left( M_{12} \right) 
\sin \left\{ 2 \phi \left( k \right) \right\} , 
\nonumber\\
= && \; 
\tilde{\omega}_{d} \left( k \right) \cos^2 \phi \left( k \right) 
+ \tilde{\omega}_{c} \left( k \right) \sin^2 \phi \left( k \right) 
- \tilde{\alpha} \sqrt{1 - 2 y^2} \sin \left\{ 2 \phi \left( k \right) \right\} 
+ A_{24} \left( k \right) , 
\end{eqnarray}

\begin{eqnarray}
A_{12} \left( k \right) \coloneqq && \; 
M_{12} \cos^2 \phi \left( k \right) 
- M^{*}_{12} \sin^2 \phi \left( k \right) 
- \frac{1}{2} \left\{ 
M_{11} \left( k \right) - M_{22} \left( k \right) 
\right\} 
\sin \left\{ 2 \phi \left( k \right) \right\} , 
\nonumber\\
= && \; 
\tilde{\alpha} \sqrt{1 - 2 y^2} \cos \left\{ 2 \phi \left( k \right) \right\} 
- \frac{1}{2} \left\{ 
\tilde{\omega}_{c} \left( k \right) - \tilde{\omega}_{d} \left( k \right)
\right\} \sin \left\{ 2 \phi \left( k \right) \right\} 
+ A_{14} \left( k \right) , 
\end{eqnarray}
$\tilde{\omega}_{c} \left( k \right) \coloneqq k^2 \xi^2_{a} + 2 \left\vert \tilde{\alpha} \right\vert y$, and 
$\tilde{\omega}_{d} \left( k \right) \coloneqq \left( k^2 \xi^2_{a} / 2 \right) + \left\vert \tilde{\alpha} \right\vert \left( 1 - 2 y^2 \right) / \left( 2 y \right) $.

Now, let 
$\tilde{\omega}_{a1} \left( k \right) \coloneqq 
\tilde{\omega}_{c} \left( k \right) \cos^2 \phi \left( k \right) 
+ \tilde{\omega}_{d} \left( k \right) \sin^2 \phi \left( k \right) 
+ \tilde{\alpha} \sqrt{1 - 2 y^2} \sin \left\{ 2 \phi \left( k \right) \right\}$, 
and 
$\tilde{\omega}_{a2} \left( k \right) \coloneqq 
\tilde{\omega}_{c} \left( k \right) \sin^2 \phi \left( k \right) 
+ \tilde{\omega}_{d} \left( k \right) \cos^2 \phi \left( k \right) 
- \tilde{\alpha} \sqrt{1 - 2 y^2} \sin \left\{ 2 \phi \left( k \right) \right\}$. 
By setting $\tan \left\{ 2 \phi \left( k \right) \right\} = 2 \tilde{\alpha} \sqrt{1 - 2 y^2} / \left\{ \tilde{\omega}_{c} \left( k \right) - \tilde{\omega}_{d} \left( k \right) \right\}$ so that $A_{12} \left( k \right) = A_{14} \left( k \right)$, the Bogoliubov $4 \times 4$ eigenproblem becomes

\begin{eqnarray}
\!\!\!\!\!\!\!\!\!\!
\left\lbrack 
\begin{array}{cccc} 
\tilde{\omega}_{a1} \left( k \right) + A_{13} \left( k \right) & A_{14} \left( k \right) & A_{13} \left( k \right) & A_{14} \left( k \right) 
\\
A_{14} \left( k \right) & \tilde{\omega}_{a2} \left( k \right) + A_{24} \left( k \right) & A_{14} \left( k \right) & A_{24} \left( k \right) 
\\
- A_{13} \left( k \right) & - A_{14} \left( k \right) & - \left\{ \tilde{\omega}_{a1} \left( k \right) + A_{13} \left( k \right) \right\} & - A_{14} \left( k \right) 
\\
- A_{14} \left( k \right) & - A_{24} \left( k \right) & - A_{14} \left( k \right) & - \left\{ \tilde{\omega}_{a2} \left( k \right) + A_{24} \left( k \right) \right\} 
\end{array}
\right\rbrack 
\left\lbrack 
\begin{array}{c} 
u_{R, p1} \left( k \right) 
\\
u_{R, p2} \left( k \right) 
\\
v_{R, p1} \left( k \right) 
\\
v_{R, p2} \left( k \right) 
\end{array}
\right\rbrack 
= \tilde{\omega}_{p} \left( k \right) 
\left\lbrack 
\begin{array}{c} 
u_{R, p1} \left( k \right) 
\\
u_{R, p2} \left( k \right) 
\\
v_{R, p1} \left( k \right) 
\\
v_{R, p2} \left( k \right) 
\end{array}
\right\rbrack , 
\qquad \quad 
\label{Rot_Bog}
\end{eqnarray}
where 
\begin{eqnarray}
\tilde{\omega}_{a1} \left( k \right) = && \; \frac{\tilde{\omega}_{c} \left( k \right) + \tilde{\omega}_{d} \left( k \right) + \sqrt{\left\{ \tilde{\omega}_{c} \left( k \right) - \tilde{\omega}_{d} \left( k \right) \right\}^2 + 4 \tilde{\alpha}^2 \left( 1 - 2 y^2 \right)}}{2} , 
\end{eqnarray}

\begin{eqnarray}
\tilde{\omega}_{a2} \left( k \right) = && \; \frac{\tilde{\omega}_{c} \left( k \right) + \tilde{\omega}_{d} \left( k \right) - \sqrt{\left\{ \tilde{\omega}_{c} \left( k \right) - \tilde{\omega}_{d} \left( k \right) \right\}^2 + 4 \tilde{\alpha}^2 \left( 1 - 2 y^2 \right)}}{2} , 
\end{eqnarray}

\begin{eqnarray}
A_{13} \left( k \right) = && \; 
\frac{1 - 2 y^2 - \left\vert \tilde{\alpha} \right\vert y + \tilde{g}_{m} y^2}{2} 
+ 
\frac{
\left( 
1 - 2 y^2 - \left\vert \tilde{\alpha} \right\vert y - \tilde{g}_{m} y^2 
\right) 
\left\{ 
\tilde{\omega}_{c} \left( k \right) - \tilde{\omega}_{d} \left( k \right) 
\right\} 
- 4 \left\vert \tilde{\alpha} \right\vert \tilde{g}_{am} y \left( 1 - 2 y^2 \right) 
}{2 \sqrt{\left\{ \tilde{\omega}_{c} \left( k \right) - \tilde{\omega}_{d} \left( k \right) \right\}^2 + 4 \tilde{\alpha}^2 \left( 1 - 2 y^2 \right)}} , 
\end{eqnarray}

\begin{eqnarray}
A_{14} \left( k \right) = && \; 
- 
\textrm{sign} \left( \alpha \right) 
\sqrt{1 - 2 y^2} 
\frac{
\tilde{g}_{am} y 
\left\{ 
\tilde{\omega}_{c} \left( k \right) - \tilde{\omega}_{d} \left( k \right) 
\right\} 
+ 
\left\vert \tilde{\alpha} \right\vert 
\left( 
1 - 2 y^2 
- \left\vert \tilde{\alpha} \right\vert y 
- \tilde{g}_{m} y^2 
\right) 
}{\sqrt{\left\{ \tilde{\omega}_{c} \left( k \right) - \tilde{\omega}_{d} \left( k \right) \right\}^2 + 4 \tilde{\alpha}^2 \left( 1 - 2 y^2 \right)}} , 
\end{eqnarray}

\begin{eqnarray}
A_{24} \left( k \right) = && \; 
\frac{1 - 2 y^2 - \left\vert \tilde{\alpha} \right\vert y + \tilde{g}_{m} y^2}{2} 
- 
\frac{
\left( 
1 - 2 y^2 - \left\vert \tilde{\alpha} \right\vert y - \tilde{g}_{m} y^2 
\right) 
\left\{ 
\tilde{\omega}_{c} \left( k \right) - \tilde{\omega}_{d} \left( k \right) 
\right\} 
- 4 \left\vert \tilde{\alpha} \right\vert \tilde{g}_{am} y \left( 1 - 2 y^2 \right) 
}{2 \sqrt{\left\{ \tilde{\omega}_{c} \left( k \right) - \tilde{\omega}_{d} \left( k \right) \right\}^2 + 4 \tilde{\alpha}^2 \left( 1 - 2 y^2 \right)}} ,
\end{eqnarray}
$U \left( k \right) = \Theta \left( k \right) U_{R} \left( k \right)$, and $V \left( k \right) = \Theta \left( k \right) V_{R} \left( k \right)$ with

\begin{eqnarray}
U_{R} \left( k \right) \coloneqq 
\left\lbrack 
\begin{array}{cc} 
u_{R, 11} \left( k \right) & u_{R, 21} \left( k \right) 
\\
u_{R, 12} \left( k \right) & u_{R, 22} \left( k \right) 
\end{array}
\right\rbrack , \quad 
V_{R} \left( k \right) \coloneqq 
\left\lbrack 
\begin{array}{cc} 
v_{R, 11} \left( k \right) & v_{R, 21} \left( k \right) 
\\
v_{R, 12} \left( k \right) & v_{R, 22} \left( k \right) 
\end{array}
\right\rbrack . 
\label{UR_VR_def}
\end{eqnarray}

Since $\Theta^{-1} \left( k \right) = \Theta^{T} \left( k \right) = \Theta^{\dagger} \left( k \right)$, $U_{R} \left( k \right)$ and $V_{R} \left( k \right)$ satisfy Bogoliubov conditions in Eqs.~\eqref{UV_cond1} and~\eqref{UV_cond2}. 
Then one can get 

\begin{eqnarray}
\tilde{\omega}_{p} \left( k \right) = 
\sqrt{\frac{
\tilde{\omega}'^2_{a1} \left( k \right) 
+ 
\tilde{\omega}'^2_{a2} \left( k \right) 
+ \left( 2 p - 3 \right) 
\sqrt{
\left\{ 
\tilde{\omega}'^2_{a1} \left( k \right) 
- 
\tilde{\omega}'^2_{a2} \left( k \right) 
\right\}^2 
+ 
16 A^2_{14} \left( k \right) 
\tilde{\omega}_{a1} \left( k \right) 
\tilde{\omega}_{a2} \left( k \right) 
}
}{2}} , 
\label{eigenvalue_omega_p}
\end{eqnarray}
which is same as~\cite{Lin} 
where $\tilde{\omega}'^2_{a1} \left( k \right) \coloneqq 
\tilde{\omega}_{a1} \left( k \right) 
\left\{ 
\tilde{\omega}_{a1} \left( k \right) + 2 A_{13} \left( k \right) 
\right\}
$ and 
$\tilde{\omega}'^2_{a2} \left( k \right) \coloneqq 
\tilde{\omega}_{a2} \left( k \right) 
\left\{ 
\tilde{\omega}_{a2} \left( k \right) + 2 A_{24} \left( k \right) 
\right\}
$. 
Note that $\tilde{\omega}_{p} \left( k \right)$ is independent of the sign of $\tilde{\alpha}$, since 
$\tilde{\omega}'^2_{a1} \left( k \right)$, $\tilde{\omega}'^2_{a2} \left( k \right)$, and $A^{2}_{14} \left( k \right)$ depend on $\left\vert \tilde{\alpha} \right\vert$.

From Eq.~\eqref{eigenvalue_omega_p}), as long as $0 \le y \le 1 / \sqrt{2}$ (physical values of $y$), 
$\tilde{\omega}_{2} \left( k \right)$ is always real and $\tilde{\omega}_{1} \left( k \right)$ is real if 
$\left\{ 
\tilde{\omega}_{a1} \left( k \right) + 2 A_{13} \left( k \right) 
\right\}
\left\{ 
\tilde{\omega}_{a2} \left( k \right) + 2 A_{24} \left( k \right) 
\right\}
\ge 4 A^2_{14} \left( k \right)
$.

Expanding in terms of $k^2$, one can get 
\begin{eqnarray}
\tilde{\omega}_{a1} \left( k \right) = 
\frac{\left\vert \tilde{\alpha} \right\vert \left( 1 + 2 y^2 \right)}{2 y} 
+ \frac{6 y^2 + 1}{2 \left( 1 + 2 y^2 \right)} k^2 \xi^2_{a} 
+ O \left( k^4 \right) , 
\quad 
\tilde{\omega}_{a2} \left( k \right) = 
\frac{1}{1 + 2 y^2} k^2 \xi^2_{a} 
+ O \left( k^4 \right) , 
\end{eqnarray}

\begin{eqnarray}
A_{13} \left( k \right) & = & 
\frac{y^2 
\left\{ 
\left( 4 - 4 \tilde{g}_{am} + \tilde{g}_{m} \right) 
\left( 1 - 2 y^2 \right) 
- 4 \left\vert \tilde{\alpha} \right\vert y 
\right\}
}{1 + 2 y^2} 
- 
\frac{
4 y^3 \left( 1 - 2 y^2 \right) 
\left\{ 
\tilde{g}_{am} - 2 
- 2 \left( 3 \tilde{g}_{am} - \tilde{g}_{m} - 2 \right) y^2 
+ 2 \left\vert \tilde{\alpha} \right\vert y 
\right\} 
}
{\left\vert \tilde{\alpha} \right\vert 
\left( 1 + 2 y^2 \right)^3 
} k^2 \xi^2_{a} 
\nonumber\\
&& 
+ O \left( k^4 \right) , 
\end{eqnarray}

\begin{eqnarray}
A_{24} \left( k \right) & = & 
\frac{
1
+ 4 \left( \tilde{g}_{am} - 1 \right) y^2 
- 4 \left( 2 \tilde{g}_{am} - 1 - \tilde{g}_{m} \right) y^4 
- \left\vert \tilde{\alpha} \right\vert y \left( 1 - 2 y^2 \right) 
}{1 + 2 y^2} 
\nonumber\\
&& + 
\frac{4 y^3 \left( 1 - 2 y^2 \right) 
\left\{ 
\tilde{g}_{am} - 2 
- 2 \left( 3 \tilde{g}_{am} - \tilde{g}_{m} - 2 \right) y^2 
+ 2 \left\vert \tilde{\alpha} \right\vert y 
\right\} 
}{\left\vert \tilde{\alpha} \right\vert 
\left( 1 + 2 y^2 \right)^3 
} 
k^2 \xi^2_{a} 
+ O \left( k^2 \right) , 
\end{eqnarray}
and

\begin{eqnarray}
A_{14} \left( k \right) & = & 
\textrm{sign} \left( \alpha \right) 
\frac{y \sqrt{1 - 2 y^2}}{1 + 2 y^2} 
\left\{ 
\tilde{g}_{am} - 2 
- 2 \left( 3 \tilde{g}_{am} - \tilde{g}_{m} - 2 \right) y^2 
+ 2 \left\vert \tilde{\alpha} \right\vert y 
\right\} 
\nonumber\\
&& 
+ 2 \frac{\textrm{sign} \left( \alpha \right)}{\left\vert \tilde{\alpha} \right\vert} 
\frac{y^2 \sqrt{1 - 2 y^2}}{\left( 1 + 2 y^2 \right)^3} 
\left\{ 
\left\vert \tilde{\alpha} \right\vert y \left( 1 - 6 y^2 \right) 
- 1 
- \left( 8 \tilde{g}_{am} - \tilde{g}_{m} - 8 \right) y^2 
+ 2 \left( 8 \tilde{g}_{am} - 3 \tilde{g}_{m} - 6 \right) y^4 
\right\} k^2 \xi^2_{a} 
+ O \left( k^4 \right) . \qquad \quad 
\end{eqnarray}

Thus, 
\begin{eqnarray}
\tilde{\omega}'^{2}_{a1} \left( k \right) & = & 
\left\vert \tilde{\alpha} \right\vert 
\frac{
\left( 1 - 2 y^2 \right) 
\left\{ 
\left\vert \tilde{\alpha} \right\vert \left( 1 + 6 y^2 \right) 
- 4 \left( 4 \tilde{g}_{am} - \tilde{g}_{m} - 4 \right) y^3 
\right\} 
}{4 y^2}
\nonumber\\
&& 
+ \frac{
2 y^3 \left( 1 - 2 y^2 \right) 
\left\{ 
12 - 8 \tilde{g}_{am} + \tilde{g}_{m} 
- 2 \left( \tilde{g}_{m} - 4 \right) y^2 
\right\} 
+ \left\vert \tilde{\alpha} \right\vert 
\left( 8 y^6 + 4 y^4 + 10 y^2 + 1 \right) 
}{2 y \left( 1 + 2 y^2 \right)^2} k^2 \xi^2_{a} 
+ O \left( k^4 \right) , 
\end{eqnarray}

\begin{eqnarray}
\tilde{\omega}'^{2}_{a2} \left( k \right) & = & 
2 \frac{
1 
+ 4 \left( \tilde{g}_{am} - 1 \right) y^2 
- 4 \left( 2 \tilde{g}_{am} - \tilde{g}_{m} - 1 \right) y^4 
- \left\vert \tilde{\alpha} \right\vert y \left( 1 - 2 y^2 \right) 
}{\left( 1 + 2 y^2 \right)^2} k^2 \xi^2_{a} 
+ O \left( k^4 \right) , 
\end{eqnarray}
and

\begin{eqnarray}
16 A^2_{14} \left( k \right) 
\tilde{\omega}_{a1} \left( k \right) 
\tilde{\omega}_{a2} \left( k \right) & = & 
8 \left\vert \tilde{\alpha} \right\vert 
\frac{y \left( 1 - 2 y^2 \right)}{\left( 1 + 2 y^2 \right)^2} 
\left\{ 
\tilde{g}_{am} - 2 
- 2 \left( 3 \tilde{g}_{am} - \tilde{g}_{m} - 2 \right) y^2 
+ 2 \left\vert \tilde{\alpha} \right\vert y 
\right\}^2 
k^2 \xi^2_{a} 
+ O \left( k^4 \right) .
\end{eqnarray}

For notational convenience, by introducing 
\begin{eqnarray}
A_1 \coloneqq 
\left\vert \tilde{\alpha} \right\vert 
\left( 1 - 2 y^2 \right) 
\frac{
\left\vert \tilde{\alpha} \right\vert \left( 1 + 6 y^2 \right) 
- 4 \left( 4 \tilde{g}_{am} - \tilde{g}_{m} - 4 \right) y^3}{4 y^2} , 
\label{A1_def}
\end{eqnarray}

\begin{eqnarray}
B_1 \coloneqq 
\frac{
2 y^3 \left( 1 - 2 y^2 \right) 
\left\{ 
12 - 8 \tilde{g}_{am} + \tilde{g}_{m} 
- 2 \left( \tilde{g}_{m} - 4 \right) y^2 
\right\} 
+ \left\vert \tilde{\alpha} \right\vert 
\left( 8 y^6 + 4 y^4 + 10 y^2 + 1 \right) 
}{2 y \left( 1 + 2 y^2 \right)^2} , 
\label{B1_def}
\end{eqnarray}

\begin{eqnarray}
B_2 \coloneqq 
2 \frac{
1 
+ 4 \left( \tilde{g}_{am} - 1 \right) y^2 
- 4 \left( 2 \tilde{g}_{am} - \tilde{g}_{m} - 1 \right) y^4 
- \left\vert \tilde{\alpha} \right\vert y \left( 1 - 2 y^2 \right) 
}{\left( 1 + 2 y^2 \right)^2} , 
\label{B2_def}
\end{eqnarray}
and

\begin{eqnarray}
B_3 \coloneqq 
8 \left\vert \tilde{\alpha} \right\vert 
\frac{y \left( 1 - 2 y^2 \right)}{\left( 1 + 2 y^2 \right)^2} 
\left\{ 
\tilde{g}_{am} - 2 
- 2 \left( 3 \tilde{g}_{am} - \tilde{g}_{m} - 2 \right) y^2 
+ 2 \left\vert \tilde{\alpha} \right\vert y 
\right\}^2 , 
\label{B3_def}
\end{eqnarray}
we get

\begin{eqnarray}
\tilde{\omega}_1 \left( k \right) & = & 
\sqrt{
\frac{A_1 - \left\vert A_1 \right\vert}{2} 
+ k^2 \xi^2_{a} 
\left( 
\frac{A_1 - \left\vert A_1 \right\vert}{2 A_1} B_1 
+ \frac{A_1 + \left\vert A_1 \right\vert}{2 A_1} B_2 
- \frac{B_3}{4 \left\vert A_1 \right\vert} 
\right) 
+ O \left( k^4 \right) 
} , 
\nonumber\\
\tilde{\omega}_2 \left( k \right) & = &
\sqrt{
\frac{A_1 + \left\vert A_1 \right\vert}{2} 
+ k^2 \xi^2_{a} 
\left( 
\frac{A_1 + \left\vert A_1 \right\vert}{2 A_1} B_1 
+ \frac{A_1 - \left\vert A_1 \right\vert}{2 A_1} B_2 
+ \frac{B_3}{4 \left\vert A_1 \right\vert} 
\right) 
+ O \left( k^4 \right) 
} . 
\label{omega_12_expression}
\end{eqnarray}

If $A_1 > 0$, 
\begin{eqnarray}
\tilde{\omega}_1 \left( k \right) = 
k \xi_{a} 
\sqrt{
\left( 
B_2 
- \frac{B_3}{4 \left\vert A_1 \right\vert} 
\right) 
+ O \left( k^2 \right) 
} , 
\quad 
\tilde{\omega}_2 \left( k \right) = 
\sqrt{
A_1 
+ k^2 \xi^2_{a} 
\left( 
B_1 
+ \frac{B_3}{4 \left\vert A_1 \right\vert} 
\right) 
+ O \left( k^4 \right) 
} . 
\label{omega_12_expression1}
\end{eqnarray}

From Eq.~\eqref{Rot_Bog}, one can get 

\begin{eqnarray}
u_{R, p2} \left( k \right) & = & 
- \frac{
\tilde{\omega}_{a2} \left( k \right) 
+ 
\tilde{\omega}_{p} \left( k \right)}
{
\tilde{\omega}_{a1} \left( k \right) 
+ 
\tilde{\omega}_{p} \left( k \right)} 
\frac{
\tilde{\omega}'^2_{a1} \left( k \right) 
- 
\tilde{\omega}^2_{p} \left( k \right)}
{2 
\tilde{\omega}_{a2} \left( k \right) 
A_{14} \left( k \right)} 
u_{R, p1} \left( k \right) 
, \quad 
v_{R, p1} \left( k \right) = 
\frac{
\tilde{\omega}_{a1} \left( k \right) 
- 
\tilde{\omega}_{p} \left( k \right)}
{
\tilde{\omega}_{a1} \left( k \right) 
+ 
\tilde{\omega}_{p} \left( k \right)} 
u_{R, p1} \left( k \right) 
, \nonumber\\
v_{R, p2} \left( k \right) & = & 
- \frac{
\tilde{\omega}_{a2} \left( k \right) 
- 
\tilde{\omega}_{p} \left( k \right)}
{
\tilde{\omega}_{a1} \left( k \right) 
+ 
\tilde{\omega}_{p} \left( k \right)} 
\frac{
\tilde{\omega}'^2_{a1} \left( k \right) 
- 
\tilde{\omega}^2_{p} \left( k \right)}
{2 
\tilde{\omega}_{a2} \left( k \right) 
A_{14} \left( k \right)} 
u_{R, p1} \left( k \right) 
. 
\end{eqnarray}

From Eqs.~\eqref{UV_cond1} and~\eqref{UV_cond2},

\begin{eqnarray}
\left\vert 
u_{R, 11} \left( k \right) 
\right\vert^2 
+ 
\left\vert 
u_{R, 21} \left( k \right) 
\right\vert^2 
- 
\left\vert 
v_{R, 11} \left( k \right) 
\right\vert^2 
- 
\left\vert 
v_{R, 21} \left( k \right) 
\right\vert^2 
= 1 
, 
\label{Bog_const_eq1}
\end{eqnarray}

\begin{eqnarray}
\left\vert 
u_{R, 12} \left( k \right) 
\right\vert^2 
+ 
\left\vert 
u_{R, 22} \left( k \right) 
\right\vert^2 
- 
\left\vert 
v_{R, 12} \left( k \right) 
\right\vert^2 
- 
\left\vert 
v_{R, 22} \left( k \right) 
\right\vert^2 
= 1 
, 
\label{Bog_const_eq2}
\end{eqnarray}

\begin{eqnarray}
\left\vert 
u_{R, p1} \left( k \right) 
\right\vert^2 
+ 
\left\vert 
u_{R, p2} \left( k \right) 
\right\vert^2 
- 
\left\vert 
v_{R, p1} \left( k \right) 
\right\vert^2 
- 
\left\vert 
v_{R, p2} \left( k \right) 
\right\vert^2 
= 1 
, 
\label{Bog_const_eq3}
\end{eqnarray}

\begin{eqnarray}
u^{*}_{R, 11} \left( k \right) 
u_{R, 12} \left( k \right) 
+ 
u^{*}_{R, 21} \left( k \right) 
u_{R, 22} \left( k \right) 
- 
v_{R, 11} \left( k \right) 
v^{*}_{R, 12} \left( k \right) 
- 
v_{R, 21} \left( k \right) 
v^{*}_{R, 22} \left( k \right) 
= 0 
, 
\label{Bog_const_eq4}
\end{eqnarray}

\begin{eqnarray}
u_{R, 11} \left( k \right) 
u^{*}_{R, 21} \left( k \right) 
+ 
u_{R, 12} \left( k \right) 
u^{*}_{R, 22} \left( k \right) 
- 
v_{R, 11} \left( k \right) 
v^{*}_{R, 21} \left( k \right) 
- 
v_{R, 12} \left( k \right) 
v^{*}_{R, 22} \left( k \right) 
= 0 
, 
\label{Bog_const_eq5}
\end{eqnarray}

\begin{eqnarray}
u_{R, 12} \left( k \right) 
v^{*}_{R, 11} \left( k \right) 
+ 
u_{R, 22} \left( k \right) 
v^{*}_{R, 21} \left( k \right) 
- 
u_{R, 11} \left( k \right) 
v^{*}_{R, 12} \left( k \right) 
- 
u_{R, 21} \left( k \right) 
v^{*}_{R, 22} \left( k \right) 
= 0 
, 
\label{Bog_const_eq6}
\end{eqnarray}
and

\begin{eqnarray}
u_{R, 21} \left( k \right) 
v_{R, 11} \left( k \right) 
+ 
u_{R, 22} \left( k \right) 
v_{R, 12} \left( k \right) 
- 
u_{R, 11} \left( k \right) 
v_{R, 21} \left( k \right) 
- 
u_{R, 12} \left( k \right) 
v_{R, 22} \left( k \right) 
= 0 
. 
\label{Bog_const_eq7}
\end{eqnarray}

For simplicity, we concentrate on a stable system where every term in Eq.~\eqref{Rot_Bog} is real. 
From Eq.~\eqref{Bog_const_eq3},

\begin{eqnarray}
u_{R, p1} \left( k \right) 
= 
\frac{\left\{ 
\tilde{\omega}_{a1} \left( k \right) 
+ 
\tilde{\omega}_{p} \left( k \right) 
\right\} 
\left\vert 
A_{14} \left( k \right) 
\right\vert 
\sqrt{\tilde{\omega}_{a2} \left( k \right)}}
{\sqrt{
\tilde{\omega}_{p} \left( k \right) 
\left\lbrack 
4 
\tilde{\omega}_{a1} \left( k \right) 
\tilde{\omega}_{a2} \left( k \right) 
A^2_{14} \left( k \right) 
+ 
\left\{ 
\tilde{\omega}'^2_{a1} \left( k \right) 
- 
\tilde{\omega}^2_{p} \left( k \right) 
\right\}^2 
\right\rbrack}} 
, 
\end{eqnarray}

\begin{eqnarray}
u_{R, p2} \left( k \right) = 
- \textrm{sign} \left\{ A_{14} \left( k \right) \right\} 
\frac{
\left\{ 
\tilde{\omega}_{a2} \left( k \right) 
+ 
\tilde{\omega}_{p} \left( k \right) 
\right\} 
\left\{ 
\tilde{\omega}'^2_{a1} \left( k \right) 
- 
\tilde{\omega}^2_{p} \left( k \right) 
\right\}}
{2 \sqrt{
\tilde{\omega}_{a2} \left( k \right) 
\tilde{\omega}_{p} \left( k \right) 
\left\lbrack 
4 
\tilde{\omega}_{a1} \left( k \right) 
\tilde{\omega}_{a2} \left( k \right) 
A^2_{14} \left( k \right) 
+ 
\left\{ 
\tilde{\omega}'^2_{a1} \left( k \right) 
- 
\tilde{\omega}^2_{p} \left( k \right) 
\right\}^2 
\right\rbrack}} 
, \qquad \quad 
\end{eqnarray}

\begin{eqnarray}
v_{R, p1} \left( k \right) = 
\frac{\left\{ 
\tilde{\omega}_{a1} \left( k \right) 
- 
\tilde{\omega}_{p} \left( k \right) 
\right\} 
\left\vert 
A_{14} \left( k \right) 
\right\vert 
\sqrt{\tilde{\omega}_{a2} \left( k \right)}}
{\sqrt{
\tilde{\omega}_{p} \left( k \right) 
\left\lbrack 
4 
\tilde{\omega}_{a1} \left( k \right) 
\tilde{\omega}_{a2} \left( k \right) 
A^2_{14} \left( k \right) 
+ 
\left\{ 
\tilde{\omega}'^2_{a1} \left( k \right) 
- 
\tilde{\omega}^2_{p} \left( k \right) 
\right\}^2 
\right\rbrack}} 
, 
\end{eqnarray}
and

\begin{eqnarray}
v_{R, p2} \left( k \right) = 
- \textrm{sign} \left\{ A_{14} \left( k \right) \right\} 
\frac{\left\{ 
\tilde{\omega}_{a2} \left( k \right) 
- 
\tilde{\omega}_{p} \left( k \right) 
\right\} 
\left\{ 
\tilde{\omega}'^2_{a1} \left( k \right) 
- 
\tilde{\omega}^2_{p} \left( k \right) 
\right\}}
{2 \sqrt{
\tilde{\omega}_{a2} \left( k \right) 
\tilde{\omega}_{p} \left( k \right) 
\left\lbrack 
4 
\tilde{\omega}_{a1} \left( k \right) 
\tilde{\omega}_{a2} \left( k \right) 
A^2_{14} \left( k \right) 
+ 
\left\{ 
\tilde{\omega}'^2_{a1} \left( k \right) 
- 
\tilde{\omega}^2_{p} \left( k \right) 
\right\}^2 
\right\rbrack}} 
. \qquad \quad 
\end{eqnarray}

From Eq.~\eqref{UR_VR_def}, 
$u_{p1} \left( k \right) = 
\cos \phi \left( k \right) 
u_{R, p1} \left( k \right) 
- 
\sin \phi \left( k \right) 
u_{R, p2} \left( k \right)$, 
$u_{p2} \left( k \right) = 
\sin \phi \left( k \right) 
u_{R, p1} \left( k \right) 
+ 
\cos \phi \left( k \right) 
u_{R, p2} \left( k \right)$, 
$v_{p1} \left( k \right) = 
\cos \phi \left( k \right) 
v_{R, p1} \left( k \right) 
- 
\sin \phi \left( k \right) 
v_{R, p2} \left( k \right)$, 
and 
$v_{p2} \left( k \right) = 
\sin \phi \left( k \right) 
v_{R, p1} \left( k \right) 
+ 
\cos \phi \left( k \right) 
v_{R, p2} \left( k \right)$.

As we already pointed out below Eq.~\eqref{UR_VR_def}, signs of $\cos \phi \left( k \right)$ and $\sin \phi \left( k \right)$ do not change the Bogoliubov conditions in Eqs.~\eqref{UV_cond1} and~\eqref{UV_cond2}. 
Thus, we will choose signs such that $u_{p1} \left( k \right) \ge 0$ for all $k > 0$. 
Then we get

\begin{eqnarray}
u_{p1} \left( k \right) 
= 
\sqrt{\frac{
u^2_{R, p1} \left( k \right) 
+ 
u^2_{R, p2} \left( k \right) 
+ 
\left\{ 
u^2_{R, p1} \left( k \right) 
- 
u^2_{R, p2} \left( k \right) 
\right\} 
\cos \left\{ 2 \phi \left( k \right) \right\} 
- 2 
u_{R, p1} \left( k \right) 
u_{R, p2} \left( k \right) 
\sin \left\{ 2 \phi \left( k \right) \right\} 
}{2}} 
, \qquad 
\label{up1_analytic}
\end{eqnarray}

\begin{eqnarray}
u_{p2} \left( k \right) 
& = & 
\sqrt{\frac{
u^2_{R, p1} \left( k \right) 
+ 
u^2_{R, p2} \left( k \right) 
- 
\left\{ 
u^2_{R, p1} \left( k \right) 
- 
u^2_{R, p2} \left( k \right) 
\right\} 
\cos \left\{ 2 \phi \left( k \right) \right\} 
+ 2 
u_{R, p1} \left( k \right) 
u_{R, p2} \left( k \right) 
\sin \left\{ 2 \phi \left( k \right) \right\} 
}{2}} 
\nonumber\\
&& \times 
\textrm{sign} 
\left\lbrack 
2 
u_{R, p1} \left( k \right) 
u_{R, p2} \left( k \right) 
\cos \left\{ 2 \phi \left( k \right) \right\} 
+ 
\left\{ 
u^2_{R, p1} \left( k \right) 
- 
u^2_{R, p2} \left( k \right) 
\right\} 
\sin \left\{ 2 \phi \left( k \right) \right\} 
\right\rbrack 
, \qquad 
\label{up2_analytic}
\end{eqnarray}

\begin{eqnarray}
v_{p1} \left( k \right) 
& = & 
\sqrt{\frac{
v^2_{R, p1} \left( k \right) 
+ 
v^2_{R, p2} \left( k \right) 
+ 
\left\{ 
v^2_{R, p1} \left( k \right) 
- 
v^2_{R, p2} \left( k \right) 
\right\} 
\cos \left\{ 2 \phi \left( k \right) \right\} 
- 2 
v_{R, p1} \left( k \right) 
v_{R, p2} \left( k \right) 
\sin \left\{ 2 \phi \left( k \right) \right\} 
}{2}} 
\nonumber\\
&& 
\times 
\textrm{sign} 
\left\lbrack 
\begin{array}{c}
u_{R, p1} \left( k \right) 
v_{R, p1} \left( k \right) 
+ 
u_{R, p2} \left( k \right) 
v_{R, p2} \left( k \right) 
+ 
\left\{ 
u_{R, p1} \left( k \right) 
v_{R, p1} \left( k \right) 
- 
u_{R, p2} \left( k \right) 
v_{R, p2} \left( k \right) 
\right\} 
\cos \left\{ 2 \phi \left( k \right) \right\} 
\\
\!\!\!\!\!\!\!\!\!\!\!\!\!\!\!\!\!\!\!\!\!\!\!\!\!\!\!\!\!\!\!\!\!\!\!\!\!\!\!\!\!\!\!\!\!\!\!\!\!\!\!\!\!\!\!\!\!\!\!\!\!\!\!\!\!\!\!\!\!\!\!\!\!\!\!\!\!\!\!\!\!\!\!\!\!\!\!\!\!\!\!\!\!\!\!
- 
\left\{ 
u_{R, p1} \left( k \right) 
v_{R, p2} \left( k \right) 
+ 
u_{R, p2} \left( k \right) 
v_{R, p1} \left( k \right) 
\right\} 
\sin \left\{ 2 \phi \left( k \right) \right\} 
\end{array}
\right\rbrack 
, \qquad \quad 
\label{vp1_analytic}
\end{eqnarray}
and

\begin{eqnarray}
v_{p2} \left( k \right) 
& = & 
\sqrt{\frac{
v^2_{R, p1} \left( k \right) 
+ 
v^2_{R, p2} \left( k \right) 
- 
\left\{ 
v^2_{R, p1} \left( k \right) 
- 
v^2_{R, p2} \left( k \right) 
\right\} 
\cos \left\{ 2 \phi \left( k \right) \right\} 
+ 2 
v_{R, p1} \left( k \right) 
v_{R, p2} \left( k \right) 
\sin \left\{ 2 \phi \left( k \right) \right\} 
}{2}} 
\nonumber\\
&& 
\times 
\textrm{sign} 
\left\lbrack 
\begin{array}{c}
u_{R, p1} \left( k \right) 
v_{R, p2} \left( k \right) 
- 
u_{R, p2} \left( k \right) 
v_{R, p1} \left( k \right) 
+ 
\left\{ 
u_{R, p1} \left( k \right) 
v_{R, p2} \left( k \right) 
+ 
u_{R, p2} \left( k \right) 
v_{R, p1} \left( k \right) 
\right\} 
\cos \left\{ 2 \phi \left( k \right) \right\} 
\\
\!\!\!\!\!\!\!\!\!\!\!\!\!\!\!\!\!\!\!\!\!\!\!\!\!\!\!\!\!\!\!\!\!\!\!\!\!\!\!\!\!\!\!\!\!\!\!\!\!\!\!\!\!\!\!\!\!\!\!\!\!\!\!\!\!\!\!\!\!\!\!\!\!\!\!\!\!\!\!\!\!\!\!\!\!\!\!\!\!\!\!\!\!\!\!
+ 
\left\{ 
u_{R, p1} \left( k \right) 
v_{R, p1} \left( k \right) 
- 
u_{R, p2} \left( k \right) 
v_{R, p2} \left( k \right) 
\right\} 
\sin \left\{ 2 \phi \left( k \right) \right\} 
\end{array}
\right\rbrack 
, \qquad \quad 
\label{vp2_analytic}
\end{eqnarray}
where $\tan \left\{ 2 \phi \left( k \right) \right\} = 2 \tilde{\alpha} \sqrt{1 - 2 y^2} / \left\{ \tilde{\omega}_{c} \left( k \right) - \tilde{\omega}_{d} \left( k \right) \right\}$. 
For small $k$ limit,

\begin{eqnarray}
u_{11} \left( k \right) = 
\frac{\sqrt{1 - 2 y^2}}{2 \sqrt{k}} 
\left( B_2 - \frac{B_3}{4 \left\vert A_1 \right\vert} \right)^{1/4} 
+ O \left( \sqrt{k} \right) 
, \quad 
u_{12} \left( k \right) = 
- \frac{y}{\sqrt{k}} 
\left( B_2 - \frac{B_3}{4 \left\vert A_1 \right\vert} \right)^{1/4} 
+ O \left( \sqrt{k} \right) 
, \quad 
\end{eqnarray}

\begin{eqnarray}
v_{11} \left( k \right) = 
- \textrm{sign} \left( 1 - 6 y^2 \right) 
\frac{\sqrt{1 - 2 y^2}}{2 \sqrt{k}} 
\left( B_2 - \frac{B_3}{4 \left\vert A_1 \right\vert} \right)^{1/4} 
+ O \left( \sqrt{k} \right) 
, \quad 
v_{12} \left( k \right) = 
\frac{y}{\sqrt{k}} 
\left( B_2 - \frac{B_3}{4 \left\vert A_1 \right\vert} \right)^{1/4} 
+ O \left( \sqrt{k} \right) 
, \qquad \quad 
\end{eqnarray}

\section{\label{Reaction_Op}Reaction Rate Operator}

In the Heisenberg picture, let 
$
\hat{N}_j \left( t \right) \coloneqq 
\int d^3 r \; 
\hat{\psi}^{\dagger}_j \left( \boldsymbol{r}, t \right) 
\hat{\psi}_j \left( \boldsymbol{r}, t \right) 
$ 
where $j = a, m$. 
From Eqs.~\eqref{hat_psi_eqofmot}, 
we get 
\begin{eqnarray}
i \hbar \frac{\partial \hat{N}_{a} \left( t \right)}{\partial t} & = & 
\int d^3 r \; 
\left\lbrack 
\hat{\psi}^{\dagger}_{a} \left( \boldsymbol{r}, t \right) 
\left\{ 
i \hbar \frac{\partial \hat{\psi}_{a} \left( \boldsymbol{r}, t \right)}{\partial t} 
\right\} 
- h. c. 
\right\rbrack 
= 
\alpha \sqrt{2} 
\int d^3 r \; 
\left\{ 
\hat{\psi}^{\dagger}_{a} \left( \boldsymbol{r}, t \right) 
\hat{\psi}^{\dagger}_{a} \left( \boldsymbol{r}, t \right) 
\hat{\psi}_{m} \left( \boldsymbol{r}, t \right) 
- h. c. 
\right\} 
\nonumber\\
& = & 
- 2 i \hbar \frac{\partial \hat{N}_{m} \left( t \right)}{\partial t} .
\end{eqnarray}
We define the dimensionless {\it reaction rate operator} $\hat{R} \left( t \right)$ by 
\begin{eqnarray}
\hat{R} \left( t \right) 
& \coloneqq & 
\frac{\hbar}{g_{a} n} \frac{\partial \hat{N}_{a} \left( t \right)}{\partial t} 
= \frac{\partial \hat{N}_{a} \left( t \right)}{\partial \tilde{t}} 
= 
i \frac{\alpha \sqrt{2}}{g_{a} n} 
\int d^3 r \; 
\left\{ 
\hat{\psi}_{a} \left( \boldsymbol{r}, t \right) 
\hat{\psi}_{a} \left( \boldsymbol{r}, t \right) 
\hat{\psi}^{\dagger}_{m} \left( \boldsymbol{r}, t \right) 
- h. c. 
\right\} , 
\end{eqnarray}
where $\tilde{t} \coloneqq g_{a} n t / \hbar$ is dimensionless time. 
If $\left\langle \psi_{R} \right\vert \hat{R} \left( t \right) \left\vert \psi_{R} \right\rangle \neq 0$ for some state $\left\vert \psi_{R} \right\rangle$, then that state describes reaction between atoms and molecules.

For a homogeneous system with $V_{a} \left( \boldsymbol{r}, t \right) = 0$ and $V_{m} \left( \boldsymbol{r}, t \right) = 0$, 
by writing
\begin{eqnarray}
\hat{\psi}_{a} \left( \boldsymbol{r}, t \right) 
= e^{- i \mu t / \hbar} 
\left\{ 
x \sqrt{n} e^{i \varphi_{a}} 
+ 
\frac{1}{\sqrt{V}} 
\sum_{\boldsymbol{k} \neq 0} 
e^{i \boldsymbol{k} \cdot \boldsymbol{r}}
\delta \hat{\Psi}_{a} \left( \boldsymbol{k}, t \right) 
\right\} , \quad 
\hat{\psi}_{m} \left( \boldsymbol{r}, t \right) 
= e^{- 2 i \mu t / \hbar} 
\left\{ 
y \sqrt{n} e^{i \varphi_{m}} 
+ 
\frac{1}{\sqrt{V}} 
\sum_{\boldsymbol{k} \neq 0} 
e^{i \boldsymbol{k} \cdot \boldsymbol{r}}
\delta \hat{\Psi}_{m} \left( \boldsymbol{k}, t \right) 
\right\}\,.
\nonumber\\
\end{eqnarray}
The reaction rate operator $\hat{R} \left( t \right)$ can be written as 
\begin{eqnarray}
\hat{R} \left( t \right) 
= && \; 
2 N 
\tilde{\alpha} x^2 y \sin \varphi_{am} 
\nonumber\\
&& - 2 i 
\tilde{\alpha} x 
\sum_{\boldsymbol{k} \neq 0} 
\left\{ 
e^{- i \varphi_{a}} 
\delta \hat{\Psi}^{\dagger}_{a} \left( \boldsymbol{k}, t \right) 
\delta \hat{\Psi}_{m} \left( \boldsymbol{k}, t \right) 
- 
e^{i \varphi_{a}} 
\delta \hat{\Psi}_{a} \left( \boldsymbol{k}, t \right) 
\delta \hat{\Psi}^{\dagger}_{m} \left( \boldsymbol{k}, t \right) 
\right\} 
\nonumber\\
&& - i 
\tilde{\alpha} y 
\sum_{\boldsymbol{k} \neq 0} 
\left\{ 
e^{i \varphi_{m}} 
\delta \hat{\Psi}^{\dagger}_{a} \left( \boldsymbol{k}, t \right) 
\delta \hat{\Psi}^{\dagger}_{a} \left( - \boldsymbol{k}, t \right) 
- 
e^{- i \varphi_{m}} 
\delta \hat{\Psi}_{a} \left( \boldsymbol{k}, t \right) 
\delta \hat{\Psi}_{a} \left( - \boldsymbol{k}, t \right) 
\right\} 
\nonumber\\
&& - i 
\frac{\tilde{\alpha}}{\sqrt{N}} 
\sum_{\boldsymbol{k}_1 \neq 0} 
\sum_{\boldsymbol{k}_2 \neq 0} 
\left\{ 
\delta \hat{\Psi}^{\dagger}_{a} \left( \boldsymbol{k}_1, t \right) 
\delta \hat{\Psi}^{\dagger}_{a} \left( \boldsymbol{k}_2, t \right) 
\delta \hat{\Psi}_{m} \left( \boldsymbol{k}_1 + \boldsymbol{k}_2, t \right) 
- 
\delta \hat{\Psi}_{a} \left( \boldsymbol{k}_1, t \right) 
\delta \hat{\Psi}_{a} \left( \boldsymbol{k}_2, t \right) 
\delta \hat{\Psi}^{\dagger}_{m} \left( \boldsymbol{k}_1 + \boldsymbol{k}_2, t \right) 
\right\} , \qquad 
\label{Reaction_Operator_k}
\end{eqnarray}
where $V$ is the volume of the system. 
Now, let 
\begin{eqnarray}
\hat{R}_2 \left( t \right) 
\coloneqq && \; 
- 2 i \tilde{\alpha} x 
\sum_{\boldsymbol{k} \neq 0} 
\left\{ 
e^{- i \varphi_{a}} 
\delta \hat{\Psi}^{\dagger}_{a} \left( \boldsymbol{k}, t \right) 
\delta \hat{\Psi}_{m} \left( \boldsymbol{k}, t \right) 
- 
e^{i \varphi_{a}} 
\delta \hat{\Psi}_{a} \left( \boldsymbol{k}, t \right) 
\delta \hat{\Psi}^{\dagger}_{m} \left( \boldsymbol{k}, t \right) 
\right\} 
\nonumber\\
&& - i 
\tilde{\alpha} y 
\sum_{\boldsymbol{k} \neq 0} 
\left\{ 
e^{i \varphi_{m}} 
\delta \hat{\Psi}^{\dagger}_{a} \left( \boldsymbol{k}, t \right) 
\delta \hat{\Psi}^{\dagger}_{a} \left( - \boldsymbol{k}, t \right) 
- 
e^{- i \varphi_{m}} 
\delta \hat{\Psi}_{a} \left( \boldsymbol{k}, t \right) 
\delta \hat{\Psi}_{a} \left( - \boldsymbol{k}, t \right) 
\right\} 
\nonumber\\
= && \; 
- i \tilde{\alpha} 
\sum_{\boldsymbol{k} \neq 0} 
\left\lbrack 
\begin{array}{cccc}
\delta \hat{\Psi}^{\dagger}_{a} \left( \boldsymbol{k}, t \right) 
& 
\delta \hat{\Psi}^{\dagger}_{m} \left( \boldsymbol{k}, t \right) 
&
\delta \hat{\Psi}_{a} \left( - \boldsymbol{k}, t \right) 
&
\delta \hat{\Psi}_{m} \left( - \boldsymbol{k}, t \right) 
\end{array}
\right\rbrack 
\left\lbrack 
\begin{array}{cc} 
M_{R, 1} & M_{R, 2} 
\\
- M^{*}_{R, 2} & - M^{*}_{R, 1} 
\end{array}
\right\rbrack 
\left\lbrack 
\begin{array}{c} 
\delta \hat{\Psi}_{a} \left( \boldsymbol{k}, t \right) 
\\
\delta \hat{\Psi}_{m} \left( \boldsymbol{k}, t \right) 
\\
\delta \hat{\Psi}^{\dagger}_{a} \left( - \boldsymbol{k}, t \right) 
\\
\delta \hat{\Psi}^{\dagger}_{m} \left( - \boldsymbol{k}, t \right) 
\end{array}
\right\rbrack ,
\label{R2_op_def}
\end{eqnarray}
where

\begin{eqnarray}
M_{R, 1} 
\coloneqq 
\left\lbrack 
\begin{array}{cc} 
0 & x e^{- i \varphi_{a}} 
\\
- x e^{i \varphi_{a}} & 0 
\end{array}
\right\rbrack , \quad 
M_{R, 2} 
\coloneqq 
\left\lbrack 
\begin{array}{cc} 
y e^{i \varphi_{m}} & 0 
\\
0 & 0 
\end{array}
\right\rbrack , 
\nonumber\\
\label{MR1_MR2_def}
\end{eqnarray}
and $x^2 + 2 y^2 = 1$ from the number conservation in Eq.~\eqref{hat_N_def}.

Inserting Eqs.~\eqref{Bog_trans} into Eq.~\eqref{R2_op_def}, one can get 
\begin{eqnarray}
\hat{R}_2 \left( t \right) 
= && \; 
- i \tilde{\alpha} 
\sum_{\boldsymbol{k} \neq 0} 
\left\lbrack 
\begin{array}{cccc} 
\hat{b}^{\dagger}_1 \left( \boldsymbol{k}, t \right) 
& 
\hat{b}^{\dagger}_2 \left( \boldsymbol{k}, t \right) 
&
\hat{b}_1 \left( - \boldsymbol{k}, t \right) 
& 
\hat{b}_2 \left( - \boldsymbol{k}, t \right) 
\end{array}
\right\rbrack 
M_{R, B} \left( k \right) 
\left\lbrack 
\begin{array}{c} 
\hat{b}_1 \left( \boldsymbol{k}, t \right) 
\\ 
\hat{b}_2 \left( \boldsymbol{k}, t \right) 
\\
\hat{b}^{\dagger}_1 \left( - \boldsymbol{k}, t \right) 
\\ 
\hat{b}^{\dagger}_2 \left( - \boldsymbol{k}, t \right) 
\end{array}
\right\rbrack ,
\label{R2_in_hatb}
\end{eqnarray}
where

\begin{eqnarray}
M_{R, B} \left( k \right) \coloneqq 
\left\lbrack 
\begin{array}{cc} 
U^{\dagger} \left( k \right) 
& 
V^{\dagger} \left( k \right) 
\\
V^{T} \left( k \right) 
& 
U^{T} \left( k \right) 
\end{array}
\right\rbrack 
\left\lbrack 
\begin{array}{cc} 
M_{R, 1} & M_{R, 2} 
\\
- M^{*}_{R, 2} & - M^{*}_{R, 1} 
\end{array}
\right\rbrack 
\left\lbrack 
\begin{array}{cc} 
U \left( k \right) 
& 
V^{*} \left( k \right) 
\\
V \left( k \right) 
& 
U^{*} \left( k \right) 
\end{array}
\right\rbrack .
\end{eqnarray}

After some lengthy calculations, one can show that $M_{R, B}$ can be written as 
\begin{eqnarray}
M_{R, B} \left( k \right) = && \; 
\left\lbrack 
\begin{array}{cc}
M_{R, B; 1} \left( k \right) & M_{R, B; 2} \left( k \right) 
\\
- M^{*}_{R, B; 2} \left( k \right) & - M^{*}_{R, B; 1} \left( k \right) 
\end{array}
\right\rbrack , 
\label{M_RB_general_form}
\end{eqnarray}
where 
\begin{eqnarray}
M_{R, B; 1} \left( k \right) \coloneqq && \; 
U^{\dagger} \left( k \right) 
M_{R, 1} 
U \left( k \right) 
- 
V^{\dagger} \left( k \right) 
M^{*}_{R, 1} 
V \left( k \right) 
+ 
U^{\dagger} \left( k \right) 
M_{R, 2} 
V \left( k \right) 
- 
V^{\dagger} \left( k \right) 
M^{*}_{R, 2} 
U \left( k \right) , 
\nonumber\\
\label{M_RB_1_def}
\end{eqnarray}
and 
\begin{eqnarray}
M_{R, B; 2} \left( k \right) \coloneqq && \; 
U^{\dagger} \left( k \right) 
M_{R, 1} 
V^{*} \left( k \right) 
- 
V^{\dagger} \left( k \right) 
M^{*}_{R, 1} 
U^{*} \left( k \right) 
+ 
U^{\dagger} \left( k \right) 
M_{R, 2} 
U^{*} \left( k \right) 
- 
V^{\dagger} \left( k \right) 
M^{*}_{R, 2} 
V^{*} \left( k \right) . 
\nonumber\\
\label{M_RB_2_def}
\end{eqnarray}

Since $M^{\dagger}_{R, 1} = - M_{R, 1}$ and 
$M^{\dagger}_{R, 2} = M^{*}_{R, 2}$, 
$M^{\dagger}_{R, B} = - M_{R, B}$. 
Thus, we may write

\begin{eqnarray}
M_{R, B; 1} \left( k \right) = 
\left\lbrack 
\begin{array}{cc} 
\Xi_{11} \left( k \right) & \Xi_{12} \left( k \right) 
\\
- \Xi^{*}_{12} \left( k \right) & \Xi_{22} \left( k \right) 
\end{array}
\right\rbrack , 
\quad 
M_{R, B; 2} \left( k \right) = 
\left\lbrack 
\begin{array}{cc} 
\Xi_{13} \left( k \right) & \Xi_{14} \left( k \right) 
\\
\Xi_{14} \left( k \right) & \Xi_{24} \left( k \right) 
\end{array}
\right\rbrack ,
\label{Xi_def_12}
\end{eqnarray}
with $\Xi_{\zeta \zeta} \left( k \right)$ being purely imaginary ($\zeta = 1, 2$) 
as 
$M^{\dagger}_{R, B; 1} \left( k \right) = - M_{R, B; 1} \left( k \right)$ and 
$M^{\dagger}_{R, B; 2} \left( k \right) = M^{*}_{R, B; 2} \left( k \right)$.

Thus, Eq.~\eqref{R2_in_hatb} can be written as 
\begin{eqnarray}
\hat{R}_2 \left( t \right) = && \; 
\tilde{\alpha} 
\sum_{\boldsymbol{k} \neq 0} 
\textrm{Im} 
\left\{ 
\Xi_{11} \left( k \right) 
\right\} 
\left\{ 
2 \hat{b}^{\dagger}_1 \left( \boldsymbol{k}, t \right) 
\hat{b}_1 \left( \boldsymbol{k}, t \right) 
+ 1
\right\} 
+ \tilde{\alpha} 
\sum_{\boldsymbol{k} \neq 0} 
\textrm{Im} 
\left\{ 
\Xi_{22} \left( k \right) 
\right\} 
\left\{ 
2 \hat{b}^{\dagger}_2 \left( \boldsymbol{k}, t \right) 
\hat{b}_2 \left( \boldsymbol{k}, t \right) 
+ 1
\right\} 
\nonumber\\
&& - 2 i \tilde{\alpha} 
\sum_{\boldsymbol{k} \neq 0} 
\left\{ 
\Xi_{12} \left( k \right) 
\hat{b}^{\dagger}_1 \left( \boldsymbol{k}, t \right) 
\hat{b}_2 \left( \boldsymbol{k}, t \right) 
- h. c. 
\right\} 
- i \tilde{\alpha} 
\sum_{\boldsymbol{k} \neq 0} 
\left\{ 
\Xi_{13} \left( k \right) 
\hat{b}^{\dagger}_1 \left( \boldsymbol{k}, t \right) 
\hat{b}^{\dagger}_1 \left( - \boldsymbol{k}, t \right) 
- h. c. 
\right\} 
\nonumber\\
&& - 2 i \tilde{\alpha} 
\sum_{\boldsymbol{k} \neq 0} 
\left\{ 
\Xi_{14} \left( k \right) 
\hat{b}^{\dagger}_1 \left( \boldsymbol{k}, t \right) 
\hat{b}^{\dagger}_2 \left( - \boldsymbol{k}, t \right) 
- h. c. 
\right\} 
- i \tilde{\alpha} 
\sum_{\boldsymbol{k} \neq 0} 
\left\{ 
\Xi_{24} \left( k \right) 
\hat{b}^{\dagger}_2 \left( \boldsymbol{k}, t \right) 
\hat{b}^{\dagger}_2 \left( - \boldsymbol{k}, t \right) 
- h. c. 
\right\} .
\label{R2_express}
\end{eqnarray}

Since we set $\cos \varphi_{am} = - \textrm{sign} \left( \alpha \right)$, 
$\sin \varphi_{am} = 0$ and $\hat{R} \left( t \right) = \hat{R}_2 \left( t \right) + O \left( \delta \hat{\Psi}^3_j \right)$. 
We also imposed $\varphi_{a} = 0$ below Eq.~\eqref{M12_to_24_hom_gen}, which leads to 
\begin{eqnarray}
&& \!\!\!\!\!
\Xi_{pp} \left( k \right) 
= 
2 i \sqrt{1 - 2 y^2} 
\textrm{Im} 
\left\{ 
u^{*}_{R, p1} \left( k \right) 
u_{R, p2} \left( k \right) 
- 
v^{*}_{R, p1} \left( k \right) 
v_{R, p2} \left( k \right) 
\right\} 
- i \textrm{sign} \left( \alpha \right) y 
\textrm{Im} 
\left\{ 
u^{*}_{R, p1} \left( k \right) 
v_{R, p1} \left( k \right) 
+ 
u^{*}_{R, p2} \left( k \right) 
v_{R, p2} \left( k \right) 
\right\} 
\nonumber\\
&& 
- i \textrm{sign} \left( \alpha \right) y 
\textrm{Im} 
\left\lbrack 
\left\{ 
u^{*}_{R, p1} \left( k \right) 
v_{R, p1} \left( k \right) 
- 
u^{*}_{R, p2} \left( k \right) 
v_{R, p2} \left( k \right) 
\right\} 
\cos \left\{ 2 \phi \left( k \right) \right\} 
- 
\left\{ 
u^{*}_{R, p1} \left( k \right) 
v_{R, p2} \left( k \right) 
+ 
u^{*}_{R, p2} \left( k \right) 
v_{R, p1} \left( k \right) 
\right\} 
\sin \left\{ 2 \phi \left( k \right) \right\} 
\right\rbrack 
, \nonumber\\
\label{Xipp_form}
\end{eqnarray}

\begin{eqnarray}
&& \!\!\!\!\!
\Xi_{12} \left( k \right) 
= 
\sqrt{1 - 2 y^2} 
\left\{ 
u^{*}_{R, 11} \left( k \right) 
u_{R, 22} \left( k \right) 
- 
v^{*}_{R, 11} \left( k \right) 
v_{R, 22} \left( k \right) 
- 
u^{*}_{R, 12} \left( k \right) 
u_{R, 21} \left( k \right) 
+ 
v^{*}_{R, 12} \left( k \right) 
v_{R, 21} \left( k \right) 
\right\} 
\nonumber\\
&& 
- \textrm{sign} \left( \alpha \right) \frac{y}{2} 
\left\{ 
u^{*}_{R, 11} \left( k \right) 
v_{R, 21} \left( k \right) 
- 
u_{R, 21} \left( k \right) 
v^{*}_{R, 11} \left( k \right) 
+ 
u^{*}_{R, 12} \left( k \right) 
v_{R, 22} \left( k \right) 
- 
u_{R, 22} \left( k \right) 
v^{*}_{R, 12} \left( k \right) 
\right\} 
\nonumber\\
&& 
- \textrm{sign} \left( \alpha \right) \frac{y}{2} 
\left\{ 
u^{*}_{R, 11} \left( k \right) 
v_{R, 21} \left( k \right) 
- 
u_{R, 21} \left( k \right) 
v^{*}_{R, 11} \left( k \right) 
- 
u^{*}_{R, 12} \left( k \right) 
v_{R, 22} \left( k \right) 
+ 
u_{R, 22} \left( k \right) 
v^{*}_{R, 12} \left( k \right) 
\right\} 
\cos \left\{ 2 \phi \left( k \right) \right\} 
\nonumber\\
&& 
+ \textrm{sign} \left( \alpha \right) \frac{y}{2} 
\left\{ 
u^{*}_{R, 11} \left( k \right) 
v_{R, 22} \left( k \right) 
+ 
u^{*}_{R, 12} \left( k \right) 
v_{R, 21} \left( k \right) 
- 
u_{R, 21} \left( k \right) 
v^{*}_{R, 12} \left( k \right) 
- 
u_{R, 22} \left( k \right) 
v^{*}_{R, 11} \left( k \right) 
\right\} 
\sin \left\{ 2 \phi \left( k \right) \right\} 
, \qquad 
\label{Xi12_form}
\end{eqnarray}

\begin{eqnarray}
&& \!\!\!\!\!\!\!\!\!\!\!\!\!\!\!\!\!\!\!\!
\Xi_{13} \left( k \right) 
= 
2 \sqrt{1 - 2 y^2} 
\left\{ 
u_{R, 11} \left( k \right) 
v_{R, 12} \left( k \right) 
- 
u_{R, 12} \left( k \right) 
v_{R, 11} \left( k \right) 
\right\}^{*} 
- \textrm{sign} \left( \alpha \right) \frac{y}{2} 
\left\lbrack 
\left\{ u_{R, 11} \left( k \right) \right\}^2 
+ 
\left\{ u_{R, 12} \left( k \right) \right\}^2 
- 
\left\{ v_{R, 11} \left( k \right) \right\}^2 
- 
\left\{ v_{R, 12} \left( k \right) \right\}^2 
\right\rbrack^{*} 
\nonumber\\
&& 
- \textrm{sign} \left( \alpha \right) \frac{y}{2} 
\left\lbrack 
\left\{ u_{R, 11} \left( k \right) \right\}^2 
- 
\left\{ u_{R, 12} \left( k \right) \right\}^2 
- 
\left\{ v_{R, 11} \left( k \right) \right\}^2 
+ 
\left\{ v_{R, 12} \left( k \right) \right\}^2 
\right\rbrack^{*} \cos \left\{ 2 \phi \left( k \right) \right\} 
\nonumber\\
&& 
+ \textrm{sign} \left( \alpha \right) y 
\left\{ 
u_{R, 11} \left( k \right) 
u_{R, 12} \left( k \right) 
- 
v_{R, 11} \left( k \right) 
v_{R, 12} \left( k \right) 
\right\}^{*} \sin \left\{ 2 \phi \left( k \right) \right\} 
, 
\label{Xi13_form}
\end{eqnarray}

\begin{eqnarray}
&& \!\!\!\!\!\!\!\!\!\!\!\!\!\!\!\!\!\!\!\!
\Xi_{24} \left( k \right) 
= 
2 \sqrt{1 - 2 y^2} 
\left\{ 
u_{R, 21} \left( k \right) 
v_{R, 22} \left( k \right) 
- 
u_{R, 22} \left( k \right) 
v_{R, 21} \left( k \right) 
\right\}^{*} 
- \textrm{sign} \left( \alpha \right) \frac{y}{2} 
\left\lbrack 
\left\{ u_{R, 21} \left( k \right) \right\}^2 
+ 
\left\{ u_{R, 22} \left( k \right) \right\}^2 
- 
\left\{ v_{R, 21} \left( k \right) \right\}^2 
- 
\left\{ v_{R, 22} \left( k \right) \right\}^2 
\right\rbrack^{*} 
\nonumber\\
&& 
- \textrm{sign} \left( \alpha \right) \frac{y}{2} 
\left\lbrack 
\left\{ u_{R, 21} \left( k \right) \right\}^2 
- 
\left\{ u_{R, 22} \left( k \right) \right\}^2 
- 
\left\{ v_{R, 21} \left( k \right) \right\}^2 
+ 
\left\{ v_{R, 22} \left( k \right) \right\}^2 
\right\rbrack^{*} \cos \left\{ 2 \phi \left( k \right) \right\} 
\nonumber\\
&& 
+ \textrm{sign} \left( \alpha \right) y 
\left\{ 
u_{R, 21} \left( k \right) 
u_{R, 22} \left( k \right) 
- 
v_{R, 21} \left( k \right) 
v_{R, 22} \left( k \right) 
\right\}^{*} \sin \left\{ 2 \phi \left( k \right) \right\} 
, 
\label{Xi24_form}
\end{eqnarray}
and

\begin{eqnarray}
&& \!\!\!\!\!\!\!\!\!\!\!\!\!\!\!\!\!\!\!\!
\Xi_{14} \left( k \right) 
= 
\sqrt{1 - 2 y^2} 
\left\{ 
u_{R, 11} \left( k \right) 
v_{R, 22} \left( k \right) 
- 
u_{R, 12} \left( k \right) 
v_{R, 21} \left( k \right) 
+ 
u_{R, 21} \left( k \right) 
v_{R, 12} \left( k \right) 
- 
u_{R, 22} \left( k \right) 
v_{R, 11} \left( k \right) 
\right\}^{*} 
\nonumber\\
&& 
- \textrm{sign} \left( \alpha \right) \frac{y}{2} 
\left\{ 
u_{R, 11} \left( k \right) 
u_{R, 21} \left( k \right) 
- 
v_{R, 11} \left( k \right) 
v_{R, 21} \left( k \right) 
+ 
u_{R, 12} \left( k \right) 
u_{R, 22} \left( k \right) 
- 
v_{R, 12} \left( k \right) 
v_{R, 22} \left( k \right) 
\right\}^{*} 
\nonumber\\
&& 
- \textrm{sign} \left( \alpha \right) \frac{y}{2} 
\left\{ 
u_{R, 11} \left( k \right) 
u_{R, 21} \left( k \right) 
- 
v_{R, 11} \left( k \right) 
v_{R, 21} \left( k \right) 
- 
u_{R, 12} \left( k \right) 
u_{R, 22} \left( k \right) 
+ 
v_{R, 12} \left( k \right) 
v_{R, 22} \left( k \right) 
\right\}^{*} \cos \left\{ 2 \phi \left( k \right) \right\} 
\nonumber\\
&& 
+ \textrm{sign} \left( \alpha \right) \frac{y}{2} 
\left\{ 
u_{R, 11} \left( k \right) 
u_{R, 22} \left( k \right) 
+ 
u_{R, 12} \left( k \right) 
u_{R, 21} \left( k \right) 
- 
v_{R, 11} \left( k \right) 
v_{R, 22} \left( k \right) 
- 
v_{R, 12} \left( k \right) 
v_{R, 21} \left( k \right) 
\right\}^{*} \sin \left\{ 2 \phi \left( k \right) \right\} 
. \qquad 
\label{Xi14_form}
\end{eqnarray}
Thus, if $x = \sqrt{1 - 2 y^2}$, $y$, $\alpha$, $U \left( k \right)$, and $V \left( k \right)$ are all real, from Eqs.~\eqref{Xipp_form},~\eqref{Xi12_form},~\eqref{Xi13_form},~\eqref{Xi24_form}, and ~\eqref{Xi14_form}, 
$\Xi_{p p} \left( k \right) = 0$ and $\Xi_{p q} \left( k \right) \in \mathbb{R}$ for $p \neq q$.

For low temperatures, 
$0 \le \hbar \omega_1\left( k \right),k_B T 
\ll \hbar \omega_2 \left( k \right)$, the massive reactons are frozen
out and the system is to good approximation 
in Bogoliubov mode 1 (states purely created by $\hat{b}^{\dagger}_1
\left( \boldsymbol{k}, t \right)$). 
$\hat{R}_2 \left( t \right)$ in Eq.~\eqref{R2_express} then simplifies as 
\begin{eqnarray}
\hat{R}_2 \left( t \right) 
= && \; 
- i 
\tilde{\alpha} 
\sum_{\boldsymbol{k} \neq 0} 
\Xi_{13} \left( k \right) 
\left\{ 
\hat{b}^{\dagger}_1 \left( \boldsymbol{k}, t \right) 
\hat{b}^{\dagger}_1 \left( - \boldsymbol{k}, t \right) 
- 
h. c. 
\right\} . \qquad 
\label{op_R_up_to_2nd}
\end{eqnarray}
We now present examples of the 
 vacuum expectation values of $\hat{R} \left( t \right)$ for two quantum-mechanical reacton 
 states.

{\bf (a)} 
We define the squeezed coherent state as 
$\left\vert \beta_{1} \left( \boldsymbol{k}_1, t \right) , 
\zeta \left( \boldsymbol{k}_1 \right) 
\right\rangle 
\coloneqq 
\hat{D} \left( \beta_{1} \left( \boldsymbol{k}_1, t \right) \right) 
\hat{D} \left( \beta_{1} \left( - \boldsymbol{k}_1, t \right) \right) 
\hat{S} \left( \zeta \left( \boldsymbol{k}_1 \right) \right) 
\left\vert \textrm{vac} \right\rangle$, where  
the displacement and squeezing operators read 
$\hat{D} \left( \beta_{1} \left( \boldsymbol{k}, t \right) \right) \coloneqq 
e^{ 
\beta_{1} \left( \boldsymbol{k}, t \right) 
\hat{b}^{\dagger}_{1} \left( \boldsymbol{k}, t \right) 
- \beta^{*}_{1} \left( \boldsymbol{k}, t \right) 
\hat{b}_{1} \left( \boldsymbol{k}, t \right) 
} $, 
$\hat{S} \left( \zeta \left( \boldsymbol{k} \right) \right) \coloneqq 
e^{ 
\zeta \left( \boldsymbol{k} \right) 
\hat{b}^{\dagger}_{1} \left( \boldsymbol{k}, t \right) 
\hat{b}^{\dagger}_{1} \left( - \boldsymbol{k}, t \right) 
- \zeta^{*} \left( \boldsymbol{k} \right) 
\hat{b}_{1} \left( \boldsymbol{k}, t \right) 
\hat{b}_{1} \left( - \boldsymbol{k}, t \right) 
} $, respectively.  
Here, $\left\vert \textrm{vac} \right\rangle$ represents the reacton 
quasi-particle vacuum at initial time $t = 0$ where $\hat{b}_{p} \left( \boldsymbol{k}, 0 \right) \left\vert \textrm{vac} \right\rangle = 0$. 
Since 
$\hat{b}_{p} \left( \boldsymbol{k}, t \right) = e^{- i \omega_{p} \left( k \right) t} \hat{b}_{p} \left( \boldsymbol{k}, 0 \right)$ from our Hamiltonian in Eq.~\eqref{Bog_H_2nd}, 
we get $\hat{b}_{p} \left( \boldsymbol{k}, t \right) \left\vert \textrm{vac} \right\rangle = 0$ {\it{for all}} $t$. 
With $\left\vert \textrm{vac} \right\rangle$, 
we obtain the reaction rate (to $ O ( \delta \hat{\Psi}^3_j ) $)

\begin{eqnarray}
\left\langle \beta_{1} \left( \boldsymbol{k}_1, t \right) , 
\zeta \left( \boldsymbol{k}_1 \right) 
\right\vert 
\hat{R} \left( t \right) 
\left\vert 
\beta_{1} \left( \boldsymbol{k}_1, t \right) , 
\zeta \left( \boldsymbol{k}_1 \right) 
\right\rangle _{\rm squeezed\,\, state}
& = & 
2 \tilde{\alpha} 
\Xi_{13} \left( k_1 \right) \left\lbrack 
\begin{array}{c}
2 
\textrm{Im} 
\left\{ 
\beta^{*}_{1} \left( \boldsymbol{k}_1, t \right) 
\beta^{*}_{1} \left( - \boldsymbol{k}_1, t \right) 
\right\} 
\qquad \qquad \qquad \qquad 
\\
+ 
\sinh \left( 2 \left\vert \zeta \left( \boldsymbol{k}_1 \right) \right\vert \right) 
\textrm{Im} 
\left\{ 
\zeta^{*} \left( \boldsymbol{k}_1 \right) /\left\vert \zeta \left( \boldsymbol{k}_1 \right) \right\vert 
\right\} 
\end{array}
\!\!\!\!
\right\rbrack 
. \qquad 
\label{reaction_squeezed_coherent}
\end{eqnarray}
The quantum mechanical nature of this average rate is manifest in the third line, containing
the squeezing parameter $ \zeta \left( \boldsymbol{k}_1 \right)$.

\noindent {\bf (b}) A reacton pair state may be defined as
\begin{eqnarray}
\!\!\!\!\!\left\vert \psi_{s, 1} \left( \boldsymbol{k}_1, t \right) \right\rangle 
\coloneqq && \; 
\frac{ 
1 + 
e^{i \theta \left( \boldsymbol{k}_1 \right)} 
\hat{b}^{\dagger}_1 \left( \boldsymbol{k}_1, t \right) 
\hat{b}^{\dagger}_1 \left( - \boldsymbol{k}_1, t \right) 
}{\sqrt{2}} 
\left\vert \textrm{vac} \right\rangle 
, \quad 
\end{eqnarray}
where $\theta \left( \boldsymbol{k}_1 \right)$ is the phase of  the excited pair
relative to the vacuum. Then the reaction rate turns out to be 
\begin{eqnarray}
\left\langle 
\psi_{s, 1} \left( \boldsymbol{k}_1, t \right) 
\right\vert 
\hat{R} \left( t \right) 
\left\vert 
\psi_{s, 1} \left( \boldsymbol{k}_1, t \right) 
\right\rangle_{\rm reacton\,\, pair}
& = & 
- 2 \tilde{\alpha} 
\Xi_{13} \left( k_1 \right) 
\sin \theta \left( \boldsymbol{k}_1 \right) 
+ O \left( \delta \hat{\Psi}^3_j \right) , \qquad 
\label{Reaction_pairstate}
\end{eqnarray}
displaying the factor $\propto \sin \theta \left( \boldsymbol{k}_1 \right)$, which is due to 
quantum interference.

\section{\label{CFI_QFI_Formalism}Small Perturbations and Hamiltonian Bilinearization}

In the Schr\"odinger picture, let the time dependent bilinear Hamiltonian be $\hat{H}_{S} \left( t \right)$ where 
\begin{eqnarray}
\hat{H}_{S} \left( t \right) = && \; 
\hbar \sum_{\boldsymbol{k} \neq 0} 
\left\{ 
\omega_{1} \left( k \right) 
\hat{b}^{\dagger}_{1} \left( \boldsymbol{k} \right) 
\hat{b}_{1} \left( \boldsymbol{k} \right) 
+ 
\omega_{2} \left( k \right) 
\hat{b}^{\dagger}_{2} \left( \boldsymbol{k} \right) 
\hat{b}_{2} \left( \boldsymbol{k} \right) 
\right\} 
+ \hat{V}_{S} \left( t \right) 
, \qquad 
\label{appendix_HS_gen_form}
\end{eqnarray}
and

\begin{eqnarray}
\hat{V}_{S} \left( t \right) & = & 
V_{ex} 
f \left( t \right) 
\sum_{\boldsymbol{k} \neq 0} 
\left\{ 
\mathbb{V}_1 \left( k \right) 
\hat{b}^{\dagger}_{1} \left( \boldsymbol{k} \right) 
\hat{b}_{1} \left( \boldsymbol{k} \right) 
+ 
\mathbb{V}_2 \left( k \right) 
\hat{b}^{\dagger}_{1} \left( \boldsymbol{k} \right) 
\hat{b}^{\dagger}_{1} \left( - \boldsymbol{k} \right) 
+ 
\mathbb{V}^{*}_2 \left( k \right) 
\hat{b}_{1} \left( \boldsymbol{k} \right) 
\hat{b}_{1} \left( - \boldsymbol{k} \right) 
\right\}  \qquad \quad 
\label{PertV_gen_form}
\end{eqnarray}
a small perturbation. 
Here, $V_{ex}$, $f \left( t \right)$, and $\mathbb{V}_1 \left( k \right)$ are real 
and $f \left( t \right) = 0$ for $t < 0$ so that our Hamiltonian is a time-independent operator at $t < 0$. 
We assume a small perturbation, i.e.~$0 \le V_{ex} \ll \hbar \omega_{1} \left( k \right)$ for $k \neq 0$, and also assume that the Bogoliubov excitation energies $\omega_{1} \left( k \right)$ and $\omega_{2} \left( k \right)$ are real and positive so that the system is in a stable state. 
For simplicity, let $\omega_{1} \left( k \right) \ll \omega_{2} \left( k \right)$ so that one may neglect effects of $\hat{b}_{2} \left( \boldsymbol{k} \right)$.

To solve this perturbation problem with time dependent perturbation $\hat{V}_{S} \left( t \right)$, we use the interaction picture instead of the Heisenberg picture. 
Let 
$\displaystyle 
\hat{H}_{0, S} \coloneqq 
\hbar \sum_{\boldsymbol{k} \neq 0} 
\omega_{1} \left( k \right) 
\hat{b}^{\dagger}_{1} \left( \boldsymbol{k} \right) 
\hat{b}_{1} \left( \boldsymbol{k} \right) 
$. By denoting states in Schr\"{o}dinger picture as $\left\vert \psi_{S} \left( t \right) \right\rangle$, states in interaction picture as 
$\left\vert \psi_{I} \left( t \right) \right\rangle 
\coloneqq 
\hat{U}_{0} \left( t \right) 
\left\vert \psi_{S} \left( t \right) \right\rangle$ 
where 
$\hat{U}_{0} \left( t \right) 
\coloneqq 
\textrm{exp} \left( i \hat{H}_{0, S} t / \hbar \right)$, 
and the time evolution operator as 
$\hat{U}_{I} \left( t \right)$ where 
$\left\vert \psi_{I} \left( t \right) \right\rangle 
= \hat{U}_{I} \left( t \right) 
\left\vert \psi_{I} \left( 0 \right) \right\rangle$, 
$i \hbar 
\partial \hat{U}_{I} \left( t \right) / \partial t 
= 
\hat{V}_{I} \left( t \right) 
\hat{U}_{I} \left( t \right)$ 
where 
$\hat{V}_{I} \left( t \right) 
\coloneqq 
\hat{U}_{0} \left( t \right) 
\hat{V}_{S} \left( t \right) 
\hat{U}^{\dagger}_{0} \left( t \right)$, we get 
\begin{eqnarray}
\hat{U}_{I} \left( t \right) & = & 
1 - \frac{i}{\hbar} \int_{0}^{t} d t_1 \; \hat{V}_{I} \left( t_1 \right) 
- \frac{1}{\hbar^2} \int_{0}^{t} d t_1 \; \hat{V}_{I} \left( t_1 \right) \int_{0}^{t_1} d t_2 \; \hat{V}_{I} \left( t_2 \right) 
+ O \left( \hat{V}^3_{I} \left( t \right) \right) 
. 
\end{eqnarray}
With our $\hat{H}_{0, S}$,  
$
\hat{U}_{0} \left( t \right) 
\hat{b}_{1} \left( \boldsymbol{k} \right) 
\hat{U}^{\dagger}_{0} \left( t \right) 
= 
e^{- i \omega_{1} \left( k \right) t} 
\hat{b}_{1} \left( \boldsymbol{k} \right)$ and hence the
time evolution operator $\hat{U}_{I} \left( \tilde{t} \right)$ is

\begin{eqnarray}
&& 
\hat{U}_{I} \left( \tilde{t} \right) 
= 
1 
- 
2 \tilde{V}^2_{ex} 
\sum_{\boldsymbol{k} \neq 0} 
T_{2, -, +} \left( k, k, \tilde{t} \right) 
\left\vert \mathbb{V}_2 \left( k \right) \right\vert^2 
\nonumber\\
&& 
- 
i \tilde{V}_{ex} \sum_{\boldsymbol{k} \neq 0} 
\left\lbrack 
T_{1, 0} \left( \tilde{t} \right) 
\mathbb{V}_1 \left( k \right) 
- 
i \tilde{V}_{ex} \left\{ 
T_{2, 0, 0} \left( \tilde{t} \right) 
\mathbb{V}^2_1 \left( k \right) 
+ 
4 T_{2, -, +} \left( k, k, \tilde{t} \right) 
\left\vert \mathbb{V}_2 \left( k \right) \right\vert^2 
\right\} 
\right\rbrack 
\hat{b}^{\dagger}_{1} \left( \boldsymbol{k} \right) 
\hat{b}_{1} \left( \boldsymbol{k} \right) 
\nonumber\\
&& 
- 
i \tilde{V}_{ex} \sum_{\boldsymbol{k} \neq 0} 
\left\{ 
T_{1, +} \left( k, \tilde{t} \right) 
- 
2 i \tilde{V}_{ex} 
T_{2, 0, +} \left( k, \tilde{t} \right) 
\mathbb{V}_1 \left( k \right) 
\right\} 
\mathbb{V}_2 \left( k \right) 
\hat{b}^{\dagger}_{1} \left( \boldsymbol{k} \right) 
\hat{b}^{\dagger}_{1} \left( - \boldsymbol{k} \right) 
\nonumber\\
&& 
- 
i \tilde{V}_{ex} \sum_{\boldsymbol{k} \neq 0} 
\left\{ 
T_{1, -} \left( k, \tilde{t} \right) 
- 
2 i \tilde{V}_{ex} 
T_{2, -, 0} \left( k, \tilde{t} \right) 
\mathbb{V}_1 \left( k \right) 
\right\} 
\mathbb{V}^{*}_2 \left( k \right) 
\hat{b}_{1} \left( \boldsymbol{k} \right) 
\hat{b}_{1} \left( - \boldsymbol{k} \right) 
\nonumber\\
&& 
- 
\tilde{V}^2_{ex} 
T_{2, 0, 0} \left( \tilde{t} \right) 
\sum_{\boldsymbol{k}_1 \neq 0} \sum_{\boldsymbol{k}_2 \neq 0} 
\mathbb{V}_1 \left( k_1 \right) 
\mathbb{V}_1 \left( k_2 \right) 
\hat{b}^{\dagger}_{1} \left( \boldsymbol{k}_1 \right) 
\hat{b}^{\dagger}_{1} \left( \boldsymbol{k}_2 \right) 
\hat{b}_{1} \left( \boldsymbol{k}_2 \right) 
\hat{b}_{1} \left( \boldsymbol{k}_1 \right) 
\nonumber\\
&& 
- 
\tilde{V}^2_{ex} 
\sum_{\boldsymbol{k}_1 \neq 0} \sum_{\boldsymbol{k}_2 \neq 0} 
\left\{ 
T_{2, +, -} \left( k_1, k_2, \tilde{t} \right) 
+ 
T_{2, -, +} \left( k_2, k_1, \tilde{t} \right) 
\right\} 
\mathbb{V}_2 \left( k_1 \right) 
\mathbb{V}^{*}_2 \left( k_2 \right) 
\hat{b}^{\dagger}_{1} \left( \boldsymbol{k}_1 \right) 
\hat{b}^{\dagger}_{1} \left( - \boldsymbol{k}_1 \right) 
\hat{b}_{1} \left( \boldsymbol{k}_2 \right) 
\hat{b}_{1} \left( - \boldsymbol{k}_2 \right) 
\nonumber\\
&& 
- 
\tilde{V}^2_{ex} 
\sum_{\boldsymbol{k}_1 \neq 0} \sum_{\boldsymbol{k}_2 \neq 0} 
\left\{ 
T_{2, +, 0} \left( k_1, \tilde{t} \right) 
+ 
T_{2, 0, +} \left( k_1, \tilde{t} \right) 
\right\} 
\mathbb{V}_2 \left( k_1 \right) 
\mathbb{V}_1 \left( k_2 \right) 
\hat{b}^{\dagger}_{1} \left( \boldsymbol{k}_1 \right) 
\hat{b}^{\dagger}_{1} \left( - \boldsymbol{k}_1 \right) 
\hat{b}^{\dagger}_{1} \left( \boldsymbol{k}_2 \right) 
\hat{b}_{1} \left( \boldsymbol{k}_2 \right) 
\nonumber\\
&& 
- 
\tilde{V}^2_{ex} 
\sum_{\boldsymbol{k}_1 \neq 0} \sum_{\boldsymbol{k}_2 \neq 0} 
\left\{ 
T_{2, -, 0} \left( k_1, \tilde{t} \right) 
+ 
T_{2, 0, -} \left( k_1, \tilde{t} \right) 
\right\} 
\mathbb{V}^{*}_2 \left( k_1 \right) 
\mathbb{V}_1 \left( k_2 \right) 
\hat{b}^{\dagger}_{1} \left( \boldsymbol{k}_2 \right) 
\hat{b}_{1} \left( \boldsymbol{k}_2 \right) 
\hat{b}_{1} \left( \boldsymbol{k}_1 \right) 
\hat{b}_{1} \left( - \boldsymbol{k}_1 \right) 
\nonumber\\
&& 
- 
\tilde{V}^2_{ex} 
\sum_{\boldsymbol{k}_1 \neq 0} \sum_{\boldsymbol{k}_2 \neq 0} 
T_{2, +, +} \left( k_1, k_2, \tilde{t} \right) 
\mathbb{V}_2 \left( k_1 \right) 
\mathbb{V}_2 \left( k_2 \right) 
\hat{b}^{\dagger}_{1} \left( \boldsymbol{k}_1 \right) 
\hat{b}^{\dagger}_{1} \left( - \boldsymbol{k}_1 \right) 
\hat{b}^{\dagger}_{1} \left( \boldsymbol{k}_2 \right) 
\hat{b}^{\dagger}_{1} \left( - \boldsymbol{k}_2 \right) 
\nonumber\\
&& 
- 
\tilde{V}^2_{ex} 
\sum_{\boldsymbol{k}_1 \neq 0} \sum_{\boldsymbol{k}_2 \neq 0} 
T_{2, -, -} \left( k_1, k_2, \tilde{t} \right) 
\mathbb{V}^{*}_2 \left( k_1 \right) 
\mathbb{V}^{*}_2 \left( k_2 \right) 
\hat{b}_{1} \left( \boldsymbol{k}_1 \right) 
\hat{b}_{1} \left( - \boldsymbol{k}_1 \right) 
\hat{b}_{1} \left( \boldsymbol{k}_2 \right) 
\hat{b}_{1} \left( - \boldsymbol{k}_2 \right) 
+ O \left( \tilde{V}^3_{ex} \right) , 
\end{eqnarray}
where $\tilde{V}_{ex} \coloneqq V_{ex} / \left( g_{a} n \right)$ and $\tilde{t} \coloneqq g_{a} n t / \hbar$. 
By introducing $\tilde{\omega}_{1} \left( k \right) \coloneqq \hbar \omega_{1} \left( k \right) / \left( g_{a} n \right)$, we get

\begin{eqnarray}
T_{1, 0} \left( \tilde{t} \right) 
\coloneqq 
\int_{0}^{\tilde{t}} d \tilde{t}_1 \; 
f \left( \tilde{t}_1 \right) 
, \quad 
T_{1, \pm} \left( k_1, \tilde{t} \right) 
\coloneqq 
\int_{0}^{\tilde{t}} d \tilde{t}_1 \; 
e^{\pm 2 i \tilde{\omega}_{1} \left( k_1 \right) \tilde{t}_1} 
f \left( \tilde{t}_1 \right) 
, \quad 
T_{2, 0, 0} \left( \tilde{t} \right) 
\coloneqq 
\int_{0}^{\tilde{t}} d \tilde{t}_1 \; 
f \left( \tilde{t}_1 \right) 
\int_{0}^{\tilde{t}_1} d \tilde{t}_2 \; 
f \left( \tilde{t}_2 \right) 
, \qquad \qquad 
\end{eqnarray}

\begin{eqnarray}
T_{2, \pm, 0} \left( k, \tilde{t} \right) 
\coloneqq 
\int_{0}^{\tilde{t}} d \tilde{t}_1 \; 
e^{\pm 2 i \tilde{\omega}_{1} \left( k \right) \tilde{t}_1} 
f \left( \tilde{t}_1 \right) 
\int_{0}^{\tilde{t}_1} d \tilde{t}_2 \; 
f \left( \tilde{t}_2 \right) 
, \quad 
T_{2, 0, \pm} \left( k, \tilde{t} \right) 
\coloneqq 
\int_{0}^{\tilde{t}} d \tilde{t}_1 \; 
f \left( \tilde{t}_1 \right) 
\int_{0}^{\tilde{t}_1} d \tilde{t}_2 \; 
e^{\pm 2 i \tilde{\omega}_{1} \left( k \right) \tilde{t}_2} 
f \left( \tilde{t}_2 \right) 
, \qquad \qquad 
\end{eqnarray}

\begin{eqnarray}
T_{2, \pm, +} \left( k_1, k_2, \tilde{t} \right) 
\coloneqq 
\int_{0}^{\tilde{t}} d \tilde{t}_1 \; 
e^{\pm 2 i \tilde{\omega}_{1} \left( k_1 \right) \tilde{t}_1} 
f \left( \tilde{t}_1 \right) 
\int_{0}^{\tilde{t}_1} d \tilde{t}_2 \; 
e^{2 i \tilde{\omega}_{1} \left( k_2 \right) \tilde{t}_2} 
f \left( \tilde{t}_2 \right) 
, 
\end{eqnarray}
and

\begin{eqnarray}
T_{2, \pm, -} \left( k_1, k_2, \tilde{t} \right) 
\coloneqq 
\int_{0}^{\tilde{t}} d \tilde{t}_1 \; 
e^{\pm 2 i \tilde{\omega}_{1} \left( k_1 \right) \tilde{t}_1} 
f \left( \tilde{t}_1 \right) 
\int_{0}^{\tilde{t}_1} d \tilde{t}_2 \; 
e^{- 2 i \tilde{\omega}_{1} \left( k_2 \right) \tilde{t}_2} 
f \left( \tilde{t}_2 \right) 
. 
\end{eqnarray}

With this perturbation, in the interaction picture, the initial Bogoliubov vacuum state $\left\vert \textrm{vac} \right\rangle$ is propagated as

\begin{eqnarray}
\left\vert \Psi_{0} \left( t \right) \right\rangle 
& \coloneqq & 
\hat{U}_{I} \left( t \right) 
\left\vert \textrm{vac} \right\rangle 
\nonumber\\
& = & 
\left\{ 
1 
- 
\tilde{V}^2_{ex} 
\psi_{0, 2} \left( \tilde{t} \right) 
\right\} 
\left\vert \textrm{vac} \right\rangle 
- 
i \tilde{V}_{ex} \sum_{\boldsymbol{k} \neq 0} 
\left\{ 
\psi_{2, 1} \left( k, \tilde{t} \right) 
- 
i \tilde{V}_{ex} 
\psi_{2, 2} \left( k, \tilde{t} \right) 
\right\} 
\mathbb{V}_2 \left( k \right) 
\left\vert \boldsymbol{k}, - \boldsymbol{k} \right\rangle 
\nonumber\\
&& 
- 
\tilde{V}^2_{ex} \sum_{\boldsymbol{k}_1 \neq 0} \sum_{\boldsymbol{k}_2 \neq 0} 
\psi_{4, 2} \left( k_1, k_2, \tilde{t} \right) 
\mathbb{V}_2 \left( k_1 \right) 
\mathbb{V}_2 \left( k_2 \right) 
\left\vert \boldsymbol{k}_1, - \boldsymbol{k}_1, \boldsymbol{k}_2, - \boldsymbol{k}_2 \right\rangle 
+ O \left( \tilde{V}^3_{ex} \right) , 
\label{psi0_st0}
\end{eqnarray}
where 

\begin{eqnarray}
&& 
\psi_{0, 2} \left( \tilde{t} \right) 
= 
2 \sum_{\boldsymbol{k} \neq 0} 
T_{2, -, +} \left( k, k, \tilde{t} \right) 
\left\vert \mathbb{V}_2 \left( k \right) \right\vert^2 
, \quad 
\psi_{2, 1} \left( k, \tilde{t} \right) 
= 
T_{1, +} \left( k, \tilde{t} \right) 
, \quad 
\psi_{2, 2} \left( k, \tilde{t} \right) 
= 
2 T_{2, 0, +} \left( k, \tilde{t} \right) 
\mathbb{V}_1 \left( k \right) 
, \nonumber\\
&& 
\psi_{4, 2} \left( k_1, k_2, \tilde{t} \right) 
= 
T_{2, +, +} \left( k_1, k_2, \tilde{t} \right) 
, 
\label{def_psis_appendix}
\end{eqnarray}

$\left\vert \boldsymbol{k}, - \boldsymbol{k} \right\rangle 
\coloneqq 
\hat{b}^{\dagger}_{1} \left( \boldsymbol{k} \right) 
\hat{b}^{\dagger}_{1} \left( - \boldsymbol{k} \right) 
\left\vert \textrm{vac} \right\rangle$, 
$\left\vert \boldsymbol{k}_1, - \boldsymbol{k}_1, \boldsymbol{k}_2, - \boldsymbol{k}_2 \right\rangle 
\coloneqq 
\hat{b}^{\dagger}_{1} \left( \boldsymbol{k}_1 \right) 
\hat{b}^{\dagger}_{1} \left( - \boldsymbol{k}_1 \right) 
\hat{b}^{\dagger}_{1} \left( \boldsymbol{k}_2 \right) 
\hat{b}^{\dagger}_{1} \left( - \boldsymbol{k}_2 \right) 
\left\vert \textrm{vac} \right\rangle$, and 
$\left\vert \boldsymbol{k}_1, - \boldsymbol{k}_1, \boldsymbol{k}_2, - \boldsymbol{k}_2, \boldsymbol{k}_3, - \boldsymbol{k}_3 \right\rangle 
\coloneqq 
\hat{b}^{\dagger}_{1} \left( \boldsymbol{k}_1 \right) 
\hat{b}^{\dagger}_{1} \left( - \boldsymbol{k}_1 \right) 
\hat{b}^{\dagger}_{1} \left( \boldsymbol{k}_2 \right) 
\hat{b}^{\dagger}_{1} \left( - \boldsymbol{k}_2 \right) 
\hat{b}^{\dagger}_{1} \left( \boldsymbol{k}_3 \right) 
\hat{b}^{\dagger}_{1} \left( - \boldsymbol{k}_3 \right) 
\left\vert \textrm{vac} \right\rangle$.

With this state, the reaction rate is 

\begin{eqnarray}
\left\langle \hat{R}_2 \left( t \right) \right\rangle 
& = & 
- 2 \tilde{\alpha} 
\sum_{\boldsymbol{k} \neq 0} 
\Xi_{13} \left( k \right) 
\textrm{Im} 
\left\{ 
e^{- 2 i \tilde{\omega}_{1} \left( k \right) \tilde{t}} 
\left\langle \Psi_{0} \left( t \right) \right\vert 
\hat{b}_{1} \left( \boldsymbol{k} \right) 
\hat{b}_{1} \left( - \boldsymbol{k} \right) 
\left\vert \Psi_{0} \left( t \right) \right\rangle 
\right\} 
\nonumber\\
& = & 
4 \tilde{\alpha} 
\tilde{V}_{ex} 
\sum_{\boldsymbol{k} \neq 0} 
\Xi_{13} \left( k \right) 
\textrm{Im} \left\{ 
e^{- 2 i \tilde{\omega}_{1} \left( k \right) \tilde{t}} 
\psi_{2, 1} \left( k, \tilde{t} \right) 
\mathbb{V}_2 \left( k \right) 
\right\} 
+ O \left( \tilde{V}^{2}_{ex} \right) 
. 
\label{R2_perturb}
\end{eqnarray}

When 
measuring $\tilde{V}_{ex}$, using formula in~\cite{Liu2014} and Eq.~\eqref{psi0_st0}, the quantum Fisher information $I_{Q} \left( \tilde{V}_{ex}, \tilde{t} \right)$ is 

\begin{eqnarray}
I_{Q} \left( \tilde{V}_{ex}, \tilde{t} \right) & = & 
8 \sum_{\boldsymbol{k} \neq 0} 
\left\vert \psi_{2, 1} \left( k, \tilde{t} \right) \right\vert^2 
\left\vert \mathbb{V}_2 \left( k \right) \right\vert^2 
+ O \left( \tilde{V}_{ex} \right) . 
\label{QFI_gen_form}
\end{eqnarray}

Let $P_{0} \left( t \right) = 
\left\vert 
\left\langle 
\Psi_{0} \left( t \right) 
\right.\! 
\left\vert 
\textrm{vac} 
\right\rangle 
\right\vert^2$ and 
$\displaystyle 
P_{2} \left( t \right) = \left( 1 / 2 \right) \sum_{\boldsymbol{k} \neq 0} 
\left\vert 
\left\langle 
\Psi_{0} \left( t \right) 
\right\vert 
\left. 
\boldsymbol{k}, - \boldsymbol{k} \right\rangle
\right\vert^2$ 
(note that $\left\vert \boldsymbol{k}, - \boldsymbol{k} \right\rangle 
= 
\left\vert - \boldsymbol{k}, \boldsymbol{k} \right\rangle$ from our definition below Eqs.~\eqref{def_psis_appendix}). Then 

\begin{eqnarray}
\!\!\!\!\!\!\!\!\!\!
P_{0} \left( t \right) 
= 1 
- 2 \tilde{V}^2_{ex} 
\textrm{Re} \left\{ \psi_{0, 2} \left( \tilde{t} \right) \right\} 
+ O \left( \tilde{V}^3_{ex} \right) 
, \quad 
P_{2} \left( t \right) 
= 
2 \tilde{V}^2_{ex} \sum_{\boldsymbol{k} \neq 0} 
\left\vert 
\psi_{2, 1} \left( k, \tilde{t} \right) 
\mathbb{V}_2 \left( k \right) 
\right\vert^2 
+ O \left( \tilde{V}^3_{ex} \right) 
, \qquad \qquad 
\end{eqnarray}
and 
$\left\langle 
\Psi_{0} \left( t \right) \right.\!\! 
\left\vert 
\Psi_{0} \left( t \right) 
\right\rangle = P_{0} \left( t \right) + P_{2} \left( t \right) + O \left( \tilde{V}^3_{ex} \right)$. 
Note that 
$I_{Q} \left( \tilde{V}_{ex}, \tilde{t} \right) = 4 P_{2} \left( t \right) / \tilde{V}^2_{ex} + O \left( \tilde{V}_{ex} \right)$.

Now, suppose that we want to measure $\hat{M}$ with the state initially in Bogoliubov vacuum where 
$
\displaystyle 
\hat{M}_{S} = 
{
\sum_{\boldsymbol{k} \neq 0} 
} 
\left\{ 
\mathbb{M}_1 \left( k \right) 
+ 
\mathbb{M}_2 \left( k \right) 
\hat{b}^{\dagger}_{1} \left( \boldsymbol{k} \right) 
\hat{b}_{1} \left( \boldsymbol{k} \right) 
+ 
\mathbb{M}_3 \left( k \right) 
\hat{b}^{\dagger}_{1} \left( \boldsymbol{k} \right) 
\hat{b}^{\dagger}_{1} \left( - \boldsymbol{k} \right) 
+ 
\mathbb{M}^{*}_3 \left( k \right) 
\hat{b}_{1} \left( \boldsymbol{k} \right) 
\hat{b}_{1} \left( - \boldsymbol{k} \right) 
\right\} 
$ in the Schr\"odinger picture. 
In the interaction picture, by denoting 
$\hat{M}_{I} \left( t \right) \coloneqq 
\hat{U}_{0} \left( t \right) 
\hat{M}_{S} 
\hat{U}^{\dagger}_{0} \left( t \right)$, we get

\begin{eqnarray}
&& \!\!\!\!\!\!\!\!\!\!\!\!\!\!\!\!\!\!\!\!\!\!\!\!\!\!\!\!\!\!
\hat{M}_{I} \left( \tilde{t} \right) 
\left\vert \hat{\Psi}_{0} \left( \tilde{t} \right) \right\rangle 
= 
\sum_{\boldsymbol{k} \neq 0} 
\left\lbrack 
\begin{array}{c}
\mathbb{M}_{1} \left( k \right) 
- 
2 i \tilde{V}_{ex} 
e^{- 2 i \tilde{\omega}_{1} \left( k \right) \tilde{t}} 
\psi_{2, 1} \left( k, \tilde{t} \right) 
\mathbb{M}^{*}_{3} \left( k \right) 
\mathbb{V}_{2} \left( k \right) 
\qquad \qquad \qquad \quad 
\\
- 
\tilde{V}^2_{ex} \left\{ 
\psi_{0, 2} \left( \tilde{t} \right) 
\mathbb{M}_{1} \left( k \right) 
+ 
2 e^{- 2 i \tilde{\omega}_{1} \left( k \right) \tilde{t}} 
\psi_{2, 2} \left( k, \tilde{t} \right) 
\mathbb{M}^{*}_{3} \left( k \right) 
\mathbb{V}_{2} \left( k \right) 
\right\} 
\end{array}
\!\!\!\!
\right\rbrack 
\left\vert \textrm{vac} \right\rangle 
\nonumber\\
&& \!\!\!\!\!\!\!\!\!\!\!\!\!\!\!\!\!\!\!\!\!\!\!\!\!\!\!\!\!\!
+ 
\sum_{\boldsymbol{k} \neq 0} 
\left\lgroup 
\begin{array}{c}
e^{2 i \tilde{\omega}_{1} \left( k \right) \tilde{t}} 
\mathbb{M}_{3} \left( k \right) 
- 
i \tilde{V}_{ex} 
\psi_{2, 1} \left( k, \tilde{t} \right) 
\left\{ 
2 \mathbb{M}_2 \left( k \right) 
+ 
\displaystyle \sum_{\boldsymbol{k}_{1} \neq 0} \mathbb{M}_1 \left( k_1 \right) 
\right\} 
\mathbb{V}_{2} \left( k \right) 
\qquad \qquad \qquad \qquad \qquad \qquad \qquad \qquad \qquad \qquad \qquad \qquad \qquad 
\\
- 
\tilde{V}^2_{ex} \left\lbrack 
\begin{array}{c} 
e^{2 i \tilde{\omega}_{1} \left( k \right) \tilde{t}} 
\psi_{0, 2} \left( \tilde{t} \right) 
\mathbb{M}_{3} \left( k \right) 
+ 
\psi_{2, 2} \left( k, \tilde{t} \right) 
\left\{ 
2 \mathbb{M}_2 \left( k \right) 
+ 
\displaystyle \sum_{\boldsymbol{k}_{1} \neq 0} \mathbb{M}_1 \left( k_1 \right) 
\right\} 
\mathbb{V}_{2} \left( k \right) 
\qquad \qquad \qquad \qquad \qquad \qquad \qquad \qquad \qquad \quad 
\\
+ 
8 e^{- 2 i \tilde{\omega}_{1} \left( k \right) \tilde{t}} 
\psi_{4, 2} \left( k, k, \tilde{t} \right) 
\mathbb{M}^{*}_{3} \left( k \right) 
\mathbb{V}^2_{2} \left( k \right) 
+ 
2 \mathbb{V}_{2} \left( k \right) 
\displaystyle \sum_{\boldsymbol{k}_{1} \neq 0} 
e^{- 2 i \tilde{\omega}_{1} \left( k_1 \right) \tilde{t}} 
\left\{ 
\psi_{4, 2} \left( k_1, k, \tilde{t} \right) 
+ 
\psi_{4, 2} \left( k, k_1, \tilde{t} \right) 
\right\} 
\mathbb{M}^{*}_{3} \left( k_1 \right) 
\mathbb{V}_{2} \left( k_1 \right) 
\end{array}
\right\rbrack 
\end{array}
\!\!\!
\right\rgroup 
\left\vert \boldsymbol{k}, - \boldsymbol{k} \right\rangle 
\nonumber\\
&& \!\!\!\!\!\!\!\!\!\!\!\!\!\!\!\!\!\!\!\!\!\!\!\!\!\!\!\!\!\!
- 
i \tilde{V}_{ex} \sum_{\boldsymbol{k}_{1} \neq 0} \sum_{\boldsymbol{k}_{2} \neq 0} 
\left\lgroup 
\begin{array}{cc} 
e^{2 i \tilde{\omega}_{1} \left( k_1 \right) \tilde{t}} 
\psi_{2, 1} \left( k_2, \tilde{t} \right) 
\mathbb{M}_{3} \left( k_1 \right) 
\qquad \qquad \qquad \qquad \qquad \qquad \qquad \qquad \qquad \qquad \qquad \qquad 
\\
- 
i \tilde{V}_{ex} \left\lbrack 
\begin{array}{c} 
e^{2 i \tilde{\omega}_{1} \left( k_1 \right) \tilde{t}} 
\psi_{2, 2} \left( k_2, \tilde{t} \right) 
\mathbb{M}_{3} \left( k_1 \right) 
+ 
\psi_{4, 2} \left( k_1, k_2, \tilde{t} \right) 
\mathbb{V}_{2} \left( k_1 \right) 
\displaystyle \sum_{\boldsymbol{k}_3 \neq 0} \mathbb{M}_1 \left( k_3 \right) 
\!\!\!
\\
+ 
2 \psi_{4, 2} \left( k_1, k_2, \tilde{t} \right) 
\left\{ 
\mathbb{M}_2 \left( k_1 \right) 
+ 
\mathbb{M}_2 \left( k_2 \right) 
\right\} 
\mathbb{V}_{2} \left( k_1 \right) 
\qquad \qquad \qquad \qquad \quad 
\end{array}
\right\rbrack 
\end{array}
\!\!\!\!\!
\right\rgroup 
\mathbb{V}_{2} \left( k_2 \right) 
\left\vert \boldsymbol{k}_1, - \boldsymbol{k}_1, \boldsymbol{k}_2, - \boldsymbol{k}_2 \right\rangle 
\nonumber\\
&& \!\!\!\!\!\!\!\!\!\!\!\!\!\!\!\!\!\!\!\!\!\!\!\!\!\!\!\!\!\!
- 
\tilde{V}^2_{ex} \sum_{\boldsymbol{k}_1 \neq 0} \sum_{\boldsymbol{k}_2 \neq 0} \sum_{\boldsymbol{k}_3 \neq 0} 
e^{2 i \tilde{\omega}_{1} \left( k_1 \right) \tilde{t}} 
\psi_{4, 2} \left( k_2, k_3, \tilde{t} \right) 
\mathbb{M}_{3} \left( k_1 \right) 
\mathbb{V}_{2} \left( k_2 \right) 
\mathbb{V}_{2} \left( k_3 \right) 
\left\vert \boldsymbol{k}_1, - \boldsymbol{k}_1, \boldsymbol{k}_2, - \boldsymbol{k}_2, \boldsymbol{k}_3, - \boldsymbol{k}_3 \right\rangle 
+ O \left( \tilde{V}^3_{ex} \right) , 
\end{eqnarray}

\begin{eqnarray}
&& \!\!\!\!\!\!\!\!\!\!\!\!\!\!\!\!
\left\langle \hat{\Psi}_{0} \left( \tilde{t} \right) \right\vert 
\hat{M}_{I} \left( \tilde{t} \right) 
\left\vert \hat{\Psi}_{0} \left( \tilde{t} \right) \right\rangle 
= 
\sum_{\boldsymbol{k} \neq 0} 
\mathbb{M}_1 \left( k \right) 
+ 
4 \tilde{V}_{ex} \sum_{\boldsymbol{k} \neq 0} 
\textrm{Im} \left\{ 
e^{- 2 i \tilde{\omega}_{1} \left( k \right) \tilde{t}} 
\psi_{2, 1} \left( k, \tilde{t} \right) 
\mathbb{M}^{*}_3 \left( k \right) 
\mathbb{V}_2 \left( k \right) 
\right\} 
+ O \left( \tilde{V}^3_{ex} \right) 
\nonumber\\
&& \!\!\!\!\!\!\!\!
+ 
\tilde{V}^2_{ex} \sum_{\boldsymbol{k} \neq 0} 
\left\lbrack 
2 
\left\vert 
\psi_{2, 1} \left( k, \tilde{t} \right) 
\right\vert^2 
\left\vert \mathbb{V}_2 \left( k \right) \right\vert^2 
\left\{ 
2 \mathbb{M}_2 \left( k \right) 
+ 
\displaystyle \sum_{\boldsymbol{k}_1 \neq 0} \mathbb{M}_1 \left( k_1 \right) 
\right\} 
- 
2 \textrm{Re} \left\{ \psi_{0, 2} \left( \tilde{t} \right) \right\} 
\mathbb{M}_1 \left( k \right) 
- 
4 \textrm{Re} \left\{ 
e^{- 2 i \tilde{\omega}_{1} \left( k \right) \tilde{t}} 
\psi_{2, 2} \left( k, \tilde{t} \right) 
\mathbb{M}^{*}_3 \left( k \right) 
\mathbb{V}_2 \left( k \right) 
\right\} 
\right\rbrack 
, \nonumber\\
\end{eqnarray}
and

\begin{eqnarray}
&& \!\!\!\!\!\!\!\!\!\!\!\!\!\!\!\!\!\!\!\!\!\!\!\!\!\!\!\!\!\!
\left\langle \hat{\Psi}_{0} \left( \tilde{t} \right) \right\vert 
\hat{M}^2_{I} \left( \tilde{t} \right) 
\left\vert \hat{\Psi}_{0} \left( \tilde{t} \right) \right\rangle 
= 
\left\lbrack 
\left\{ 
\sum_{\boldsymbol{k} \neq 0} \mathbb{M}_1 \left( k \right) 
\right\}^2 
+ 
2 \sum_{\boldsymbol{k} \neq 0} \left\vert \mathbb{M}_3 \left( k \right) \right\vert^2 
+ 
8 \tilde{V}_{ex} \sum_{\boldsymbol{k} \neq 0} 
\left\{ 
\mathbb{M}_2 \left( k \right) 
+ 
\displaystyle 
\sum_{\boldsymbol{k}_1 \neq 0} \mathbb{M}_1 \left( k_1 \right) 
\right\} 
\textrm{Im} \left\{ 
e^{- 2 i \tilde{\omega}_{1} \left( k \right) \tilde{t}} 
\psi_{2, 1} \left( k, \tilde{t} \right) 
\mathbb{M}^{*}_3 \left( k \right) 
\mathbb{V}_2 \left( k \right) 
\right\} 
\right\rbrack 
\nonumber\\
&& \!\!\!\!\!\!\!\!\!\!\!\!\!\!\!
+ 
2 \tilde{V}^2_{ex} 
\left\lgroup \!\!\!
\begin{array}{c} 
\displaystyle 
4 
\left\vert 
\sum_{\boldsymbol{k} \neq 0} 
e^{- 2 i \tilde{\omega}_{1} \left( k \right) \tilde{t}} 
\psi_{2, 1} \left( k, \tilde{t} \right) 
\mathbb{M}^{*}_3 \left( k \right) 
\mathbb{V}_2 \left( k \right) 
\right\vert^2 
+ 
\sum_{\boldsymbol{k} \neq 0} 
\left\{ 
2 \mathbb{M}_2 \left( k \right) 
+ 
\sum_{\boldsymbol{k} \neq 0} \mathbb{M}_1 \left( k \right) 
\right\}^2 
\left\vert 
\psi_{2, 1} \left( k, \tilde{t} \right) 
\right\vert^2 
\left\vert 
\mathbb{V}_2 \left( k \right) 
\right\vert^2 
\qquad \qquad \qquad \qquad \qquad \quad 
\\ \displaystyle 
+ 
8 \sum_{\boldsymbol{k} \neq 0} 
\left\vert 
\psi_{2, 1} \left( k, \tilde{t} \right) 
\right\vert^2 
\left\vert 
\mathbb{M}_3 \left( k \right) 
\right\vert^2 
\left\vert 
\mathbb{V}_2 \left( k \right) 
\right\vert^2 
+ 
2 \sum_{\boldsymbol{k}_1 \neq 0} 
\left\vert 
\psi_{2, 1} \left( k_1, \tilde{t} \right) 
\right\vert^2 
\left\vert 
\mathbb{V}_2 \left( k_1 \right) 
\right\vert^2 
\sum_{\boldsymbol{k}_2 \neq 0} 
\left\vert 
\mathbb{M}_3 \left( k_2 \right) 
\right\vert^2 
\qquad \qquad \qquad \qquad \qquad \qquad \qquad \qquad \; 
\\ \displaystyle 
- 
\textrm{Re} \left\{ \psi_{0, 2} \left( \tilde{t} \right) \right\} \!\!
\left\lbrack 
\left\{ 
\sum_{\boldsymbol{k} \neq 0} \mathbb{M}_1 \left( k \right) 
\right\}^2 
+ 
2 \sum_{\boldsymbol{k} \neq 0} \left\vert \mathbb{M}_3 \left( k \right) \right\vert^2 
\right\rbrack 
- 
4 \sum_{\boldsymbol{k} \neq 0} 
\left\{ 
\mathbb{M}_2 \left( k \right) 
+ 
\sum_{\boldsymbol{k} \neq 0} \mathbb{M}_1 \left( k \right) 
\right\} \! 
\textrm{Re} \left\{ 
e^{- 2 i \tilde{\omega}_{1} \left( k \right) \tilde{t}} 
\psi_{2, 2} \left( k, \tilde{t} \right) 
\mathbb{M}^{*}_3 \left( k \right) 
\mathbb{V}_2 \left( k \right) 
\right\} 
\\ \displaystyle 
- 
8 \sum_{\boldsymbol{k}_1 \neq 0} \sum_{\boldsymbol{k}_2 \neq 0} 
\textrm{Re} \left\lbrack 
e^{- 2 i \left\{ \tilde{\omega}_1 \left( k_1 \right) + \tilde{\omega}_1 \left( k_2 \right) \right\} \tilde{t}} 
\psi_{4, 2} \left( k_1, k_2, \tilde{t} \right) 
\mathbb{M}^{*}_3 \left( k_1 \right) 
\mathbb{M}^{*}_3 \left( k_2 \right) 
\mathbb{V}_2 \left( k_1 \right) 
\mathbb{V}_2 \left( k_2 \right) 
\right\rbrack 
\qquad \qquad \qquad \qquad \qquad \qquad \qquad \qquad \qquad 
\\ \displaystyle 
- 
16 \sum_{\boldsymbol{k} \neq 0} 
\textrm{Re} \left\lbrack 
e^{- 4 i \tilde{\omega}_{1} \left( k \right) \tilde{t}} 
\psi_{4, 2} \left( k, k, \tilde{t} \right) 
\left\{ \mathbb{M}^{*}_3 \left( k \right) \right\}^2 
\mathbb{V}^2_2 \left( k \right) 
\right\rbrack 
\qquad \qquad \qquad \qquad \qquad \qquad \qquad \qquad \qquad \qquad \qquad \qquad \qquad \qquad \qquad \quad 
\end{array}
\right\rgroup \!\!
\nonumber\\
&& \!\!\!\!\!\!\!\!\!\!\!\!\!\!\!
+ O \left( \tilde{V}^3_{ex} \right) 
. 
\end{eqnarray}

Let 
$
\left\langle \hat{M} \left( k, \tilde{t} \right) \right\rangle' 
\coloneqq 
\textrm{Im} \left\{ 
e^{- 2 i \tilde{\omega}_{1} \left( k \right) \tilde{t}} 
\psi_{2, 1} \left( k, \tilde{t} \right) 
\mathbb{M}^{*}_3 \left( k \right) 
\mathbb{V}_2 \left( k \right) 
\right\} 
$ and

\begin{eqnarray}
\!\!\!\!\!\!\!\!\!\!\!\!\!\!\!\!\!\!
\left\langle \hat{M} \left( k, \tilde{t} \right) \right\rangle'' 
& \coloneqq & 
2 
\left\vert 
\psi_{2, 1} \left( k, \tilde{t} \right) 
\right\vert^2 
\left\vert \mathbb{V}_2 \left( k \right) \right\vert^2 
\left\{ 
2 \mathbb{M}_2 \left( k \right) 
+ 
\displaystyle \sum_{\boldsymbol{k}_1 \neq 0} \mathbb{M}_1 \left( k_1 \right) 
\right\} 
- 
2 \textrm{Re} \left\{ \psi_{0, 2} \left( \tilde{t} \right) \right\} 
\mathbb{M}_1 \left( k \right) 
- 
4 \textrm{Re} \left\{ 
e^{- 2 i \tilde{\omega}_{1} \left( k \right) \tilde{t}} 
\psi_{2, 2} \left( k, \tilde{t} \right) 
\mathbb{M}^{*}_3 \left( k \right) 
\mathbb{V}_2 \left( k \right) 
\right\} 
. 
\nonumber\\
\end{eqnarray}

Then the lower bound of the Fisher information $I_{C} \left( \tilde{V}_{ex}, \tilde{t} \right)$ when measuring $\tilde{V}_{a}$ is~\cite{Stein2014} 
\begin{eqnarray}
\!\!\!\!\!\!\!\!\!\!\!\!\!\!\!\!\!\!\!\!\!\!\!\!\!\!\!\!\!\!
I_{C} \left( \tilde{V}_{ex}, \tilde{t} \right) & = & 
8 \!\!
\left\lgroup 
\frac{\displaystyle 
\left\{ 
\sum_{\boldsymbol{k} \neq 0} \left\langle \hat{M} \left( k, \tilde{t} \right) \right\rangle' 
\right\}^2 
}{\displaystyle 
\sum_{\boldsymbol{k} \neq 0} \left\vert \mathbb{M}_{3} \left( k \right) \right\vert^2} 
+ 
\tilde{V}_{ex} 
\left\lbrack 
\frac{\displaystyle 
\sum_{\boldsymbol{k} \neq 0} \left\langle \hat{M} \left( k, \tilde{t} \right) \right\rangle' 
\sum_{\boldsymbol{k} \neq 0} \left\langle \hat{M} \left( k, \tilde{t} \right) \right\rangle'' 
}{\displaystyle 
\sum_{\boldsymbol{k} \neq 0} \left\vert \mathbb{M}_{3} \left( k \right) \right\vert^2} 
- 
\frac{\displaystyle 
\left\{ 
\sum_{\boldsymbol{k} \neq 0} \left\langle \hat{M} \left( k, \tilde{t} \right) \right\rangle' 
\right\}^2 
\sum_{\boldsymbol{k} \neq 0} \mathbb{M}_2 \left( k \right) 
\left\langle \hat{M} \left( k, \tilde{t} \right) \right\rangle' 
}{\displaystyle 
\left\{ 
\sum_{\boldsymbol{k} \neq 0} \left\vert \mathbb{M}_{3} \left( k \right) \right\vert^2 
\right\}^2} 
\right\rbrack 
\right\rgroup 
+ O \left( \tilde{V}^2_{ex} \right) 
\nonumber\\
& = & 
8 \frac{\displaystyle 
\left\lbrack 
\sum_{\boldsymbol{k} \neq 0} 
\textrm{Im} \left\{ 
e^{- 2 i \tilde{\omega}_{1} \left( k \right) \tilde{t}} 
\psi_{2, 1} \left( k, \tilde{t} \right) 
\mathbb{M}^{*}_3 \left( k \right) 
\mathbb{V}_2 \left( k \right) 
\right\} 
\right\rbrack^2 
}{\displaystyle 
\sum_{\boldsymbol{k} \neq 0} \left\vert \mathbb{M}_{3} \left( k \right) \right\vert^2} 
+ O \left( \tilde{V}_{ex} \right) 
. 
\label{CFI_gen_form}
\end{eqnarray}

To summarize, 
\begin{eqnarray}
I_{Q} \left( \tilde{V}_{ex}, \tilde{t} \right) 
& = & 
8 \sum_{\boldsymbol{k} \neq 0} 
\left\vert 
e^{- 2 i \tilde{\omega}_{1} \left( k \right) \tilde{t}} 
\int_{0}^{\tilde{t}} d \tilde{t}_1 \; 
e^{2 i \tilde{\omega}_{1} \left( k \right) \tilde{t}_1} 
f \left( \tilde{t}_1 \right) 
\right\vert^2 
\left\vert \mathbb{V}_2 \left( k \right) \right\vert^2 
+ O \left( \tilde{V}_{ex} \right) , 
\nonumber\\
I_{C} \left( \tilde{V}_{ex}, \tilde{t} \right) 
& = & 
8 \frac{\displaystyle 
\left\lbrack 
\sum_{\boldsymbol{k} \neq 0} 
\textrm{Im} \left\{ 
e^{- 2 i \tilde{\omega}_{1} \left( k \right) \tilde{t}} 
\int_{0}^{\tilde{t}} d \tilde{t}_1 \; 
e^{2 i \tilde{\omega}_{1} \left( k \right) \tilde{t}_1} 
f \left( \tilde{t}_1 \right) 
\mathbb{M}^{*}_3 \left( k \right) 
\mathbb{V}_2 \left( k \right) 
\right\} 
\right\rbrack^2 
}{\displaystyle 
\sum_{\boldsymbol{k} \neq 0} \left\vert \mathbb{M}_{3} \left( k \right) \right\vert^2} 
+ O \left( \tilde{V}_{ex} \right) 
,  
\label{QFICFI_Dyson_expansion}
\end{eqnarray}
where

\begin{eqnarray}
&& \!\!\!\!\!\!\!\!\!\!\!\!\!\!\!
\textrm{Re} 
\left\{ 
e^{- 2 i \tilde{\omega}_{1} \left( k \right) \tilde{t}} 
\int_{0}^{\tilde{t}} d \tilde{t}_1 \; 
e^{2 i \tilde{\omega}_{1} \left( k \right) \tilde{t}_1} 
f \left( \tilde{t}_1 \right) 
\right\} 
= 
\textrm{Re} 
\left\{ 
\int_{0}^{\tilde{t}} d \tilde{t}_1 \; 
e^{2 i \tilde{\omega}_{1} \left( k \right) \tilde{t}_1} 
f \left( \tilde{t}_1 \right) 
\right\} 
\cos \left\{ 2 \tilde{\omega}_{1} \left( k \right) \tilde{t} \right\} 
+ 
\textrm{Im} 
\left\{ 
\int_{0}^{\tilde{t}} d \tilde{t}_1 \; 
e^{2 i \tilde{\omega}_{1} \left( k \right) \tilde{t}_1} 
f \left( \tilde{t}_1 \right) 
\right\} 
\sin \left\{ 2 \tilde{\omega}_{1} \left( k \right) \tilde{t} \right\} 
, \nonumber\\ 
&& \!\!\!\!\!\!\!\!\!\!\!\!\!\!\!
\textrm{Im} 
\left\{ 
e^{- 2 i \tilde{\omega}_{1} \left( k \right) \tilde{t}} 
\int_{0}^{\tilde{t}} d \tilde{t}_1 \; 
e^{2 i \tilde{\omega}_{1} \left( k \right) \tilde{t}_1} 
f \left( \tilde{t}_1 \right) 
\right\} 
= 
\textrm{Im} 
\left\{ 
\int_{0}^{\tilde{t}} d \tilde{t}_1 \; 
e^{2 i \tilde{\omega}_{1} \left( k \right) \tilde{t}_1} 
f \left( \tilde{t}_1 \right) 
\right\} 
\cos \left\{ 2 \tilde{\omega}_{1} \left( k \right) \tilde{t} \right\} 
- 
\textrm{Re} 
\left\{ 
\int_{0}^{\tilde{t}} d \tilde{t}_1 \; 
e^{2 i \tilde{\omega}_{1} \left( k \right) \tilde{t}_1} 
f \left( \tilde{t}_1 \right) 
\right\} 
\sin \left\{ 2 \tilde{\omega}_{1} \left( k \right) \tilde{t} \right\} 
. \nonumber\\
\end{eqnarray}

Suppose that $f \left( t \right) = \delta \left( t - t_p \right)$ for $t \ge 0$ (assuming $t_p \ge 0$). Then 

\begin{eqnarray}
&& \!\!\!\!\!\!\!\!\!\!
\textrm{Re} 
\left\{ 
e^{- 2 i \tilde{\omega}_{1} \left( k \right) \tilde{t}} 
\int_{0}^{\tilde{t}} d \tilde{t}_1 \; 
e^{2 i \tilde{\omega}_{1} \left( k \right) \tilde{t}_1} 
f \left( \tilde{t}_1 \right) 
\right\} 
= 
\theta \left( \tilde{t} - \tilde{t}_p \right) 
\cos \left\{ 2 \tilde{\omega}_{1} \left( k \right) \left( \tilde{t} - \tilde{t}_p \right) \right\} 
, \nonumber\\
&& \!\!\!\!\!\!\!\!\!\!
\textrm{Im} 
\left\{ 
e^{- 2 i \tilde{\omega}_{1} \left( k \right) \tilde{t}} 
\int_{0}^{\tilde{t}} d \tilde{t}_1 \; 
e^{2 i \tilde{\omega}_{1} \left( k \right) \tilde{t}_1} 
f \left( \tilde{t}_1 \right) 
\right\} 
= 
- 
\theta \left( \tilde{t} - \tilde{t}_p \right) 
\sin \left\{ 2 \tilde{\omega}_{1} \left( k \right) \left( \tilde{t} - \tilde{t}_p \right) \right\} 
, \nonumber\\
\end{eqnarray}
where $\theta \left( \tilde{t} \right) = 0$ if $\tilde{t} < 0$ and 1 otherwise, 
and thus

\begin{eqnarray}
&& \!\!\!\!\!\!\!\!\!\!\!\!\!\!\!\!\!\!
I_{Q} \left( \tilde{V}_{ex}, \tilde{t} \right) 
= 
8 \theta \left( \tilde{t} - \tilde{t}_p \right) 
\sum_{\boldsymbol{k} \neq 0} 
\left\vert \mathbb{V}_2 \left( k \right) \right\vert^2 
+ O \left( \tilde{V}_{ex} \right) 
, \nonumber\\
&& \!\!\!\!\!\!\!\!\!\!\!\!\!\!\!\!\!\!
I_{C} \left( \tilde{V}_{ex}, \tilde{t} \right) 
= 
8 \theta \left( \tilde{t} - \tilde{t}_p \right) 
\frac{\displaystyle 
\left\lgroup 
\sum_{\boldsymbol{k} \neq 0} 
\left\lbrack 
\textrm{Re} \left\{ 
\mathbb{M}^{*}_3 \left( k \right) 
\mathbb{V}_2 \left( k \right) 
\right\} 
\sin \left\{ 2 \tilde{\omega}_{1} \left( k \right) \left( \tilde{t} - \tilde{t}_p \right) \right\} 
- 
\textrm{Im} \left\{ 
\mathbb{M}^{*}_3 \left( k \right) 
\mathbb{V}_2 \left( k \right) 
\right\} 
\cos \left\{ 2 \tilde{\omega}_{1} \left( k \right) \left( \tilde{t} - \tilde{t}_p \right) \right\} 
\right\rbrack 
\right\rgroup^2 
}{\displaystyle 
\sum_{\boldsymbol{k} \neq 0} \left\vert \mathbb{M}_{3} \left( k \right) \right\vert^2} 
+ O \left( \tilde{V}_{ex} \right) 
. \nonumber\\
\end{eqnarray}
Note that $I_{Q} \left( \tilde{V}_{ex}, \tilde{t} \right)$ is constant in scaled time $\tilde{t}$ for $\tilde{t} \ge \tilde{t}_p$.

If $f \left( t \right) = \theta \left( t - t_p \right)$ for $t \ge 0$ (assuming $t_p \ge 0$).

\begin{eqnarray}
&& \!\!\!\!\!\!\!\!\!\!
\textrm{Re} 
\left\{ 
e^{- 2 i \tilde{\omega}_{1} \left( k \right) \tilde{t}} 
\int_{0}^{\tilde{t}} d \tilde{t}_1 \; 
e^{2 i \tilde{\omega}_{1} \left( k \right) \tilde{t}_1} 
f \left( \tilde{t}_1 \right) 
\right\} 
= 
\theta \left( \tilde{t} - \tilde{t}_p \right) 
\frac{\sin \left\{ \tilde{\omega}_{1} \left( k \right) \left( \tilde{t} - \tilde{t}_p \right) \right\}}{\tilde{\omega}_{1} \left( k \right)} 
\cos \left\{ \tilde{\omega}_{1} \left( k \right) \left( \tilde{t} - \tilde{t}_p \right) \right\} 
, \nonumber\\
&& \!\!\!\!\!\!\!\!\!\!
\textrm{Im} 
\left\{ 
e^{- 2 i \tilde{\omega}_{1} \left( k \right) \tilde{t}} 
\int_{0}^{\tilde{t}} d \tilde{t}_1 \; 
e^{2 i \tilde{\omega}_{1} \left( k \right) \tilde{t}_1} 
f \left( \tilde{t}_1 \right) 
\right\} 
= 
- 
\theta \left( \tilde{t} - \tilde{t}_p \right) 
\frac{\sin \left\{ \tilde{\omega}_{1} \left( k \right) \left( \tilde{t} - \tilde{t}_p \right) \right\}}{\tilde{\omega}_{1} \left( k \right)} 
\sin \left\{ \tilde{\omega}_{1} \left( k \right) \left( \tilde{t} - \tilde{t}_p \right) \right\} 
, \nonumber\\
\end{eqnarray}
and thus

\begin{eqnarray}
&& \!\!\!\!\!\!\!\!\!\!\!\!\!\!\!\!\!\!\!\!\!\!\!\!\!\!
I_{Q} \left( \tilde{V}_{ex}, \tilde{t} \right) 
= 
8 \theta \left( \tilde{t} - \tilde{t}_p \right) 
\sum_{\boldsymbol{k} \neq 0} 
\frac{\sin^2 \left\{ \tilde{\omega}_{1} \left( k \right) \left( \tilde{t} - \tilde{t}_p \right) \right\}}{\tilde{\omega}^2_{1} \left( k \right)} 
\left\vert \mathbb{V}_2 \left( k \right) \right\vert^2 
+ O \left( \tilde{V}_{ex} \right) 
, \nonumber\\
&& \!\!\!\!\!\!\!\!\!\!\!\!\!\!\!\!\!\!\!\!\!\!\!\!\!\!
I_{C} \left( \tilde{V}_{ex}, \tilde{t} \right) 
= 
8 \theta \left( \tilde{t} - \tilde{t}_p \right) 
\frac{\displaystyle 
\left\lgroup \!
\sum_{\boldsymbol{k} \neq 0} 
\frac{\sin \left\{ \tilde{\omega}_{1} \left( k \right) \left( \tilde{t} - \tilde{t}_p \right) \right\}}{\tilde{\omega}_{1} \left( k \right)} 
\left\lbrack 
\textrm{Re} \left\{ 
\mathbb{M}^{*}_3 \left( k \right) 
\mathbb{V}_2 \left( k \right) 
\right\} 
\sin \left\{ \tilde{\omega}_{1} \left( k \right) \left( \tilde{t} - \tilde{t}_p \right) \right\} 
- 
\textrm{Im} \left\{ 
\mathbb{M}^{*}_3 \left( k \right) 
\mathbb{V}_2 \left( k \right) 
\right\} 
\cos \left\{ \tilde{\omega}_{1} \left( k \right) \left( \tilde{t} - \tilde{t}_p \right) \right\} 
\right\rbrack 
\!
\right\rgroup^2 \!
}{\displaystyle 
\sum_{\boldsymbol{k} \neq 0} \left\vert \mathbb{M}_{3} \left( k \right) \right\vert^2} 
\nonumber\\
&& \qquad \quad 
+ O \left( \tilde{V}_{ex} \right) 
. 
\end{eqnarray}

If $f \left( t \right) = \cos \left( \omega_{a} t + \delta_{a} \right)$ for $t \ge 0$,

\begin{eqnarray}
&& \!\!\!\!\!\!\!\!\!\!\!\!\!\!\!\!\!\!\!\!\!\!\!
I_{Q} \left( \tilde{V}_{ex}, \tilde{t} \right) 
= 
8 \sum_{\boldsymbol{k} \neq 0} 
\left\vert 
e^{- 2 i \tilde{\omega}_{1} \left( k \right) \tilde{t}} 
\psi_{2, 1} \left( k, \tilde{t} \right) 
\right\vert^2 
\left\vert \mathbb{V}_2 \left( k \right) \right\vert^2 
+ O \left( \tilde{V}_{ex} \right) , 
\nonumber\\
&& \!\!\!\!\!\!\!\!\!\!\!\!\!\!\!\!\!\!\!\!\!\!\!
I_{C} \left( \tilde{V}_{ex}, \tilde{t} \right) 
= 
8 \frac{\displaystyle 
\left\lbrack 
\sum_{\boldsymbol{k} \neq 0} 
\textrm{Im} \left\{ 
e^{- 2 i \tilde{\omega}_{1} \left( k \right) \tilde{t}} 
\psi_{2, 1} \left( k, \tilde{t} \right) 
\right\} 
\textrm{Re} \left\{ 
\mathbb{M}^{*}_3 \left( k \right) 
\mathbb{V}_2 \left( k \right) 
\right\} 
+ 
\textrm{Re} \left\{ 
e^{- 2 i \tilde{\omega}_{1} \left( k \right) \tilde{t}} 
\psi_{2, 1} \left( k, \tilde{t} \right) 
\right\} 
\textrm{Im} \left\{ 
\mathbb{M}^{*}_3 \left( k \right) 
\mathbb{V}_2 \left( k \right) 
\right\} 
\right\rbrack^2 
}{\displaystyle 
\sum_{\boldsymbol{k} \neq 0} \left\vert \mathbb{M}_{3} \left( k \right) \right\vert^2} 
+ O \left( \tilde{V}_{ex} \right) 
, \qquad \qquad 
\label{QFICFI_sinusoidal}
\end{eqnarray}
where

\begin{eqnarray}
\textrm{Re} 
\left\{ 
e^{- 2 i \tilde{\omega}_{1} \left( k \right) \tilde{t}} 
\psi_{2, 1} \left( k, \tilde{t} \right) 
\right\} 
& = & 
\frac{
\cos \delta_{a} 
\sin \left\lbrack 
\left\{ 
\tilde{\omega}_{a} 
+ \epsilon \left( k \right) 
\right\} \tilde{t} 
\right\rbrack}
{2 \tilde{\omega}_{a} + \epsilon \left( k \right)} 
+ 
\frac{\tilde{\omega}_{a} 
\cos \left\lbrack 
\left\{ 
\tilde{\omega}_{a} 
+ \frac{\epsilon \left( k \right)}{2} 
\right\} \tilde{t} 
+ \delta_{a} 
\right\rbrack}{2 \tilde{\omega}_{a} + \epsilon \left( k \right)} 
\frac{
\sin \left\{ \epsilon \left( k \right) \tilde{t} / 2 \right\} 
}{\epsilon \left( k \right) / 2} 
, \qquad \qquad 
\label{ReIntPsi21}
\end{eqnarray}

\begin{eqnarray}
&& \!\!\!\!\!\!\!\!\!\!
\textrm{Im} 
\left\{ 
e^{- 2 i \tilde{\omega}_{1} \left( k \right) \tilde{t}} 
\psi_{2, 1} \left( k, \tilde{t} \right) 
\right\} 
= 
\frac{
\cos \delta_{a} 
\cos \left\lbrack 
\left\{ 
\tilde{\omega}_{a} 
+ \epsilon \left( k \right) 
\right\} \tilde{t} 
\right\rbrack 
- 
\cos \left( \tilde{\omega}_{a} \tilde{t} + \delta_{a} \right) 
}
{2 \tilde{\omega}_{a} + \epsilon \left( k \right)} 
- 
\frac{\tilde{\omega}_{a} 
\sin \left\lbrack 
\left\{ 
\tilde{\omega}_{a} 
+ \frac{\epsilon \left( k \right)}{2} 
\right\} \tilde{t} 
+ \delta_{a} 
\right\rbrack}{2 \tilde{\omega}_{a} + \epsilon \left( k \right)} 
\frac{
\sin \left\{ \epsilon \left( k \right) \tilde{t} / 2 \right\} 
}{\epsilon \left( k \right) / 2} 
, \qquad \qquad 
\label{ImIntPsi21}
\end{eqnarray}
$\tilde{\omega}_{a} \coloneqq \omega_{a} / \left( g_{a} n \right)$, and $\epsilon \left( k \right) \coloneqq 2 \tilde{\omega}_{1} \left( k \right) - \tilde{\omega}_{a}$. 
Note that both 
$\textrm{Im} 
\left\{ 
e^{- 2 i \tilde{\omega}_{1} \left( k \right) \tilde{t}} 
\psi_{2, 1} \left( k, \tilde{t} \right) 
\right\}$ and $\left\vert e^{- 2 i \tilde{\omega}_{1} \left( k \right) \tilde{t}} 
\psi_{2, 1} \left( k, \tilde{t} \right) \right\vert$ do not diverge at $2 \tilde{\omega}_{1} \left( k \right) = \tilde{\omega}_{a}$ (where $\epsilon \left( k \right) = 0$).

From now on, we focus on measuring the 
number of created molecules where 
$\hat{M}_{S} = \int d^3 r \; 
\delta \hat{\psi}^{\dagger}_{m} \left( \boldsymbol{r} \right) 
\delta \hat{\psi}_{m} \left( \boldsymbol{r} \right) 
$. Since we neglected effects of $\hat{b}_2 \left( k \right)$,

\begin{eqnarray}
\hat{M}_{S} = \sum_{\boldsymbol{k} \neq 0} 
\left\lbrack 
v^2_{12} \left( k \right) 
+ 
\left\{ 
u^2_{12} \left( k \right) 
+ 
v^2_{12} \left( k \right) 
\right\} 
\hat{b}^{\dagger}_{1} \left( \boldsymbol{k} \right) 
\hat{b}_{1} \left( \boldsymbol{k} \right) 
+ 
u_{12} \left( k \right) 
v_{12} \left( k \right) 
\left\{ 
\hat{b}^{\dagger}_{1} \left( \boldsymbol{k} \right) 
\hat{b}^{\dagger}_{1} \left( - \boldsymbol{k} \right) 
+ 
\hat{b}_{1} \left( \boldsymbol{k} \right) 
\hat{b}_{1} \left( - \boldsymbol{k} \right) 
\right\} 
\right\rbrack 
, \qquad 
\end{eqnarray}
where we get $\mathbb{M}_3 \left( k \right) = u_{12} \left( k \right) v_{12} \left( k \right)$, which is real in the stable system.

\subsection{Density Perturbation}

Neglecting effects of $\hat{b}_2 \left( k \right)$ (for a system with large gap between $\tilde{\omega}_{1} \left( k \right)$ and $\tilde{\omega}_{2} \left( k \right)$), when we perturb 
the mass densities of 
the atoms and molecules in the state, 
i.e., when we impose 
$\hat{V}_{S} \left( t \right) = \left( V_{ex} / m_a \right) f \left( t \right) 
\int d^3 r \; 
\left\{ 
m_a 
\hat{\psi}^{\dagger}_{a} \left( \boldsymbol{r} \right) 
\hat{\psi}_{a} \left( \boldsymbol{r} \right) 
+ 
m_m 
\hat{\psi}^{\dagger}_{m} \left( \boldsymbol{r} \right) 
\hat{\psi}_{m} \left( \boldsymbol{r} \right) 
\right\} 
$ with $m_m \simeq 2 m_a$, we get 

\begin{eqnarray}
\!\!\!\!\!\!\!\!\!\!\!\!\!\!\!
\hat{V}_{S} \left( t \right) 
& = & 
V_{ex} f \left( t \right) 
\left\lgroup 
N 
+ 
\sum_{\boldsymbol{k} \neq 0} 
\left\lbrack 
v^2_{11} \left( k \right) 
+ 
2 v^2_{12} \left( k \right) 
+ 
\left\{ 
u^2_{11} \left( k \right) 
+ 
v^2_{11} \left( k \right) 
+ 
2 u^2_{12} \left( k \right) 
+ 
2 v^2_{12} \left( k \right) 
\right\} 
\hat{b}^{\dagger}_{1} \left( \boldsymbol{k} \right) 
\hat{b}_{1} \left( \boldsymbol{k} \right) 
\right\rbrack
\right\rgroup 
\nonumber\\
&& + 
V_{ex} f \left( t \right) 
\sum_{\boldsymbol{k} \neq 0} 
\left\{ 
u_{11} \left( k \right) 
v_{11} \left( k \right) 
+ 
2 
u_{12} \left( k \right) 
v_{12} \left( k \right) 
\right\} 
\left\{ 
\hat{b}^{\dagger}_{1} \left( \boldsymbol{k} \right) 
\hat{b}^{\dagger}_{1} \left( - \boldsymbol{k} \right) 
+ 
\hat{b}_{1} \left( \boldsymbol{k} \right) 
\hat{b}_{1} \left( - \boldsymbol{k} \right) 
\right\} 
, \qquad \qquad 
\label{density_V2}
\end{eqnarray}
which makes 
$\mathbb{V}_2 \left( k \right) 
= 
u_{11} \left( k \right) v_{11} \left( k \right) 
+ 
2 u_{12} \left( k \right) v_{12} \left( k \right)$.

\subsection{\label{comp_Dennis_sec}Density Perturbation on Purely Atomic BEC}
For scalar BEC, it is known that~\cite{pitaevskii2003bose}
\begin{eqnarray}
\tilde{\omega}_1 \left( k \right) 
= 
k \xi_a 
\sqrt{\left( k \xi_a \right)^2 + 2}
, \quad 
u_{11} \left( k \right) 
= 
\sqrt{\frac{1}{2}} 
\sqrt{\frac{\left( k \xi_a \right)^2 + 1}{\tilde{\omega}_1 \left( k \right)} + 1}
, \quad \textrm{and} \quad 
v_{11} \left( k \right) 
= 
- \sqrt{\frac{1}{2}} 
\sqrt{\frac{\left( k \xi_a \right)^2 + 1}{\tilde{\omega}_1 \left( k \right)} - 1} 
\quad . 
\end{eqnarray}
\\
Note that 
in the Heisenberg picture with purely atomic BEC, Eqs.~\eqref{homogeneous_box_delta_hat_psi_am},~\eqref{Bog_trans}, and~\eqref{def_ujvj} can be written as 
\begin{eqnarray}
\delta \hat{\psi}_a \left( \boldsymbol{r}, t \right) 
= 
e^{- i \int^{t} d t_1 \; V_a \left( \boldsymbol{r}, t_1 \right) / \hbar} 
e^{- i \mu t / \hbar} 
\sum_{\boldsymbol{k} \neq 0} 
\mathcal{B} \left( \boldsymbol{r}, \boldsymbol{k} \right) 
\left\{ 
u_{11} \left( k \right) 
\hat{b}_1 \left( \boldsymbol{k}, t \right) 
+ 
v^{*}_{11} \left( k \right) 
\hat{b}^{\dagger}_1 \left( - \boldsymbol{k}, t \right) 
\right\} 
, \quad 
\mathcal{B} \left( \boldsymbol{r}, \boldsymbol{k} \right) 
= 
\frac{1}{\sqrt{V}} e^{i \boldsymbol{k} \cdot \boldsymbol{r}} 
, \qquad \qquad 
\end{eqnarray}
where the $\mathcal{B} \left( \boldsymbol{r}, \boldsymbol{k} \right)$ 
span an orthonormal basis 
of the single-particle Hilbert space 
that satisfies
$\int d^3 r \; 
\mathcal{B}^{*} \left( \boldsymbol{r}, \boldsymbol{k}_1 \right) 
\mathcal{B} \left( \boldsymbol{r}, \boldsymbol{k}_2 \right) 
= 
\delta_{\boldsymbol{k}_1, \boldsymbol{k}_2}$. 
\\
When $f \left( t \right) = \sin \left( \omega_a t \right)$, from Eqs.~\eqref{QFICFI_sinusoidal},~\eqref{ReIntPsi21}, and~\eqref{ImIntPsi21}, we get 
\begin{eqnarray}
I_Q \left( \tilde{V}_{ex}, t \right) 
\rightarrow 
4 \tilde{t}^2 
\left\vert 
u_{11} \left( k_{r} \right) 
v_{11} \left( k_{r} \right) 
\right\vert^2 
= 
\left\vert 
2 g_a n 
\frac{t}{\hbar} 
u_{11} \left( k_{r} \right) 
v_{11} \left( k_{r} \right) 
\right\vert^2 
\textrm{ for $t \gg 1$,}
\end{eqnarray}
where $\omega_1 \left( k_{r} \right) = \omega_a / 2$ (Remember that we define $\tilde{\omega}_1 \left( k \right) \coloneqq \hbar \omega_1 \left( k \right) / \left( g_a n \right)$, $\tilde{t} \coloneqq g_a n t / \hbar$, and $\tilde{V}_{ex} \coloneqq V_{ex} / \left( g_a n \right)$ in the main text). 
For $V_{ex} = \left( m_a \mathcal{G} / 2 \right) \left( L^2 / 12 \right)$ (our approximation of the density perturbation due to gravitational wave in Eq.~\eqref{eq:hintgw}), we get 
\begin{eqnarray}
I_Q \left( \mathcal{G}, t \right) 
= 
\left( 
\frac{\partial \tilde{V}_{ex}}{\partial \mathcal{G}} 
\right)^2  
I_Q \left( \tilde{V}_{ex}, t \right) 
\rightarrow 
\left\vert 
2 
\frac{V_{ex}}{\mathcal{G}} 
\frac{t}{\hbar} 
u_{11} \left( k_{r} \right) 
v_{11} \left( k_{r} \right) 
\right\vert^2 
\textrm{ for $t \gg 1$.} 
\label{ourQFI_BasisNotation}
\end{eqnarray}
\\
For $\hat{V}_{S} \left( t \right) 
= 
\sin \left( \omega_a t \right) 
\int d^3 r \; 
\left( m_a \mathcal{G} r^2_1 / 2 \right) 
\hat{\psi}^{\dagger}_a \left( \boldsymbol{r} \right) 
\hat{\psi}_a \left( \boldsymbol{r} \right)$ where 
$\displaystyle \boldsymbol{r} = \sum_{j = 1}^3 r_j \boldsymbol{e}_j$ ($\boldsymbol{e}_j$ are Cartesian basis 
vectors), 
\cite{Dennis} independently calculated $I_Q \left( \mathcal{G}, t \right)$ for a quasi-1D system with $- L/2 < r_1 < L/2$ and $t \gg 1$ by taking 
$\boldsymbol{B} \left( \boldsymbol{r}, \boldsymbol{k} \right) 
= 
\sqrt{2 / \left( A L \right)} 
\cos \left\{ k_{n_1} \left( r_1 + L / 2 \right) \right\}$ with $k_{n_1} = n_1 \pi / L$ ($n_1$ are positive integers).
Their results for mode $k_{r}$ where $\omega_1 \left( k_{r} \right) = \omega_a / 2$ can be written as 
\begin{eqnarray}
I_Q \left( \mathcal{G}, \tilde{t} \right) 
& \rightarrow & 
\left\vert 
2 \int d^3 r \; 
\frac{m_a}{2} \mathcal{G} r^2_1 
\mathcal{B}^{*} \left( \boldsymbol{r}, \boldsymbol{k}_r \right) 
u^{*}_{11} \left( k_{r} \right) 
\mathcal{B} \left( \boldsymbol{r}, \boldsymbol{k}_r \right) 
v_{11} \left( k_{r} \right) 
\frac{t}{\mathcal{G} \hbar} 
\right\vert^2 
\nonumber\\
& \rightarrow & 
\left\vert 
2 \left\{ 
\frac{m_a}{2} 
\int d^3 r \; 
r^2_1 
\mathcal{B}^{*} \left( \boldsymbol{r}, \boldsymbol{k}_r \right) 
\mathcal{B} \left( \boldsymbol{r}, \boldsymbol{k}_r \right) 
\right\} 
\frac{t}{\hbar} 
u_{11} \left( k_{r} \right) 
v_{11} \left( k_{r} \right) 
\right\vert^2 
. 
\label{DennisQFI_BasisNotation}
\end{eqnarray}
One can see that Eq.~\eqref{DennisQFI_BasisNotation} becomes Eq.~\eqref{ourQFI_BasisNotation} if one takes 
$\mathcal{B} \left( \boldsymbol{r}, \boldsymbol{k} \right) 
= 
\left( 1 / \sqrt{V} \right) e^{i \boldsymbol{k} \cdot \boldsymbol{r}}$ as we do. 
Since BEC atoms are confined in $-L/2 < r_1 < L/2$, $\delta \hat{\psi}_a \left( \boldsymbol{r}, t \right)$ should be zero at $r_1 = \pm L/2$ and our choice of $\mathcal{B} \left( \boldsymbol{r}, \boldsymbol{k} \right)$ can satisfy this condition by taking $\boldsymbol{k} \rightarrow \left( 2 \pi / L \right) \boldsymbol{e}_1$ which make $\mathcal{B} \left( \left( -L/2, r_2, r_3 \right), \boldsymbol{k} \right) = \mathcal{B} \left( \left( L/2, r_2, r_3 \right), \boldsymbol{k} \right)$.

\section{\label{Symplectic_Fisher}Symplectic Formalism and Fisher Information}

Suppose that the Hamiltonian $\hat{H}_{S} \left( t \right)$ in Schr\"odinger picture is given as 
$\displaystyle 
\hat{H}_{S} \left( t \right) 
=  
\sum_{\boldsymbol{k} \neq 0} 
\hat{H}_{c} \left( \boldsymbol{k}, t \right) 
$.
The time evolution operator $\hat{U}_{S} \left( t \right)$ in the Schr\"odinger picture satisfies
$\displaystyle 
i \hbar \frac{\partial \hat{U}_{S} \left( t \right)}{\partial t} 
= 
\hat{H}_{S} \left( t \right) 
\hat{U}_{S} \left( t \right) 
$.

From Appendix A of~\cite{Qvarfort_2020}, let 
$\displaystyle 
\hat{U}_{S} \left( t \right) 
= 
\hat{U}_1 \left( t \right) 
\hat{U}_2 \left( t \right) 
\hat{U}_3 \left( t \right) 
$ with 
$\displaystyle 
\hat{U}_j \left( t \right) 
= 
\exp \left\{ 
- i \sum_{\boldsymbol{k} \neq 0} 
F_j \left( k, t \right) 
\hat{G}_j \left( \boldsymbol{k} \right) 
\right\} 
$ for $j = 1, 2, 3$ where $k \coloneqq \left\vert \boldsymbol{k} \right\vert$. 
Then we get

\begin{eqnarray}
\hat{H}_{S} \left( t \right) 
& = & 
\hbar \sum_{\boldsymbol{k} \neq 0} 
\left\{ 
\frac{\partial F_1 \left( k, t \right)}{\partial t} 
\hat{G}_1 \left( \boldsymbol{k} \right) 
+ 
\frac{\partial F_2 \left( k, t \right)}{\partial t} 
\hat{U}_1 \left( t \right) 
\hat{G}_2 \left( \boldsymbol{k} \right) 
\hat{U}^{\dagger}_1 \left( t \right) 
+ 
\frac{\partial F_3 \left( k, t \right)}{\partial t} 
\hat{U}_1 \left( t \right) 
\hat{U}_2 \left( t \right) 
\hat{G}_3 \left( \boldsymbol{k} \right) 
\hat{U}^{\dagger}_2 \left( t \right) 
\hat{U}^{\dagger}_1 \left( t \right) 
\right\} 
. \qquad \qquad 
\end{eqnarray}

Let $n_b \left( \boldsymbol{k} \right)$ be the number of 
reactons with momentum $\boldsymbol{k}$. 
From now on, we will denote the state with $n_b \left( \boldsymbol{k}_j \right) = n_b \left( - \boldsymbol{k}_j \right)$ for all $j$ which satisfy $\left\vert \boldsymbol{k}_j \right\vert \le \left\vert \boldsymbol{k}_c \right\vert$ as 
$\left\vert 
n_b \left( \boldsymbol{k}_1 \right) , 
n_b \left( \boldsymbol{k}_2 \right) , 
\cdots 
n_b \left( \boldsymbol{k}_c \right) 
\right\rangle 
$ 
where $\boldsymbol{k}_c$ is the cutoff. 
By defining

\begin{eqnarray}
\hat{K}_{z} \left( \boldsymbol{k} \right) 
& \coloneqq & 
\frac{1}{2} 
\left\{ 
\hat{b}^{\dagger}_1 \left( \boldsymbol{k} \right) 
\hat{b}_1 \left( \boldsymbol{k} \right) 
+ 
\hat{b}_1 \left( - \boldsymbol{k} \right) 
\hat{b}^{\dagger}_1 \left( - \boldsymbol{k} \right) 
\right\} 
, 
\quad 
\hat{K}_{x} \left( \boldsymbol{k} \right)
\coloneqq 
\frac{1}{2} \left\{ 
\hat{b}^{\dagger}_1 \left( \boldsymbol{k} \right) 
\hat{b}^{\dagger}_1 \left( - \boldsymbol{k} \right) 
+ 
\hat{b}_1 \left( \boldsymbol{k} \right) 
\hat{b}_1 \left( - \boldsymbol{k} \right) 
\right\} 
, 
\nonumber\\
\hat{K}_{y} \left( \boldsymbol{k} \right) 
& \coloneqq & 
\frac{1}{2 i} \left\{ 
\hat{b}^{\dagger}_1 \left( \boldsymbol{k} \right) 
\hat{b}^{\dagger}_1 \left( - \boldsymbol{k} \right) 
- 
\hat{b}_1 \left( \boldsymbol{k} \right) 
\hat{b}_1 \left( - \boldsymbol{k} \right) 
\right\} 
, 
\end{eqnarray}
and 
$\hat{K}_{\pm} \left( \boldsymbol{k} \right) 
\coloneqq 
\hat{K}_{x} \left( \boldsymbol{k} \right)
\pm i 
\hat{K}_{y} \left( \boldsymbol{k} \right)
$, 
we get 
$\hat{K}_{\zeta} \left( \boldsymbol{k} \right) = \hat{K}_{\zeta} \left( - \boldsymbol{k} \right)$ for $\zeta = x, y, z, \pm$,

\begin{eqnarray}
\hat{K}_{+} \left( \boldsymbol{k}_j \right\rangle 
\left\vert 
n_b \left( \boldsymbol{k}_1 \right) , 
\cdots , 
n_b \left( \boldsymbol{k}_j \right), 
\cdots , 
n_b \left( \boldsymbol{k}_c \right) 
\right\rangle 
& = & 
\hat{b}^{\dagger}_1 \left( \boldsymbol{k}_j \right) 
\hat{b}^{\dagger}_1 \left( - \boldsymbol{k}_j \right) 
\left\vert 
n_b \left( \boldsymbol{k}_1 \right) , 
\cdots , 
n_b \left( \boldsymbol{k}_j \right), 
\cdots , 
n_b \left( \boldsymbol{k}_c \right) 
\right\rangle 
\nonumber\\
& = & 
\underbrace{\sqrt{n_b \left( \boldsymbol{k}_j \right) + 1}}_{\textrm{with } \hat{b}^{\dagger}_{1} \left( \boldsymbol{k}_{j} \right)}  
\underbrace{\sqrt{n_b \left( \boldsymbol{k}_j \right) + 1}}_{\textrm{with } \hat{b}^{\dagger}_{1} \left( - \boldsymbol{k}_{j} \right)}  
\left\vert 
n_b \left( \boldsymbol{k}_1 \right) , 
\cdots , 
n_b \left( \boldsymbol{k}_j \right) + 1 , 
\cdots , 
n_b \left( \boldsymbol{k}_c \right) 
\right\rangle 
\nonumber\\
& = & 
\left\{ 
n_b \left( \boldsymbol{k}_j \right) + 1 
\right\} 
\left\vert 
n_b \left( \boldsymbol{k}_1 \right) , 
\cdots , 
n_b \left( \boldsymbol{k}_j \right) + 1 , 
\cdots , 
n_b \left( \boldsymbol{k}_c \right) 
\right\rangle 
, 
\end{eqnarray}

\begin{eqnarray}
\hat{K}_{-} \left( \boldsymbol{k}_j \right\rangle 
\left\vert 
n_b \left( \boldsymbol{k}_1 \right) , 
\cdots , 
n_b \left( \boldsymbol{k}_j \right), 
\cdots , 
n_b \left( \boldsymbol{k}_c \right) 
\right\rangle 
& = & 
\hat{b}_1 \left( \boldsymbol{k}_j \right) 
\hat{b}_1 \left( - \boldsymbol{k}_j \right) 
\left\vert 
n_b \left( \boldsymbol{k}_1 \right) , 
\cdots , 
n_b \left( \boldsymbol{k}_j \right), 
\cdots , 
n_b \left( \boldsymbol{k}_c \right) 
\right\rangle 
\nonumber\\
& = & 
\underbrace{\sqrt{n_b \left( \boldsymbol{k}_j \right)}}_{\textrm{with } \hat{b}_{1} \left( \boldsymbol{k}_{j} \right)}  
\underbrace{\sqrt{n_b \left( \boldsymbol{k}_j \right)}}_{\textrm{with } \hat{b}_{1} \left( - \boldsymbol{k}_{j} \right)}  
\left\vert 
n_b \left( \boldsymbol{k}_1 \right) , 
\cdots ,
n_b \left( \boldsymbol{k}_j \right) - 1 , 
\cdots ,
n_b \left( \boldsymbol{k}_c \right) 
\right\rangle 
\nonumber\\
& = & 
n_b \left( \boldsymbol{k}_j \right) 
\left\vert 
n_b \left( \boldsymbol{k}_1 \right) , 
\cdots , 
n_b \left( \boldsymbol{k}_j \right) - 1 , 
\cdots , 
n_b \left( \boldsymbol{k}_c \right) 
\right\rangle 
, 
\end{eqnarray}
and

\begin{eqnarray}
\hat{K}_{z} \left( \boldsymbol{k}_j \right\rangle 
\left\vert 
n_b \left( \boldsymbol{k}_1 \right) , 
\cdots , 
n_b \left( \boldsymbol{k}_j \right), 
\cdots , 
n_b \left( \boldsymbol{k}_c \right) 
\right\rangle 
& = & 
\frac{1}{2} 
\left\{ 
\hat{b}^{\dagger}_1 \left( \boldsymbol{k}_j \right) 
\hat{b}_1 \left( \boldsymbol{k}_j \right) 
+ 
\hat{b}_1 \left( - \boldsymbol{k}_j \right) 
\hat{b}^{\dagger}_1 \left( - \boldsymbol{k}_j \right) 
\right\} 
\left\vert 
n_b \left( \boldsymbol{k}_1 \right) , 
\cdots , 
n_b \left( \boldsymbol{k}_j \right), 
\cdots , 
n_b \left( \boldsymbol{k}_c \right) 
\right\rangle 
\nonumber\\
& = & 
\frac{1}{2} 
\{ 
\underbrace{n_b \left( \boldsymbol{k}_j \right)}_{\textrm{with } 
\hat{b}^{\dagger}_1 \left( \boldsymbol{k}_j \right) 
\hat{b}_1 \left( \boldsymbol{k}_j \right)} 
+ 
\underbrace{n_b \left( \boldsymbol{k}_j \right) + 1}_{\textrm{with } 
\hat{b}_1 \left( - \boldsymbol{k}_j \right) 
\hat{b}^{\dagger}_1 \left( - \boldsymbol{k}_j \right)}  
\} 
\left\vert 
n_b \left( \boldsymbol{k}_1 \right) , 
\cdots , 
n_b \left( \boldsymbol{k}_j \right), 
\cdots , 
n_b \left( \boldsymbol{k}_c \right) 
\right\rangle 
\nonumber\\
& = & 
\left\{ 
n_b \left( \boldsymbol{k}_j \right) 
+ 
\frac{1}{2} 
\right\} 
\left\vert 
n_b \left( \boldsymbol{k}_1 \right) , 
\cdots , 
n_b \left( \boldsymbol{k}_j \right), 
\cdots , 
n_b \left( \boldsymbol{k}_c \right) 
\right\rangle 
.
\end{eqnarray}
Therefore, we get

\begin{eqnarray}
\left\vert 
n_b \left( \boldsymbol{k}_1 \right) , 
\cdots 
n_b \left( \boldsymbol{k}_c \right) 
\right\rangle 
& = & 
\prod_{j = 1} 
\frac{1}{n_b \left( \boldsymbol{k}_j \right) !} 
\left\{ 
\hat{K}_{+} \left( \boldsymbol{k}_j \right) 
\right\}^{n_b \left( \boldsymbol{k}_j \right)} 
\left\vert \textrm{vac} \right\rangle 
, 
\end{eqnarray}
where $\left\vert \textrm{vac} \right\rangle$ is Bogoliubov vacuum state.

From now on, for simplicity, we will consider $\hat{H}_{S} \left( t \right)$ in Eq.~\eqref{appendix_HS_gen_form} and~\eqref{PertV_gen_form} without $\hat{b}_2 \left( \boldsymbol{k} \right)$ terms as we did in section~\ref{CFI_QFI_Formalism}. 
Additionally, we will assume that $\mathbb{V}_2 \left( k \right) \in \mathbb{R}$ like Eq.~\eqref{density_V2}. 
By choosing $\hat{G}_1 \left( \boldsymbol{k} \right) = \hat{K}_z \left( \boldsymbol{k} \right)$, 
$\hat{G}_2 \left( \boldsymbol{k} \right) = \hat{K}_x \left( \boldsymbol{k} \right)$, and 
$\hat{G}_3 \left( \boldsymbol{k} \right) = \hat{K}_y \left( \boldsymbol{k} \right)$, 
we get  
\begin{eqnarray}
\hat{H}_c \left( \boldsymbol{k}, t \right) 
& = & 
\left\{ 
\hbar \omega_1 \left( k \right) 
+ 
V_{ex} 
f \left( t \right) 
\mathbb{V}_1 \left( k \right) 
\right\} 
\frac{1}{2} \left\{ 
\hat{b}^{\dagger}_1 \left( \boldsymbol{k} \right) 
\hat{b}_1 \left( \boldsymbol{k} \right) 
+ 
\hat{b}_1 \left( - \boldsymbol{k} \right) 
\hat{b}^{\dagger}_1 \left( - \boldsymbol{k} \right) 
\right\} 
+ 
V_{ex} f \left( t \right) 
\mathbb{V}_2 \left( k \right) 
\left\{ 
\hat{b}^{\dagger}_1 \left( \boldsymbol{k} \right) 
\hat{b}^{\dagger}_1 \left( - \boldsymbol{k} \right) 
+ 
\hat{b}_1 \left( \boldsymbol{k} \right) 
\hat{b}_1 \left( - \boldsymbol{k} \right) 
\right\} 
\nonumber\\
& = & 
\left\lbrack 
\hbar \omega_1 \left( k \right) 
+ 
V_{ex} 
f \left( t \right) 
\mathbb{V}_1 \left( k \right) 
\right\rbrack 
\hat{K}_z \left( \boldsymbol{k} \right) 
+ 
V_{ex} f \left( t \right) 
\mathbb{V}_2 \left( k \right) 
\left\{ 
\hat{K}_{+} \left( \boldsymbol{k} \right) 
+ 
\hat{K}_{-} \left( \boldsymbol{k} \right) 
\right\} 
, 
\end{eqnarray}
which leads to 

\begin{eqnarray}
&& 
\frac{\partial \left\{ 2 F_1 \left( k, \tilde{t} \right) \right\}}{\partial \tilde{t}}  
= 
2 \tilde{g}_1 \left( k, \tilde{t} \right) 
- 4 \tilde{g}_2 \left( k, \tilde{t} \right) 
\sin \left\{ 
2 F_1 \left( k, \tilde{t} \right) 
\right\} 
\tanh \left\{ 2 F_2 \left( k, \tilde{t} \right) \right\} 
, \quad 
\frac{\partial \left\{ 2 F_2 \left( k, \tilde{t} \right) \right\}}{\partial \tilde{t}}  
= 
4 \tilde{g}_2 \left( k, \tilde{t} \right) 
\cos \left\{ 
2 F_1 \left( k, \tilde{t} \right) 
\right\} 
, 
\nonumber\\
&& 
\frac{\partial \left\{ 2 F_3 \left( k, \tilde{t} \right) \right\}}{\partial \tilde{t}}  
= 
- 4 \tilde{g}_2 \left( k, \tilde{t} \right) 
\sin \left\{ 
2 F_1 \left( k, \tilde{t} \right) 
\right\} 
/ \cosh \left\{ 2 F_2 \left( k, \tilde{t} \right) \right\} 
, 
\label{F1F2F3_diffeqs}
\end{eqnarray}
and

\begin{eqnarray}
\frac{\partial}{\partial \tilde{t}}  
\frac{\partial \left\{ 2 F_1 \left( k, \tilde{t} \right) \right\}}{\partial \tilde{V}_{ex}}  
& = & 
2 \frac{\partial \tilde{g}_1 \left( k, \tilde{t} \right)}{\partial \tilde{V}_{ex}} 
- 
4 \frac{\partial \tilde{g}_2 \left( k, \tilde{t} \right)}{\partial \tilde{V}_{ex}} 
\sin \left\{ 
2 F_1 \left( k, \tilde{t} \right) 
\right\} 
\tanh \left\{ 2 F_2 \left( k, \tilde{t} \right) \right\} 
\nonumber\\
&& \quad 
- 
4 \tilde{g}_2 \left( k, \tilde{t} \right) 
\cos \left\{ 
2 F_1 \left( k, \tilde{t} \right) 
\right\} 
\tanh \left\{ 2 F_2 \left( k, \tilde{t} \right) \right\} 
\frac{\partial \left\{ 2 F_1 \left( k, \tilde{t} \right) \right\}}{\partial \tilde{V}_{ex}}  
- 
4 \tilde{g}_2 \left( k, \tilde{t} \right) 
\frac{\sin \left\{ 
2 F_1 \left( k, \tilde{t} \right) 
\right\}}
{\cosh^2 \left\{ 
2 F_2 \left( k, \tilde{t} \right) 
\right\}} 
\frac{\partial \left\{ 2 F_2 \left( k, \tilde{t} \right) \right\}}{\partial \tilde{V}_{ex}}  
, 
\nonumber\\
\frac{\partial}{\partial \tilde{t}}  
\frac{\partial \left\{ 2 F_2 \left( k, \tilde{t} \right) \right\}}{\partial \tilde{V}_{ex}} 
& = & 
4 \frac{\partial \tilde{g}_2 \left( k, \tilde{t} \right)}{\partial \tilde{V}_{ex}} 
\cos \left\{ 
2 F_1 \left( k, \tilde{t} \right) 
\right\} 
- 
4 \tilde{g}_2 \left( k, \tilde{t} \right) 
\sin \left\{ 
2 F_1 \left( k, \tilde{t} \right) 
\right\} 
\frac{\partial \left\{ 2 F_1 \left( k, \tilde{t} \right) \right\}}{\partial \tilde{V}_{ex}}  
, 
\nonumber\\
\frac{\partial}{\partial \tilde{t}}  
\frac{\partial \left\{ 2 F_3 \left( k, \tilde{t} \right) \right\}}{\partial \tilde{V}_{ex}} 
& = & 
- 4 \frac{\partial \tilde{g}_2 \left( k, \tilde{t} \right)}{\partial \tilde{V}_{ex}} 
\frac{\sin \left\{ 
2 F_1 \left( k, \tilde{t} \right) 
\right\}}
{\cosh \left\{ 2 F_2 \left( k, \tilde{t} \right) \right\}} 
- 
4 \tilde{g}_2 \left( k, \tilde{t} \right) 
\frac{\cos \left\{ 
2 F_1 \left( k, \tilde{t} \right) 
\right\}}
{\cosh \left\{ 2 F_2 \left( k, \tilde{t} \right) \right\}} 
\frac{\partial \left\{ 2 F_1 \left( k, \tilde{t} \right) \right\}}{\partial \tilde{V}_{ex}}  
\nonumber\\
&& \quad 
+ 
4 \tilde{g}_2 \left( k, \tilde{t} \right) 
\sin \left\{ 
2 F_1 \left( k, \tilde{t} \right) 
\right\} 
\frac{\tanh \left\{ 2 F_2 \left( k, \tilde{t} \right) \right\}}
{\cosh \left\{ 2 F_2 \left( k, \tilde{t} \right) \right\}} 
\frac{\partial \left\{ 2 F_2 \left( k, \tilde{t} \right) \right\}}{\partial \tilde{V}_{ex}}  
, 
\end{eqnarray}
where 
$\tilde{t} \coloneqq g_a n t / \hbar$, 
$\tilde{\omega}_1 \left( k \right) \coloneqq \hbar \omega_1 \left( k \right) / \left( g_a n \right)$, 
$\tilde{V}_{ex} \coloneqq V_{ex} / \left( g_a n \right)$, 
$\displaystyle 
\tilde{g}_1 \left( k, \tilde{t} \right) 
\coloneqq 
\tilde{\omega}_1 \left( k \right) 
+ 
\tilde{V}_{ex} 
f \left( \tilde{t} \right) 
\mathbb{V}_1 \left( k \right) 
$, and 
$\displaystyle 
\tilde{g}_2 \left( k, t \right) 
\coloneqq 
\tilde{V}_{ex} 
f \left( \tilde{t} \right) 
\mathbb{V}_2 \left( k \right) 
$.

Note that the operator $\hat{\boldsymbol{X}} \left( \boldsymbol{k} \right)$ in~\cite{Fabre2012} satisfies
\begin{eqnarray}
&& \!\!\!\!\!\!\!\!\!\!\!\!\!\!\!\!\!\!\!\!\!\!\!\!
\left\{ 
\prod_{\boldsymbol{k}_1 \neq 0} 
e^{i F_3 \left( k_1, t \right) 
\hat{K}_y \left( \boldsymbol{k}_1 \right)} 
e^{i F_2 \left( k_1, t \right) 
\hat{K}_x \left( \boldsymbol{k}_1 \right)} 
e^{i F_1 \left( k_1, t \right) 
\hat{K}_z \left( \boldsymbol{k}_1 \right)} 
\right\} 
\hat{\boldsymbol{X}} \left( \boldsymbol{k} \right) 
\left\{ 
\prod_{\boldsymbol{k}_2 \neq 0} 
e^{- i F_1 \left( k_2, t \right) 
\hat{K}_z \left( \boldsymbol{k}_2 \right)} 
e^{- i F_2 \left( k_2, t \right) 
\hat{K}_x \left( \boldsymbol{k}_2 \right)} 
e^{- i F_3 \left( k_2, t \right) 
\hat{K}_y \left( \boldsymbol{k}_2 \right)} 
\right\} 
= 
\Lambda \left( k, t \right) 
\hat{\boldsymbol{X}} \left( \boldsymbol{k} \right) 
, 
\nonumber\\
\label{X_Udagg_U}
\end{eqnarray}
which gives 
$\displaystyle 
\bar{\boldsymbol{X}} \left( \boldsymbol{k} \right) 
\coloneqq 
\left\langle 
\hat{\boldsymbol{X}} \left( \boldsymbol{k} \right) 
\right\rangle 
= 0$ for our state 
$\left\vert 
\Psi_{S} \left( t \right) 
\right\rangle 
\coloneqq 
\hat{U}_{S} \left( t \right) \left\vert \textrm{vac} \right\rangle$. 
Here,

\begin{eqnarray}
\Lambda \left( k, t \right) 
\coloneqq 
\left\lbrack 
\begin{array}{cccc} 
\Lambda_{1, 1} \left( k, t \right) 
& 
\Lambda_{1, 2} \left( k, t \right) 
& 
\Lambda_{1, 3} \left( k, t \right) 
& 
\Lambda_{1, 4} \left( k, t \right) 
\\
- \Lambda_{1, 2} \left( k, t \right) 
& 
\Lambda_{1, 1} \left( k, t \right) 
& 
\Lambda_{1, 4} \left( k, t \right) 
& 
- \Lambda_{1, 3} \left( k, t \right) 
\\
\Lambda_{1, 3} \left( k, t \right) 
& 
\Lambda_{1, 4} \left( k, t \right) 
& 
\Lambda_{1, 1} \left( k, t \right) 
& 
\Lambda_{1, 2} \left( k, t \right) 
\\
\Lambda_{1, 4} \left( k, t \right) 
& 
- \Lambda_{1, 3} \left( k, t \right) 
& 
- \Lambda_{1, 2} \left( k, t \right) 
& 
\Lambda_{1, 1} \left( k, t \right) 
\label{Lambda_matrix_def}
\end{array}
\right\rbrack 
, 
\end{eqnarray}
where

\begin{eqnarray}
&& \!\!\!\!\!\!\!\!\!\!\!\!\!\!\!\!
\Lambda_{1, 1} \left( k, t \right) 
= 
\cos \left\{ 
F_1 \left( k, t \right) 
\right\} 
\cosh \left\{ 
F_2 \left( k, t \right) 
\right\} 
\cosh \left\{ 
F_3 \left( k, t \right) 
\right\} 
+ 
\sin \left\{ 
F_1 \left( k, t \right) 
\right\} 
\sinh \left\{ 
F_2 \left( k, t \right) 
\right\} 
\sinh \left\{ 
F_3 \left( k, t \right) 
\right\} 
, 
\nonumber\\
&& \!\!\!\!\!\!\!\!\!\!\!\!\!\!\!\!
\Lambda_{1, 2} \left( k, t \right) 
= 
\sin \left\{ 
F_1 \left( k, t \right) 
\right\} 
\cosh \left\{ 
F_2 \left( k, t \right) 
\right\} 
\cosh \left\{ 
F_3 \left( k, t \right) 
\right\} 
- 
\cos \left\{ 
F_1 \left( k, t \right) 
\right\} 
\sinh \left\{ 
F_2 \left( k, t \right) 
\right\} 
\sinh \left\{ 
F_3 \left( k, t \right) 
\right\} 
, 
\nonumber\\
&& \!\!\!\!\!\!\!\!\!\!\!\!\!\!\!\!
\Lambda_{1, 3} \left( k, t \right) 
= 
- \left\lbrack 
\cos \left\{ 
F_1 \left( k, t \right) 
\right\} 
\cosh \left\{ 
F_2 \left( k, t \right) 
\right\} 
\sinh \left\{ 
F_3 \left( k, t \right) 
\right\} 
+ 
\sin \left\{ 
F_1 \left( k, t \right) 
\right\} 
\sinh \left\{ 
F_2 \left( k, t \right) 
\right\} 
\cosh \left\{ 
F_3 \left( k, t \right) 
\right\} 
\right\rbrack 
, 
\nonumber\\
&& \!\!\!\!\!\!\!\!\!\!\!\!\!\!\!\!
\Lambda_{1, 4} \left( k, t \right) 
= 
\sin \left\{ 
F_1 \left( k, t \right) 
\right\} 
\cosh \left\{ 
F_2 \left( k, t \right) 
\right\} 
\sinh \left\{ 
F_3 \left( k, t \right) 
\right\} 
- 
\cos \left\{ 
F_1 \left( k, t \right) 
\right\} 
\sinh \left\{ 
F_2 \left( k, t \right) 
\right\} 
\cosh \left\{ 
F_3 \left( k, t \right) 
\right\} 
. \qquad \qquad 
\label{Lambda_jl_def}
\end{eqnarray}

Hence, from~\cite{Fabre2012}, the formal expression of the quantum Fisher information $I_{Q, \rm exact} \left( \tilde{V}_{ex}, t \right)$ is 
\begin{eqnarray}
I_{Q, \rm exact} \left( \tilde{V}_{ex}, t \right) 
= 
\frac{1}{4} 
\sum_{\boldsymbol{k} \neq 0} 
\textrm{Tr} \left\lbrack 
\left\{ 
\frac{\partial \Gamma \left( k, t \right)}{\partial \tilde{V}_{ex}} 
\Gamma^{-1} \left( k, t \right) 
\right\}^2 
\right\rbrack 
\quad \textrm{where } \; 
\Gamma \left( k, t \right) = 
\Lambda \left( k, t \right) 
\Lambda^{T} \left( k, t \right)
. 
\label{QFI_symplectic}
\end{eqnarray}

Now, let 
\begin{eqnarray}
\mathbb{B}_2 \left( k, t \right) 
& \coloneqq & 
- 
\frac{1}{2} 
\mathbb{M}_2 \left( k \right) 
\left\lbrack 
\cosh \left\{ 
2 F_2 \left( k, \tilde{t} \right) 
\right\} 
\sinh \left\{ 
2 F_3 \left( k, \tilde{t} \right) 
\right\} 
+ 
i \sinh \left\{ 
2 F_2 \left( k, \tilde{t} \right) 
\right\} 
\right\rbrack 
\nonumber\\
&& \quad 
+ 
\mathbb{M}_3 \left( k \right)
\left\lbrack 
\begin{array}{c}
\sin \left\{ 
2 F_1 \left( k, t \right) 
\right\} 
\sinh \left\{ 2 F_2 \left( k, t \right) \right\} 
\sinh \left\{ 
2 F_3 \left( k, \tilde{t} \right) 
\right\} 
\qquad \qquad \qquad \qquad \qquad \quad 
\\
+ 
\cos \left\{ 
2 F_1 \left( k, t \right) 
\right\} 
\cosh \left\{ 
2 F_3 \left( k, t \right) 
\right\} 
+ 
i \sin \left\{ 
2 F_1 \left( k, t \right) 
\right\} 
\cosh \left\{ 2 F_2 \left( k, t \right) \right\} 
\end{array}
\!\!\!\!
\right\rbrack 
, 
\end{eqnarray}
and

\begin{eqnarray}
\mathbb{B}_3 \left( k, t \right) 
& \coloneqq & 
\mathbb{M}_2 \left( k \right) 
\cosh \left\{ 
2 F_2 \left( k, \tilde{t} \right) 
\right\} 
\cosh \left\{ 
2 F_3 \left( k, \tilde{t} \right) 
\right\} 
\nonumber\\
&& \; 
- 
2 \mathbb{M}_3 \left( k \right) 
\left\lbrack 
\sin \left\{ 
2 F_1 \left( k, t \right) 
\right\} 
\sinh \left\{ 2 F_2 \left( k, t \right) \right\} 
\cosh \left\{ 
2 F_3 \left( k, \tilde{t} \right) 
\right\} 
+ 
\cos \left\{ 
2 F_1 \left( k, t \right) 
\right\} 
\sinh \left\{ 
2 F_3 \left( k, t \right) 
\right\} 
\right\rbrack 
. \qquad 
\end{eqnarray}
Then we get

\begin{eqnarray}
&& 
\left\langle 
\Psi_{S} \left( t \right) 
\right\vert 
\hat{M}_S 
\left\vert 
\Psi_{S} \left( t \right) 
\right\rangle 
= 
\sum_{\boldsymbol{k} \neq 0} 
\left\{ 
\mathbb{M}_1 \left( k \right) 
- 
\frac{1}{2} 
\mathbb{M}_2 \left( k \right) 
+ 
\frac{1}{2} 
\mathbb{B}_3 \left( k, t \right) 
\right\} 
, \nonumber\\
&&
\left\langle 
\Psi_{S} \left( t \right) 
\right\vert 
\hat{M}^2_S 
\left\vert 
\Psi_{S} \left( t \right) 
\right\rangle 
= 
\left\{ 
\left\langle 
\Psi_{S} \left( t \right) 
\right\vert 
\hat{M}_S 
\left\vert 
\Psi_{S} \left( t \right) 
\right\rangle 
\right\}^2 
+ 
2 \sum_{\boldsymbol{k} \neq 0} 
\left\vert 
\mathbb{B}_2 \left( k, t \right) 
\right\vert^2 
, 
\qquad \qquad 
\end{eqnarray}
and the formal expression of the lower bound of classical Fisher information $I_{C, \rm exact} \left( \tilde{V}_{ex}, t \right)$ is 
\begin{eqnarray}
I_{C, \rm exact} \left( \tilde{V}_{ex}, t \right)
= 
\frac{1}{\displaystyle 
8 \sum_{\boldsymbol{k} \neq 0} 
\left\vert 
\mathbb{B}_2 \left( k, t \right) 
\right\vert^2} 
\left\{ 
\sum_{\boldsymbol{k} \neq 0} 
\frac{\partial \mathbb{B}_3 \left( k, t \right)}{\partial \tilde{V}_{ex}} 
\right\}^2 
. 
\label{CFI_symplectic}
\end{eqnarray}

\end{widetext}
\end{document}